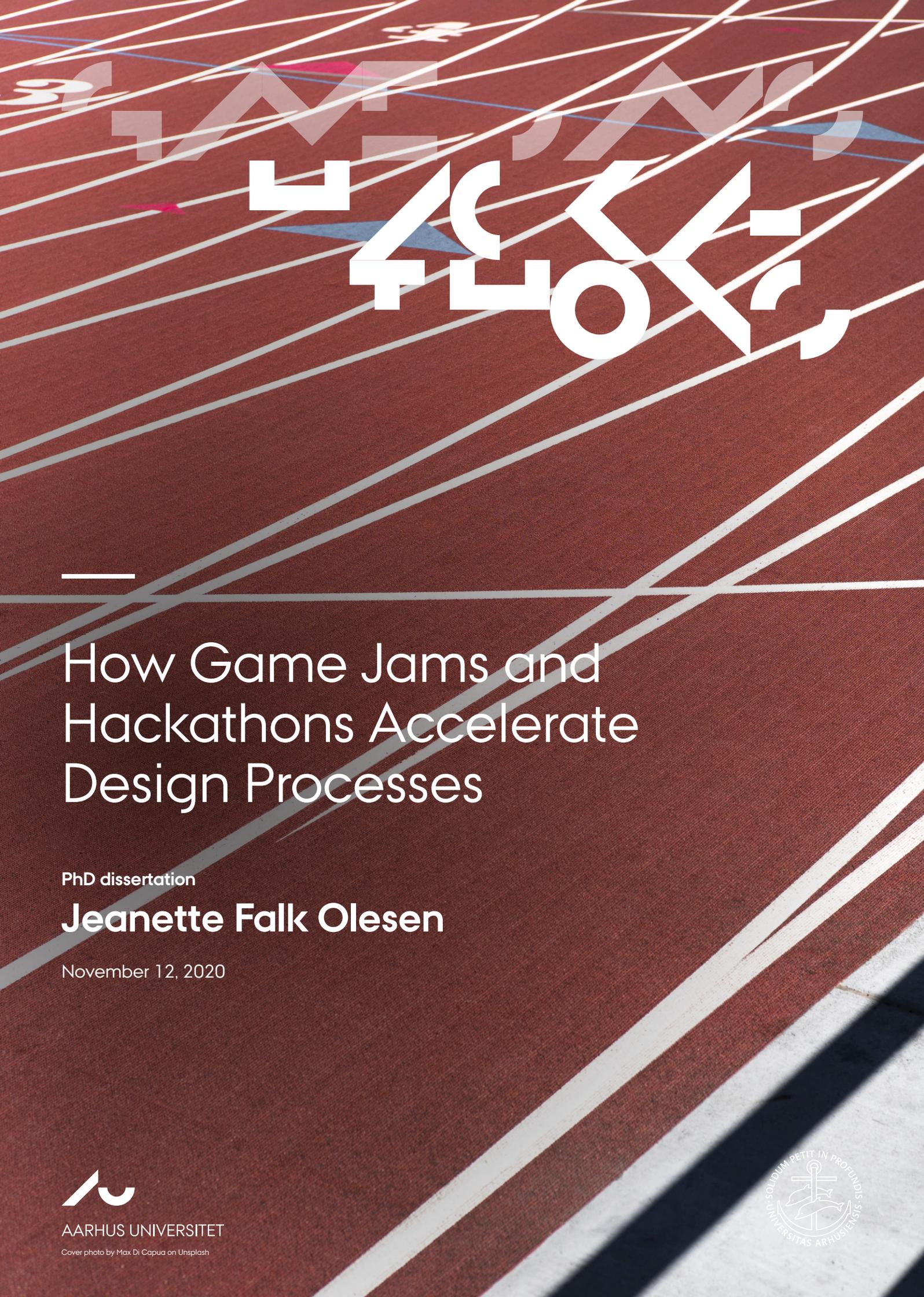

# How Game Jams and Hackathons Accelerate Design Processes

**PhD dissertation**

## Jeanette Falk Olesen

November 12, 2020

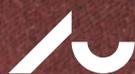

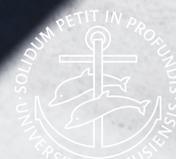

"...we found a problem between the backend and the frontend where they can't communicate, because of some internet security basically...And it's not good for us that it's safe. It's good for the rest of the world."
(P2, Saturday, status update, T:23:25:34)

# Contents

# How Game Jams and Hackathons Accelerate Design Processes

# ACKNOWLEDGEMENTS | 0

Thanks to Kim Halskov, who has been everything one could wish for in a supervisor, and thanks for teaching me how to navigate academia in a pragmatic and down–to–earth way.

Thanks to Rikke Toft Nørgård, my co-supervisor, for your inspiring mentorship, collaboration, and support.

A thank you to everyone who was part of the CIBIS group, for the discussions, feedback, inspiration, and collaborations.

Thanks to my great colleagues at Aarhus University, who I have learned so much from, including, but not limited to: Asnath Paula Kambunga, Banu Saatçi, Bjarke Vognstrup Fog, Christian Dindler, Christian Remy, Claus Toft-Nielsen, Clemens Nylandsted Klokmose, Eva Eriksson, Germán Leiva, Gökçe Elif Baykal, Jonas Frich Pedersen, Kasper Skov Christensen, Lindsay MacDonald Vermeulen, Lone Koefoed Hansen, Maarten Pieter P Van Mechelen, Michael Mose Biskjær, Midas Nouwens, Nicolai Brodersen Hansen, Peter Dalsgaard, Peter Lyle, Susanne Bødker, and Søren Rasmussen. And to Nanna Inie Strømberg-Derczynski for the emotional support, pomodoros, and great music taste!

Thank you to everyone who was part of the PhD group Kosagruppa for Rasende Unge Forskere (VrUF). Especially to: Jiyoung Kim, Joeb Høfdinghoff Grønborg, Raune Frankjær, and Rui Xu.

Thank you to Annakaisa Kultima, for hosting my visit at Aalto University, for great discussions and collaborations.

Thank you to my family and friends for always being there.

*This PhD project has been fully funded by the Faculty of Arts, Aarhus University.*

# ABSTRACT | 0

This dissertation presents three years of research on how design processes in game jams and hackathons can be understood as *accelerated*. Hackathons and game jams can both be described as formats where participants engage in designing and developing prototypes during an intentionally short time frame, such as 48 hours, which is meant to facilitate creativity, and encourage fast decision making and rapid prototyping. The main difference between hackathons and game jams is that participants in hackathons typically engage with general interaction design, while participants in game jams typically engage with game design.

Game jams and hackathons are organised in many different contexts and for many different purposes as well, such as: internally in companies to spark new ideas; for fostering citizen innovation for municipalities; in cultural and governmental agencies; integral parts of education; entry points for developers wanting to enter especially the game industry (Falk Olesen and Halskov 2020; Kultima 2015a). During the recent decade, game jams and hackathons have been introduced to academia as well, as formats for teaching and learning, and as research platforms as well.

Only few research contributions engage with understanding how accelerated design processes in game jams and hackathons unfold, or how the organisation of game jam and hackathon formats influence these accelerated design processes. A proposition of my PhD project is that a meticulously developed understanding of how game jam and hackathon formats accelerate design processes can provide a base for advancing how the formats are adapted and organised for different purposes. During my PhD project I have therefore addressed the following three research questions:

**RQ1: How can we understand the accelerated design processes during hackathons and game jams?**

**RQ2: How have people organised hackathons and game jams in academia and for supporting creativity?**

**RQ3: How may we explore alternative ways of organising hackathons and game jams for academia and for supporting creativity?**

In order to address the research questions I have gathered a large amount of data consisting of: case studies; participant observations; in situ visual, textual, and auditory documentation; survey responses; interviews; and literature reviews. The data is the result of using a range of the following primarily exploratory and qualitative methods: autobiographical design case studies; an intervention with an exploratory prototype; survey; and an extensive literature review. I have primarily used pragmatist design theory, drawing on Schön's work, to scaffold the analysis of the conducted studies. This approach enabled me to contribute to advancing how accelerated design processes can be understood, and furthermore explore how game jams and hackathons may be organised in academia and for supporting creativity.

The main contributions of my PhD project are:

▶ Descriptive process-level knowledge, which contextualise and solidify how accelerated design processes unfold under the circumstances of a game jam and a hackathon.
▶ Overviews of how game jams have been organised for supporting participants' creativity and of how hackathons have been used as means and as research focus within academia.
▶ Exploring how game jam and hackathon formats may be organised in order to support knowledge generation such as within academia, and in order to support creativity.

The dissertation consists of this overview article, which includes eight chapters, and five appended research papers, where four are peer-reviewed and published, and one is submitted for review. The purpose of this article is to provide an overview of my PhD project, and relate the contributions of the five papers to the three research questions.

# RESUME | 0

Denne PhD afhandling præsenterer tre års forskning af hvordan de-signprocesser i game jams og hackathons kan forstås som værende accelererede. Hackathons og game jams kan begge beskrives som for-mater hvor deltagere engagerer sig i at designe og udvikle prototyper under en intentionel kort tidsramme, såsom 48 timer, som bruges til at facilitere kreativitet, og opfordre til hurtige beslutningstagninger og prototyping. Hovedforskellen mellem hackathons og game jams er at delt-agere i hackathons typisk beskæftiger sig med generel interaction design, mens deltagere i game jams typisk beskæftiger sig med spildesign.

Game jams og hackathons bliver organiseret indenfor mange forskel-lige kontekster, og med mange forskellige formål, såsom: internt i virk-somheder for at skabe nye idéer; til at støtte borgerinnovation for offentlige organisationer; i kulturelle og statslige instanser; som integrerede dele af uddannelse; som entrypoint for udviklere som ønsker en karriere særligt indenfor spilindustrien. I løbet af det seneste årti, er game jams og hackathons også blevet introduceret til den akademiske verden, som formater for undervisning og læring, samt som forskningsplatforme.

Kun få forskningsbidrag beskæftiger sig med forståelsen af hvordan accelererede designprocesser i game jams og hackathons udfolder sig, eller med hvordan organiseringen af game jam og hackathon formater påvirker disse accelererede designprocesser. En antagelse for mit PhD projekt er at en omhyggeligt udviklet forståelse af hvordan game jam og hackathon formater accelererer designprocesser kan skabe en base for at udvikle hvordan formaterne adapteres og organiseres i forhold til forskellige formål. I løbet af mit PhD projekt har jeg derfor beskæftiget mig med følgende tre forskningsspørgsmål:

**RQ1: Hvordan kan vi forstå den accelererede designproces i hackathons og game jams?**

**RQ2: Hvordan har man før organiseret hackathons og game jams i den akademiske verden og for at understøtte kreativitet?**

**RQ3: Hvordan kan vi udforske alternative måder hvormed hackathons og game jams kan organiseres indenfor den akademiske verden og for at understøtte kreativitet?**

For at tilgå forskningsspørgsmålene har jeg samlet en større mængde data bestående af: case studier; deltagerobservationer; in situ visuel, tekstbaseret og auditiv dokumentation; spørgeskemabesvarelser; inter-views; og litteratur reviews. Dataene er resultatet af en række af følgende primært eksplorative og kvalitative metoder: autobiografiske design case studier; en intervention med en eksplorativ prototype; spørgeskemaun-dersøgelse; og et omfattende litteratur review. Jeg har anvendt pragmatisk designteori, hvor jeg trækker på Schön's værk, for at understøtte analysen af de udførte studier. Denne tilgang muliggjorde at jeg kunne bidrage til at avancere hvordan accelererede designprocesser kan forstås, og ydermere udforske hvordan game jams og hackathon kan organiseres indenfor den akademiske verden og for at understøtte kreativitet.

Hovedbidragene for mit PhD projekt er:

- ▶ Deskriptiv procesniveau viden, som kontekstualiserer hvordan accelererede designprocesser udfoldes under forholdene i et game jam og et hackathon.
- ▶ Overblik over hvordan game jams har været organiseret for at understøtte deltageres kreativitet, og over hvordan hackathons har være brugt som forskningsmiddel og -fokus i den akademiske verden.
- ▶ Udforskning af hvordan game jams og hackathons kan organiseres for at understøtte vidensgenerering, såsom indenfor den akademiske verden, og for at understøtte kreativitet.

Afhandlingen består af denne overbliksartikel, som inkluderer otte kapitler, og fem vedhæftede forskningsartikler, hvor fire er fagfællebedømt og en er indsendt til bedømmelse. Formålet med denne artikel er at give et overblik af mit PhD projekt, og relatere bidragene fra de fem artikler til de tre forskningsspørgsmål.

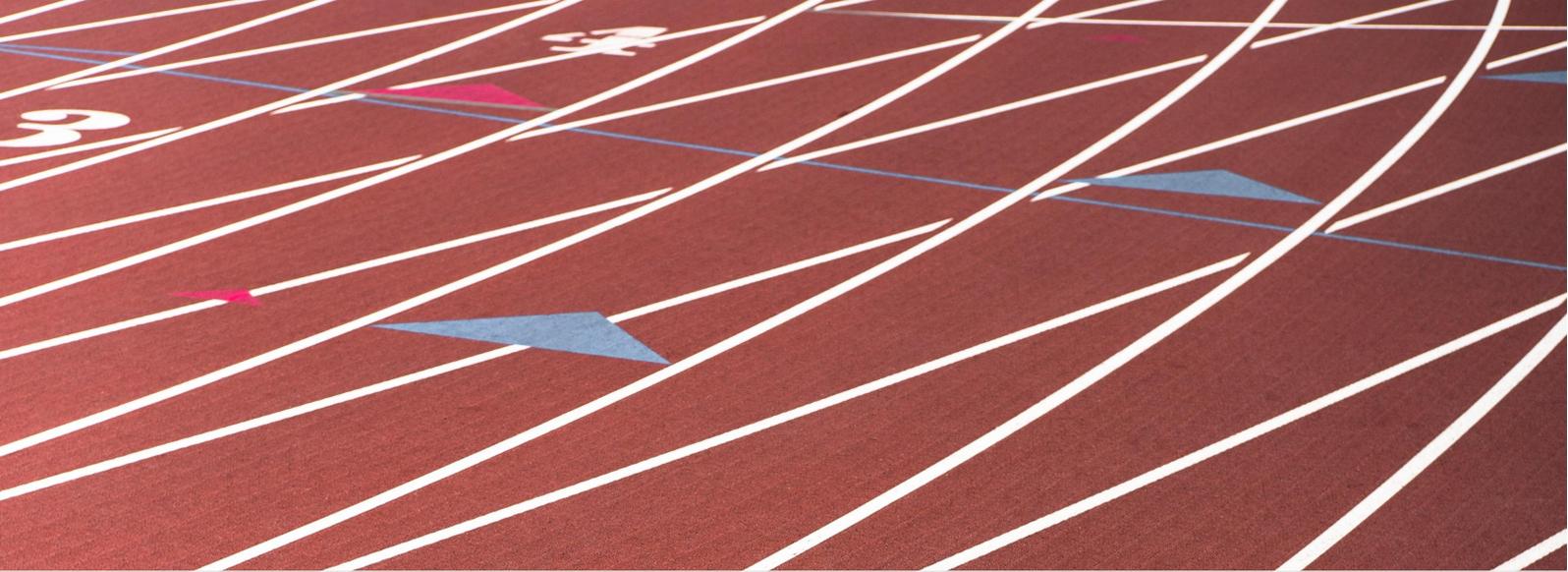

# 1 INTRODUCTION

This dissertation presents three years of research, with the goal of improving the understanding of *accelerated design processes*. I use the term *accelerated design processes* to denote the particular kind of design process which unfolds in *hackathons* and *game jams*, which are the focus of my PhD project. Hackathons and game jams have gained a significant role, however it is mainly during the recent decade that we have begun researching these hackathons and game jams. The dissertation contributes to a growing body of research which study accelerated design processes, and addresses the following research questions:

**RQ1: How can we understand the accelerated design processes during hackathons and game jams?**

**RQ2: How have people organised hackathons and game jams in academia and for supporting creativity?**

**RQ3: How may we explore alternative ways of organising hackathons and game jams for academia and for supporting creativity?**

In addressing the three research questions, I aim at advancing the understanding of how game jam and hackathon formats accelerate design processes in particular ways.[1] I am furthermore focusing on how game jam and hackathon formats are organised and can be organised for purposes in academia and for supporting creativity. The *proposition* of my PhD project is that a profound and meticulously developed understanding of how game jam and hackathon formats accelerate design processes, provide a base for better adapting and utilising these formats for particular purposes. The research questions are motivated by the observation that only a few research contributions take on a processual focus on the inner workings of the design processes in hackathons and game jams, despite an increasing research interest into them. My PhD project approach this gap in research by operating with a focus on the accelerated design *processes* and how these are shaped by the *formats*.

Based on a comprehensive amount of data in the form of: case studies, participant observations, in situ visual, textual and auditory documenta-

[1]: I understand format in this context, as the *particular setup* of a game jam or a hackathon, such as the chosen theme for a game jam, selected cases at a hackathon, or the location. In the paper (Olesen, Hansen, and Halskov 2018), we describe format as encompassing for example: The imposed time frame, case selections, themes, tools and materials provided, and judging criteria.

tion, surveys, interviews, and an extensive literature review, my research has resulted in five papers (four peer-reviewed and published papers and one paper to be submitted for review) which together contribute to addressing the three research questions.

I unpack the motivation for the focus on academia and creativity in the next section where I define and describe game jams and hackathons, and the overarching umbrella-term *accelerated design processes*. Subsequently, I present a brief overview of how I have approached the research questions and present the main contributions of the PhD project. The chapter concludes with a description of the dissertation structure.

## 1.1  Presentation of Key Terms

In this section, I present the key terms of the PhD project, since an early common understanding of the terms is essential for the overview article. First, I present working definitions for the two selected accelerated design processes; *hackathons* and *game jams*. Then I present the umbrella-term *accelerated design processes*.

### Hackathons

I understand hackathons in the same way as Lodato and DiSalvo describe them: "[...] rapid design and development events at which volunteer participants come together to conceptualize, prototype, and make (mostly digital) products and services [...] The structure of most hackathons is similar: they occur over the span of [a] day or two, challenges are presented to participants, teams form around these challenges, the teams engage in a fervor of activity to produce solutions of (varying completeness)[sic], and at the end of the event, the teams present their work, in some cases judges are brought in and awards are given." (Lodato and DiSalvo 2016).

### Game Jams

Based on an analysis of selected academic papers discussing game jams, Kultima synthesised how the papers conceptualise *game jams*, and during my PhD project I have ascribed to her definition: "A game jam is an accelerated opportunistic game creation event where a game is created in a relatively short timeframe exploring given design constraint(s) and end results are shared publically." (Kultima 2015a).

### Similarities

The definition of game jams is similar to how hackathons can be described, and format-wise hackathons and game jams can be very much alike. An immediate aspect which game jams and hackathons share, is that they both impose a relatively short and predetermined **time frame** on participants which is meant to *accelerate* the design process, since participants are encouraged to develop a functioning prototype within that time frame. Both hackathons and game jams can be attended by a wide range of people, such as: programmers, graphic artists, audio designers, designers, writers, students, professionals, amateurs, and researchers. Oftentimes, both hackathons and game jams can be open for anyone who are just curious to participate.

To support the argument that game jams and hackathons have gained a significant role today, and provide the reader with an idea of what these formats can result in, I now provide some examples on commercial successes which resulted from a game jam or a hackathon. For example, Facebook often run 43-hours hackathons for its employees, who must hack on something that is not part of their day-to-day job (Chang 2012). The "Like-button" is one example of a successful hackathon outcome (Chang 2012).[2] Some games started in game jams and have since been developed into commercial successes as well. For example, the popular game Fortnite is the result of an internal game jam at the game company Epic Games (McWhertor 2014). Other award winning games which began their development in game jams are the indie game Celeste (Hudgins 2016), Goat Simulator (Ibrisagic 2014), and Surgeon Simulator (Rose 2013).[3]

Game jams and hackathons share several aspects besides being similar format-wise and some hackathons and game jams resulting in commercial products: Both game jams and hackathons have been introduced to academia, and for example hackathons have been widely embraced in academia for teaching, learning and research (Falk Olesen and Halskov 2020). The major annual game jam *Global Game Jam* (GGJ) even have their own research committee, where online surveys can be distributed to thousands of participants (*Global Game Jam* 2020). Furthermore, both hackathons and game jams have been widely commended for facilitating creativity. However, there is little research into the relationship between accelerated design process formats and creativity. In the following two sections, I motivate the focus on game jams and hackathons in academia and for supporting creativity in my research.

**Game Jams and Hackathons in Academia**

Hackathons are today used as both a means and subject for research, as well as an integral part of education (Olesen, Hansen, and Halskov 2018; Falk Olesen and Halskov 2020). Since hackathons are widely available, open for the public and are often attended by big crowds, they can be used as research platforms, where researchers can participate in, observe and even facilitate their own hackathons (Falk Olesen and Halskov 2020). Hackathons are interesting for research since they are being used in a range of different contexts such as in companies to spark new ideas, for fostering citizen innovation for municipalities, and in cultural and governmental agencies to name a few (Falk Olesen and Halskov 2020). Game jams have likewise been used in education, for scientific purposes, and as research platforms (Kultima 2015a). Like hackathons, game jams are also used in a range of different contexts and for different purposes, such as entry points for aspiring game developers wanting to enter the game industry, as internal creativity boosts for game companies, and for articulating sensitive social issues (Kultima 2015a). As such, the contexts in which hackathons and game jams are employed are quite similar, and both have been embraced within academia. Though hackathons and game jams can be used as means to research topics in these different contexts,

2: Some of these hackathon projects have big impacts: "There have been hackathon projects that have changed the direction of the company [Facebook] in terms of what we focus on, and how we think about our strategy, and what we realize is possible[...] The thesis is, if 100 people have an idea that could be awesome – the next big thing – try all of them. 99 of them are going to be terrible, but there's going to be that one idea that is going to be awesome." (Chang 2012).

3: For example the premise of Goat Simulator might at first glance seem like a silly idea, but it however turned out to be very successful: "There were a lot of great ideas going around at our office for game jam games. A lot of them were deep and complex games. I wanted to go in another direction - something totally stupid and not serious at all. After all, creativity and outside-of-the-box games was the purpose of the game jam." (Ibrisagic 2014). As of 2015 the game had sold 2.5 million copies (Minotti 2015) and as of 2016 the game had made $12 million in revenue (Wawro 2016).

there is not much research on the proccessual aspects of hackathons and game jams, such as how design processes unfold during hackathon and game jam formats, or how design decisions are influenced by the particularity of a game jam or hackathon format.

## Game Jams and Hackathons for Creativity

Game developers generally perceive game jams as events which *foster creativity* (Guevara-Villalobos 2011). Preston et al. ascribed this quality of fostering creativity as a characteristic of game jams by: "[...] inducing creativity through constraint" (Preston et al. 2012). Organisers of game jams also emphasise this characteristic of game jams, and find it important that participants get to be creative and that organisers *support* participants' creativity (Falk et al. 2021). In order to promote their game jams, organisers often mention the game jams as inducing creativity. For example, one of the first game jams encouraged participation in this way: "[...] Come design something weird and new in a fun, creative, and collaborative atmosphere! Who knows what people will come up with?" (Hecker 2002). Recently, the Global Game Jam encouraged people to participate in order to: "[...] contribute to this global spread of game development and creativity." (*Global Game Jam* 2020). Some game companies also facilitate game jams for their employees; DoubleFine is one example on a game company which regularly have two-week game jams internally (Wallace 2017).[4] Hackathons are likewise described as facilitating creativity (Briscoe and Mulligan 2014; Taylor and L. Clarke 2018) as well as innovation (Soltani et al. 2014). Hackathons are for example used to facilitate participants' creativity in education (Brookes 2018; Nandi and Mandernach 2016), and within companies to harvest: "[...] untapped creativity in their own employee pool" (Taylor and L. Clarke 2018). Companies such as Facebook, Google, Yahoo! (Trainer et al. 2016), Microsoft (Whitney-Morris 2018), and Hasbro (Elias 2014) frequently run hackathons for their employees. There are, however, not much research on *creativity* in neither game jams (Falk et al. 2021) or hackathons (Falk Olesen and Halskov 2020).

## Inclusive Definitions

As I have argued and described so far, game jams and hackathons share several aspects. In this section I discuss how I during my PhD project then have approached hackathons and game jams. Throughout the PhD project I have not sought to set strict boundaries as to what entails a hackathon or a game jam, and the working definitions above serve to introduce how game jams and hackathons commonly are described. Rather, I have aimed to describe in detail the particular hackathon or game jam which I have studied. When describing game jams *in general* in the PhD project, I use Kultima's definition (Kultima 2015a). However, Kultima herself points out that the aspects which constitute game jams are constantly changing, as game jams are still a relatively new and still developing phenomenon. As such, what constitute a game jam is a "moving target". To my knowledge, there have not been a similar effort to Kultima's to synthesise a definition of *hackathons*, however I will argue that hackathons are also a "moving target" in terms of strictly defining what aspects constitute them.

4: Founder of DoubleFine, Tim Schafer, explains why the company run game jams internally: "Schafer also noted that beyond being a creative refresher, the game jam also serves as a sort of two-week meditation, in which the team can re-focus and, when it comes down to it, have some fun. 'It helps you to kind of de-stress,' Schafer admitted. 'You're like, 'Well, this is only for two weeks, so even if we come up with nothing, even if it's just a lark, at least we had this break, this vacation.' So it does all these positive creative things to your brain; it's an exercise in raw, free-form creativity, relative to these giant triple-A games." (Wallace 2017).

To complicate matters further, aspects, which are typically ascribed to game jams, can be found in hackathons and vice versa. There are for example hackathons which encourage a *playful* approach to the design process, an aspect usually ascribed to game jams (Grace 2016), such as the Stupid Hackathon (Lavigne and Winger-Bearskin 2020). There are also examples of game jams, which specifically seek to create games that can potentially create positive social impacts by solving real-world problems, an aspect often ascribed to hackathons (Falk Olesen and Halskov 2020), such as the collaboration between GGJ and Games for Change: "GGJ has proven itself to be an incubator for innovative experiments and even new startups and products. We want to support a growing interest in game design to tackle and solve real-world problems through their creative work[...] GGJ is the place to test new ideas, collaborate, experiment and innovate." (Alhadeff 2013).

As part of my PhD project, we reviewed 381 papers mentioning the term "hackathon", however we did not develop a definition based on the review (Falk Olesen and Halskov 2020).[5] On the contrary, we aimed for an inclusive understanding of hackathons. In practice, this meant that if an event in a paper was labelled with the term "hackathon", we included the paper in the review and then inquired *how* the event was described and conceptualised on the authors' premise.

5: Whenever I write the pronoun "we" I refer to my co-authors of the pertinent paper and myself.

The boundaries between hackathons and game jams are hence not clear cut. I argue that a more fruitful approach to framing game jams and hackathons is taking into account how game jam and hackathon formats are organised, which community they address (for example a community which is concerned with game making or *making* in general), and how the community itself name and conceptualise the format. Furthermore, I argue that though hackathons and game jams differs in some aspects, such as origin and communities, the study of *accelerated design processes* can benefit from studying both hackathons and game jams since both accelerate design processes because of several shared aspects, such as the intentional short time frames, and limited resources intended to accelerate the creative design processes. Findings from studying for example hackathons can also potentially benefit the research, organisation of and participation in game jams and vice versa. In my PhD project the focus is therefore on how the findings from studying *both* game jams and hackathons generate insights on the *shared* aspect of accelerating design processes.

## Accelerated Design Processes

Game jams have been described as "compressed development processes" (Zook and Riedl 2013), because participants generally go through many different steps of game development (Kultima 2015a) such as: collaboration, ideation, and designing, developing, playtesting, and polishing a functional prototype, which is often demonstrated in the end. A similar design process occurs in hackathons, though in hackathons the design process is typically characterised by a broader *interaction design* perspective in contrast to game jams' *game design* perspective. In my PhD project I approach the way in which people design and develop prototypes in game jams and hackathons as particular and similar kinds of *design processes*.

On a conceptual level, Löwgren and Stolterman define design processes as beginning when: "[. . . ] the initial ideas concerning a possible future take shape. The process goes on all the way to a complete and final specification that can function as a basis for construction or production. In some cases, the final specification is identical to the final product. We do not distinguish between processes that lead to construction of new technology and processes that lead to the composition of an artifact by assembling readymade components or configuring an off-the-shelf product. In both cases, the work constitutes a design process." (Löwgren and Erik Stolterman 2004). Their definition applies to design processes in game jams and hackathons as well. Though, as I exemplified earlier, some hackathon and game jam outcomes have gone through production, most outcomes stay at the construction or prototyping phase. Löwgren and Stolterman further use Donald Schön's pragmatist design perspective to describe how the design process can be seen as a form of *conversation* between a designer and a design situation (Löwgren and Erik Stolterman 2004; Schön 1987). In Chapter 5, I elaborate on the notion of design processes as informed by pragmatist design theory. Framing the processes in hackathons and game jams as *design processes* grounds the study of them within a well-established field which offers a detailed vocabulary to study the dynamic transformations which takes places in these particular formats.

In order to emphasise the particular character of design processes in game jams and hackathons, I use the umbrella-term *accelerated design processes*. Using the term *accelerated* to describe these kinds of design processes is inspired by how game jams have been described as *accelerated game development* (Turner, Thomas, and Owen 2013) and "[...] *accelerated opportunistic game creation event[s]*" (Kultima 2015a). Another inspiration for the term, is the notion of *creativity constraints*, which informs a part of the theoretical foundation for my PhD project, which I elaborate on in Chapter 5 (Biskjaer and Halskov 2014). All design processes entail certain constraints and it is furthermore not a new concept that constraints *accelerate* design processes. Biskjaer and Halskov elaborate that constraints accelerate a design process: "[...] by pushing it forward in the form of an expected leap." (Biskjaer and Halskov 2014). I use the term *accelerated design processes* to emphasise that hackathons and game jams are exceptionally accelerated, and particularly the time-frame speeds up ideation, design decisions and prototyping. The dual role of creativity constraints in the form of the time-frame becomes clear in accelerated design processes: It prompts some design decisions to be made over others, since it is imperative that an appropriate outcome, which may illustrate aspects of an overarching idea, is developed within the time frame.

Applying the notion of design processes as *accelerated* must necessarily entail that some design processes are *not accelerated*. I argue that *accelerated* design processes are discerned from *non-accelerated* design processes in the sense that accelerated design processes, such as those occurring in hackathons and game jams, are *intentionally* accelerated in order to promote creativity among participants. This means that an intentionally short time-frame is imposed as a *constraint* in order to actively progress the creative design process, hence prompting fast decision making and rapid prototyping. In non-accelerated design processes there may of course also be a defined time frame, even a short one. However, in those cases I argue that a time frame most likely is decided by other factors than an intention to accelerate a creative design process. Other factors deciding the time frame in non-accelerated design processes could instead be: a budget with limited resources, or a client or stakeholder who decide on a time frame. In non-accelerated design processes creativity can be related to the time frame, however it is not the main reason for how long or short the time frame is.

## 1.2 Approach

I have operated with two levels of research questions during my PhD project: The first level consists of five research questions which are addressed by a corresponding paper. The second level is the overarching three research questions of the PhD project. The contributions to the first level of the five research questions are related to and contribute to answering the second level of research questions of the PhD project. In Chapter 2 I present an overview of each of the five appended papers and a summary of the main contributions pertaining to each paper's research question. Chapter 7 elaborates on the accumulated contributions of the five papers and I relate them to the three overarching research questions of my PhD project. In this section I briefly summarise how I have approached the three overarching research questions and the main contributions.

**RQ1: How can we understand the accelerated design processes during hackathons and game jams?**

In addressing RQ1, we engaged with the inner workings of the accelerated design process in a game jam and a hackathon from a first-person perspective by conducting *autobiographical design case studies*. The accelerated design processes were documented meticulously in order to capture in situ knowledge and to support later analysis.

The contribution is *descriptive processual knowledge* which supports how we can understand the *accelerated design processes* in game jams and hackathons. Particularly, I identify and analyse how *four specific events* during a game jam were contributing to transforming the *design space* (Olesen and Halskov 2018). I furthermore identify and analyse how, in particular, *four factors* in a hackathon were contributing to how *design judgements* and thereby *design decisions* were made (Olesen, Hansen, and Halskov 2018). Together, the four events and the four factors describe *particularities* of how the game jam and hackathon format had a *profound influence* on how the accelerated design processes unfolded. Because the two studies are autobiographical design case studies, the findings are not generalisable, however, the two studies serve as rich *examples* which *contextualise* and *solidify* how accelerated design processes might unfold under the circumstances of a game jam and a hackathon.

**RQ2: How have people organised game jams and hackathons in academia and for creativity?**

We conducted an *extensive literature review* of 381 papers of how researchers have utilised hackathons as research platforms and as research subjects (Falk Olesen and Halskov 2020). The contribution is *an overview of key examples* that demonstrate the many different ways research has been conducted with and on hackathons. We *identify three overarching motivations* for employing hackathon formats as part of research: Structuring learning, structuring processes, and enabling participation, We furthermore *synthesise the benefits and challenges* researchers have reported in conducting research *with* and *on* hackathons.

We conducted a *deductive thematic analysis* of survey responses of 27 game jam organisers (Falk et al. 2021). The contribution is insights into how game jam organisers *understand* and *promote participants' creativity* in game jams. We conducted this study as a *first step* into researching the relationship between game jams and creativity. The motivation for studying the perspective of game jam organisers is because their role entails potentially significant influence on how participants' creativity may be organised for.

The two studies provided a base for a broader understanding of how people have organised and utilised game jam and hackathon formats before. The two studies further informed my answer to RQ3:

**RQ3: How may we explore alternative ways of organising hackathons and game jams for academia and for supporting creativity?**

In approaching RQ3 we conducted a form of *intervention*, or an in-the-wild study (Rogers 2012), in an academic game jam with 25 participants in order to explore how a real-time annotation tool can support the *documentation* of accelerated design processes (Rasmussen, Olesen, and Halskov 2019). After conducting the two autobiographical studies (Olesen and Halskov 2018; Olesen, Hansen, and Halskov 2018), I was interested in further exploring a method to document an accelerated design process in order to leverage *in situ* knowledge and insights for later analysis. Documenting accelerated design processes can be challenging due to the pace, but the real time annotation tool, Co-notate, had potential for documenting in a non-intrusive way which can be important when the design process is accelerated. The contribution is the *identification and analysis of benefits and challenges* of using a real time annotations tool for leveraging in situ knowledge in a fast-paced game jam (Rasmussen, Olesen, and Halskov 2019).

My findings for RQ3 is furthermore informed by the findings for RQ2: The findings of how researchers and game jam organisers have organised and employed hackathons and game jams (respectively) provided a *base* for at the same time *discussing* how formats for game jams and hackathons then *may be organised* in academia and for supporting creativity. Specifically, we discuss how the overview of exemplars and the accumulated benefits and challenges of using hackathons in research can serve as a *point of departure* for informing future and *more systematic* research *with* and *on* hackathons (Falk Olesen and Halskov 2020). In analysing deductively how game jam organisers understand and promote participants' creativity, we discuss how the organisers might *advance their practices* with insights based on recent creativity research (Falk et al. 2021). Furthermore, we discuss directions for future research on the relationship between creativity and game jam formats based on insights from our findings.

## 1.3 Structure of the Dissertation

The dissertation is constituted by this overview article, consisting of 8 chapters, and five appended research papers. *Details* regarding data and methods of the PhD project are presented and discussed in each paper, which furthermore present and discuss individual contributions pertaining to the conducted study of the paper. This overview article sums up the five contributions of the five papers and connects contributions of the papers to the overarching three research questions of my PhD project. The main purpose of the article is to provide an overview of the PhD project as a whole based on the contributions of the five papers.

Following this Introduction, Chapter 1, the overview article consists of the following chapters:

Chapter 2 on the following page is an overview of the appended five papers, where each paper is summarised in regards to main data, method and contribution.

In Chapter 3 on page 22 I describe the research environment and field which have informed the theoretical foundation as well as the methodology of the PhD project.

In Chapter 4 on page 25 I outline related works which serve as the context for the research conducted during my PhD project.

In Chapter 5 on page 37 I outline the theoretical foundation of the PhD project, which consist of pragmatist design research, and the theoretical notion of creativity constraints.

In Chapter 6 on page 48 I discuss the methodology, which is informed by the research context and the theoretical foundation.

In Chapter 7 on page 61 I summarise the contributions of the five papers and discuss how they answer the three research questions. I furthermore discuss the impact of the findings as well as future work.

In Chapter 8 on page 91 I conclude the dissertation.

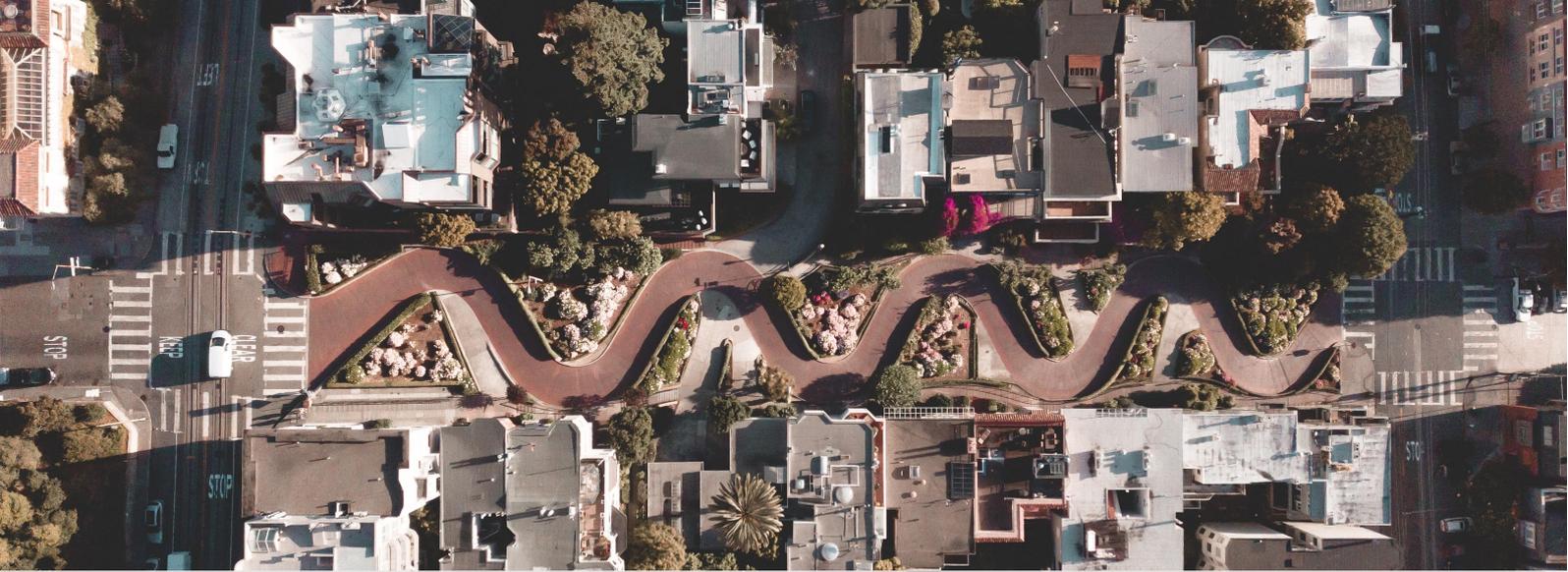

# 2  OVERVIEW OF APPENDED PAPERS

The five papers of my PhD project all contribute with addressing the overall research questions:

**RQ1: How can we understand the accelerated design processes during hackathons and game jams?**

**RQ2: How have people organised hackathons and game jams in academia and for supporting creativity?**

**RQ3: How may we explore alternative ways of organising hackathons and game jams for academia and for supporting creativity?**

This chapter presents an overview of the appended papers which I have chosen to include in the dissertation.[1]  Table 2.1 presents the papers in a chronological order and shows that there has been a progression in the PhD project from: first, researching game jam and hackathon formats from a "bottom-up" participant perspective; over exploring how a real-time annotations tool may leverage situational knowledge gained in a game jam; to a "top-down" perspective where I explore how game jam organisers understand creativity in game jams and how researchers use hackathons in their processes.

For each paper in Table 2.1 I sum up title, my co-authors, the research question of the paper, and main contributions. In Chapter 7 I elaborate on how the contributions of the papers address the three overarching research questions of the PhD project. In the remaining parts of this overview article I refer to each paper by number, so when I refer to paper 1, I refer to the paper "The Dynamic Design Space During a Game Jam", and so on. In the next section I describe how I approached the research question pertaining to each paper and describe the main contributions of the paper.

1: For a full overview of my publications, including some supplementary publications written during my PhD project, see: https://tinyurl.com/qtrz5y4.

---

Photo by Will Truettner on Unsplash

**Table 2.1:** Chronological overview of selected appended papers.

| | Research question | Main contributions |
| --- | --- | --- |
| **Paper 1: The Dynamic Design Space During a Game Jam** Jeanette Falk Olesen and Kim Halskov | How can the complex *inner workings* of a game jam be captured and analysed as a dynamic transformation of a conceptual, constraint-based *design space*? | Autobiographical design case study contributing with detailed, process-level knowledge on how a design process was shaped by a game jam format, and how this can be analysed by using the notion of design spaces |
| **Paper 2: Four Factors Informing Design Judgement at a Hackathon** Jeanette Falk Olesen, Nicolai Brodersen Hansen, and Kim Halskov | How is *design judgement* influenced and driven by the hackathon format? | Autobiographical design case study contributing with detailed, process-level insights on how design decisions in a hackathon is shaped by and focused on the situation of the hackathon, and not on the context beyond the hackathon itself and what implications this may have |
| **Paper 3: Co-notate: Exploring Real-time Annotations to Capture Situational Design Knowledge** Søren Rasmussen, Jeanette Falk Olesen, and Kim Halskov | How can *situational knowledge* in a game jam be *captured for later analysis*? | In-the-wild study which explores how a real time annotation tool can capture immediate responses to design activities in situ, in order to inform subsequent analysis |
| **Paper 4: How Organisers Understand and Promote Participants' Creativity in Game Jams** Jeanette Falk Olesen, Michael Mose Biskjær, Annakaisa Kultima, and Kim Halskov | How do game jam organisers *understand creativity*, and how do and can they *support participants' creativity*? | Maps out and analyses how game jam organisers currently organise game jams and understand creativity in relation to game jams. Discusses how game jam organisers can advance their practices with insights from creativity research |
| **Paper 5: 10 Years of Research With and On Hackathons** Jeanette Falk Olesen, and Kim Halskov | How do researchers engage with hackathons, and what kinds of *challenges and benefits* do hackathons entail? | An extensive and detailed review of how researchers have engaged with hackathons. Discusses some directions for future incorporation of hackathons in research processes |

## 2.1 Appended Papers

### Paper 1

#### The Dynamic Design Space During a Game Jam

*Jeanette Falk Olesen and Kim Halskov. 2018. In Proceedings of the 22nd International Academic Mindtrek Conference (Mindtrek '18). Association for Computing Machinery, Tampere, Finland, 30–38.*

This paper presents an autobiographical design case study of a single game jam team, where I participated in the team as both the observing researcher and the level designer with the responsibility of implementing assets into Unity 3D, a game engine. The study was conducted during the Nordic Game Jam 2016, a 48 hour game jam with the theme "Leak". Building on the notion of *constraint-based design spaces* (Biskjaer, Dalsgaard, and Halskov 2014), I used design space schemas to meticulously document how certain factors during the game jam influenced our accelerated design process. Scaffolding the analysis with the vocabulary of design spaces allow for detailed process-level insights on how the design process was shaped and transformed by the game jam format. The contribution are detailed, and process-level descriptive accounts of the temporal transformation of the design space. We outline four events which in particular had a transforming effect on the design space:

1. Establishing the design space
2. Elaborating the design space
3. Inquiry into gameplay options
4. Breakdown of movement and gameplay

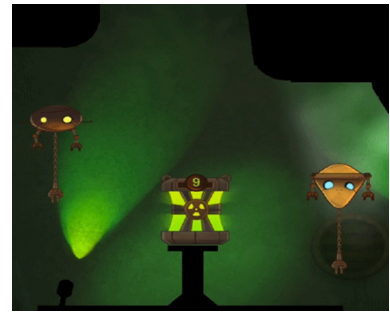

**Figure 2.1:** A screenshot from the final game, Cobots, developed during the autobiographical design case study in paper 1. Our final game prototype can be accessed online: `https://bit.ly/3fZzGSL`. Figure published in (Olesen and Halskov 2018).

### Paper 2

#### Four Factors Informing Design Judgement at a Hackathon

*Jeanette Falk Olesen, Nicolai Brodersen Hansen and Kim Halskov. 2018. In Proceedings of OzCHI'18. ACM, Melbourne, Australia, 473–483.*

Based on the method in paper 1, this paper presents an autobiographical case study of a single team in a hackathon. I, together with my coauthor Nicolai Brodersen Hansen, participated in the team as both the observing researchers and contributing designers. The study was conducted during the hackathon, AU Hack 2017, which lasted 36 hours. Throughout the hackathon we, the researchers, would prompt and record frequent status updates, as a way of documenting cross sections of the unfolding, accelerated design process. Our team addressed the case given by the customer experience agency Creuna (Creuna 2020): "How can we enrich the offline shopping experience by taking all the best from online retail and new technologies and bringing it to life in stores?" During the study, we developed a functioning prototype, see Figure 2.2. Similarly to paper 1, the goal with the study was to provide detailed, process-level insights into how an accelerated design process in a hackathon setting is shaped. Still based on the notion of constraint-based design spaces, the paper identifies how particularly four factors during the hackathon shaped our design judgement, and thereby which design decisions were

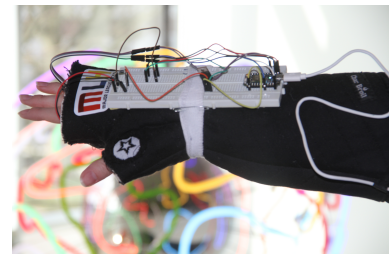

**Figure 2.2:** During the autobiographical design case study in paper 2, we developed a prototype intended to warn for example lactose intolerant users if they were reaching for a product containing lactose. A description of our final prototype can be read here: `https://bit.ly/2WC7ttC`. Figure published in (Olesen, Hansen, and Halskov 2018).

made. We identified four factors which in particular informed design judgements:

1. **The hackathon format:** Frames and imposes certain elements to the design process. Prompts prioritising of resources.
2. **Tools and materials**: Played a central role in idea generation, since the design process depends on what tools and materials are available.
3. **Domain knowledge:** A main driver for idea generation and user insights. Sought internally and highly dependent on the composition of team members.
4. **Technical knowledge:** The team depend on their internal technical knowledge and what is actually realisable within the team.

Though the four factors may be evident in other design processes, the factors prompts an inward oriented focus for the design judgement on the situation of the hackathon itself and not on long-term implications beyond the hackathon. Despite the focus and vision of the design process in the hackathon in this sense is very short-sighted, we argue that the hackathons have value in their potential to exemplify the situatedness and messiness of design situations while engaging people in creative technology development.

## Paper 3

### Co-notate: Exploring Real-time Annotations to Capture Situational Design Knowledge

*Søren Rasmussen, Jeanette Falk Olesen, and Kim Halskov. 2019. In Proceedings of the 2019 on Designing Interactive Systems Conference (DIS '19). ACM, San Diego, California, USA 161–172*

This paper presents the findings of an intervention, or an in-the-wild study, with a real time annotation tool, Co-notate, which enables users to annotate video and audio recordings in real-time during a design process. The tool enables multiple users in a design process to use a range of user- and predefined tags for annotating recordings in real time, see also Figure 2.3. The annotations are automatically synchronised with the video or audio recordings. Co-notate is organised around three phases: 1) preparation of tags, 2) recording and annotating activities with the predefined tags, and 3) analysing the annotated recordings. For a more detailed description of the functionality and user interface of Co-notate, I refer to paper 3.

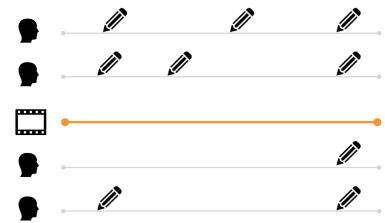

**Figure 2.3:** The picture illustrates how multiple people can annotate the same recording in real time during an activity. Figure published in (Rasmussen, Olesen, and Halskov 2019).

Co-notate seemed promising as a way to document accelerated design processes unobtrusively, which can be particularly important in order to avoid slowing down the pace of the process and thereby ensure authenticity in the documentation of the process. Previous research has shown that participants in game jams report that the participation contribute to their learning and acquiring of skills (Preston et al. 2012; Steinke et al. 2016). We therefore organised a 48 hour game jam as part of a university summer school course. The game jam played an important part in the university course and can be described as an *academic game jam*, since the participants were required to reflect on the course syllabus during the game jam, as they had to motivate their design decisions based on theory from the syllabus. This setting allowed us to study the prototype in an accelerated design process which emphasised practice-based knowledge generation. For the analysis, we focused on the participants' experience with using Co-notate and we identified the following challenges and potentials of using real-time annotations for documenting and reflecting on design activities during a game jam:

▶ **Finding appropriate tags:** A challenge is defining tags that are specific enough to gain insights from the design process in retrospective analysis and general enough to be applicable in the situational flow.

▶ **Supporting capture of immediate impressions:** A potential for real-time annotation during design processes is expressive rather than descriptive annotation, which indicates that something of immediate importance was noted and deemed relevant for subsequent reflection.

▶ **Distributing responsibility:** Though Co-notate enables simultaneous annotating performed on multiple device, having a single shared device for tagging during the design process may provide a sense of distributed and shared responsibility for tagging.

▶ **Instantly searchable documentation:** Though Co-notate is designed with a focus on retrospective analysis, the instantly searchable annotation enables a channel for inspiration and argumentation done in the immediate situation while the design process is going on.

▶ **Supporting retrospective analysis:** A key potential of Co-notate is the invitation for participants to capture in situ cursory reflection and transfer some of this situational reflection to retrospective analysis.

## Paper 4

### How Organisers Understand and Promote Participants' Creativity in Game Jams

*Jeanette Falk Olesen, Michael Mose Biskjær, Kim Halskov and Annakaisa Kultima. Manuscript. In Proceedings of the Sixth Annual International Conference on Game Jams, Hackathons, and Game Creation Events (ICGJ'21). ACM, online*

Though game jams have been commended for potentially facilitating participants' creativity, an aspect often ascribed to the short time frame, not much research have been conducted on the relation between game jam formats and creativity Furthermore, to our knowledge, no research has been conducted on the relation between game jams and creativity from the game jam organiser's perspective. In this paper, we invited 27 game jam organisers to answer an online survey revolving around how they understood creativity in relation to game jams. We furthermore asked them whether they supported participants' creativity during the game jams, and what concrete initiatives they took to support participants' creativity. Since the organisers' answers were quite brief, perhaps because their understanding of creativity builds on colloquial knowledge, we conducted a deductive thematic analysis. We selected four key aspects from creativity research as propositions for the analysis: *Novelty*, *risk-taking*, *combinational creativity*, and *creativity constraints*. According to recent creativity research, we assessed that these four aspects capture prominent and essential themes for creativity, while we acknowledge that the four aspects do not capture all parts of creativity as a phenomenon. Choosing a deductive analytical approach enabled a comparison of whether and how the organisers' answers reflected the four aspects. We identified five kinds of initiatives where the organisers in different ways sought to support participants' creativity:

▶ Establishing the physical surroundings
▶ Supplying tools and materials
▶ Selecting a theme
▶ Providing talks and facilitating discussion
▶ Organising activities

The paper takes the first steps towards discussing how organisers of game jams can leverage knowledge from creativity research and advance how they support participants' creativity. Specifically, we discuss how organisers can:

▶ Induce novelty
▶ Encourage and reward risk-taking
▶ Identify and utilise creativity constraints
▶ Apply combinational creativity

## Paper 5

### 10 Years of Research With and On Hackathons

*Jeanette Falk Olesen, and Kim Halskov. 2020 In Proceedings of the 2020 on Designing Interactive Systems Conference (DIS '20). ACM, Eindhoven, Netherlands*

Acknowledging that hackathons have been widely embraced by academia, we wanted to map how researchers have engaged with hackathons. We reviewed an extensive sample of 381 publications from the ACM Digital Library published during a 10 year time span. Aiming for breadth over depth, the contribution of the paper is an overview of the various ways researchers have used hackathons as means for their research, or even as the focus for their research. We identified three motivations for using hackathons as part of research: structuring learning, structuring processes, and enabling participation. We furthermore identified two different kinds of hackathon research: research *with* hackathons and research *on* hackathons. In the paper we include several key examples from the reviewed papers in order to demonstrate the three motivations and the two kinds of research. Another contribution of the paper is a condensed overview of the salient benefits and challenges of hackathons, which we identified in the reviewed papers. We argue that the paper can serve as a stepping stone for further advancing the use and discussion of hackathons for research purposes.

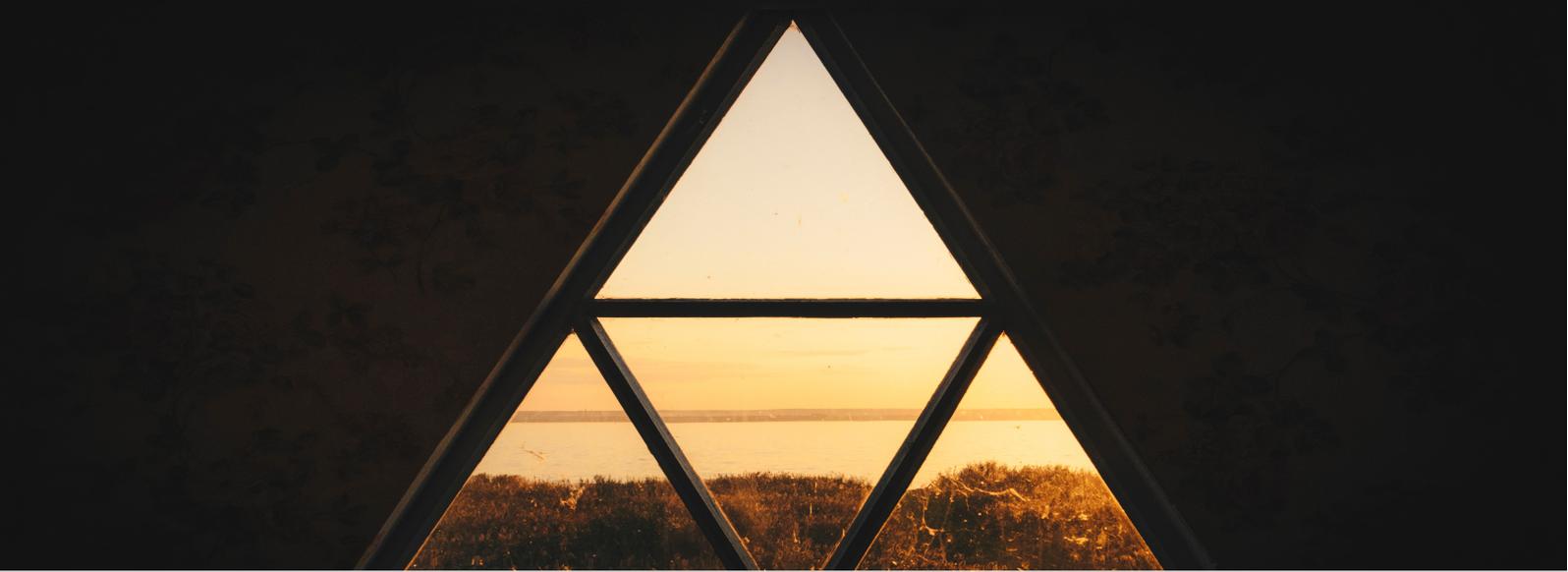

# 3   RESEARCH ENVIRONMENT & FIELD

In this chapter, I first outline the research environment which has constituted the context of my PhD project. Secondly, I outline the research field which I base my research work within. Together, the research environment and field both informed how I built the theoretical foundation, described in Chapter 5, and how I developed the methodology of the PhD project, described in Chapter 6.

## 3.1  Research Environment

In this section I outline the research environment which I was part of at Aarhus University. The research environment was constituted by the Department of Digital Design and Information Studies (Dept. 2020), and the following four research centres, group, and project: The Human-Computer-Interaction group (HCI 2020), the Participatory Information Technology Center (PIT 2020), the Creativity in Blended Interaction Spaces project (CIBIS 2020), and Centre for Digital Creativity (CDC 2020). This particular research environment has a strong design research profile with historical ties to the development of participatory design.

The research environment enabled access to opportunities for inspiration, feedback, discussions, and collaborations, which all informed and advanced my PhD project. In the following I outline the focus points for the different centres, group, and project. The HCI group at Aarhus University has since the 1980's maintained a strong focus on users and how to empower them through design processes (HCI 2020). The PIT Center's mission is to: "[. . .] develop the foundation for understanding alternative forms of thinking and supporting participation through IT [. . .]" (PIT 2020). The objectives for the CIBIS project was to: 1) "[. . .] demonstrate the potential for integrating multiple digital devices and analog materials in a shared environment, to support individual and group creativity." and 2) "[. . .] develop the theoretical foundation for the study of constraints on creativity, design ideas, generative design materials, and creative methods in design processes." (CIBIS 2020). The Centre for Digital Creativity explores how: "[...] digital technologies support, enable, and transform creative activities." and is constituted by projects which focus on how: "[...] people come together to create new knowledge and products, and how can interactive systems scaffold these processes." (CDC 2020). Summarising, the research environment has a strong focus on design processes, and on supporting participation and creativity. A consequence of being part of this research environment, is that the theoretical foundation of my PhD project in general is informed by theories on design processes and creativity.

## 3.2 Research Field

As the above mentioned description of my research environment indicate, my research has been conducted in the research fields of HCI and Interaction Design. Positioning ones research as either being predominantly within HCI or Interaction Design is not a straightforward or binary task, as the fields are highly diverse and intertwined, both internally and mutually. In the following section, I describe how I frame and understand this complex relation between HCI and Interaction Design.

### HCI and Interaction Design

Traditionally, the HCI field had a narrower focus than Interaction Design and was concerned with: "[...] the design evaluation, and implementation of interactive computing systems for human use and with the study of major phenomena surrounding them." (Preece, Rogers, et al. 2015). However, today HCI is a highly diverse and intertwined field as demonstrated by Liu et al. who conducted a co-word analysis of 3152 papers published at the flagship HCI conference, SIGCHI Conference on Human Factors in Computing Systems (CHI) (Liu et al. 2014). For example, "design" and "interaction design" was hence only two out of a total of 28 identified popular keywords for papers published during 2004-2013.[1]

Despite HCI emerging as a highly diverse research field, Fallman argued how HCI can be described as a design-*oriented* research field, since researchers are frequently involved in for example designing research prototypes (Fallman 2003). In regards to the field of HCI, I position the PhD project as what Fallman defines *design-oriented research*, which engages with design in order to contribute with knowledge (Fallman 2003). Here, the resulting artefact is considered: "[...] more a means than an end" (Fallman 2003). However, though design processes generally are assumed in HCI they are rarely acknowledged (Wakkary 2004) and tends to become: "[...] concealed under conservative covers of theory dependence, fieldwork data, user testing, and rigorous evaluations." (Fallman 2003). Wakkary argues for research to focus on the design process, as there is a tendency to analyse design retrospectively rather than prospectively and thereby: "[...] overinterpret for rational attributes such as logic and symmetry." (Wakkary 2004).

Interaction Design generally implies a greater focus on the practice of designing, and can be described as an interdisciplinary convergence

1: The other 26 popular keywords were: mobile phone, ubicomp, visualization, handheld devices, CMC, gestures, user studies, collaboration, privacy, CSCW, children, sustainability, ethnography, evaluation, infoviz, mobile, TUI, games, Fitt's Law, online communities, HCI4D/ICTD, augmented reality, participatory design, social networks, usability, crowdsourcing (Liu et al. 2014).

of design and HCI, including aspects of: "[. . . ] interactive art, performance, computing science, cognitive science, psychology and sociology." (Wakkary 2004). Löwgren and Stolterman defines Interaction Design as: "[...] the process that is arranged within existing resource constraints to create, shape, and decide all use-oriented qualities (structural, functional, ethical, and aesthetic) of a digital artifact for one or many clients." (Löwgren and Erik Stolterman 2004). According to Preece, Sharp and Rogers the difference between Interaction Design and other fields such as HCI are down to which: "[. . . ] methods, philosophies, and lenses they use to study, analyze, and design computers. Another way they vary is in terms of the scope and problems they address." (Preece, Rogers, et al. 2015).

With this delineation of a complex relation between two interchangeably and intertwined research fields, I will emphasise that the contributions of the papers in the dissertation do not strictly fall under either the one or the other research field. Rather, the research in my PhD project are mainly conducted where the fields of design-oriented HCI and Interaction Design overlap. This overlap is specifically where the two fields share the focus on how people design and develop technology, focusing on hackathons and game jams.

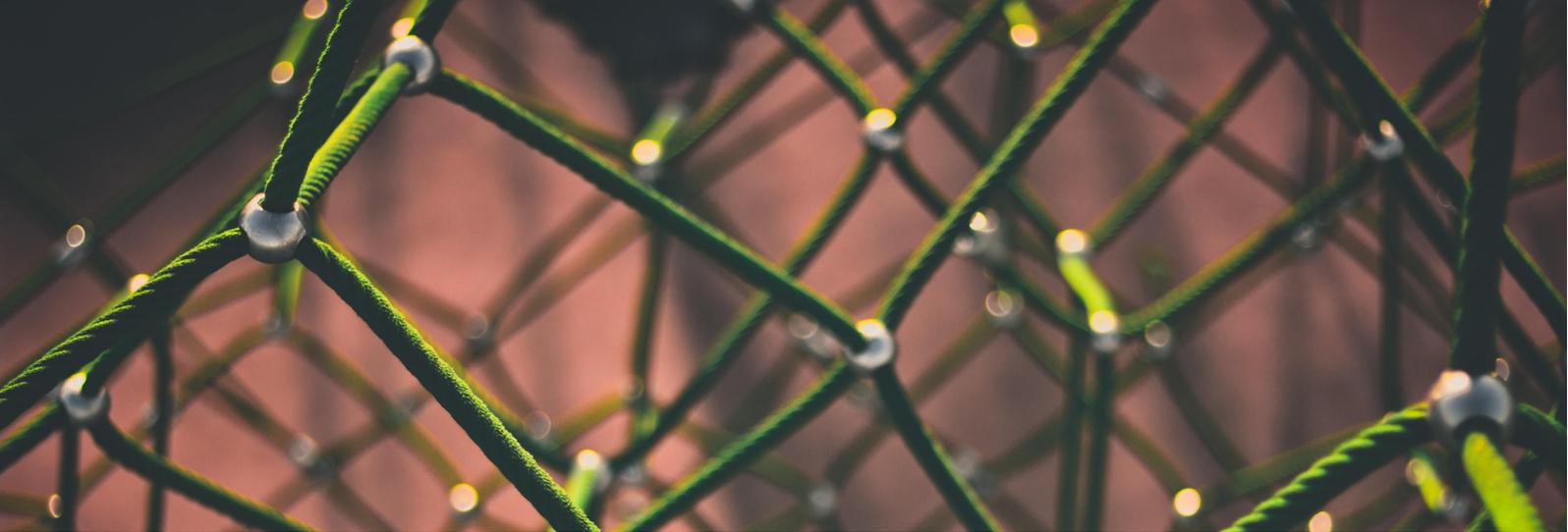

# 4   RELATED WORK

In this chapter, I first outline maker culture and how it relates to hackathons and game jams. Then, I sum up salient critique points which have been directed at maker culture, and which can be applied to hackathons and game jams as well. Next, the research field of *game studies* is outlined, since much game jam research have roots in this research field. Lastly, I outline the general research context of game jam and hackathon research.

## 4.1  Maker Culture

In Chapter 1 I introduced how I base the notion of design processes on especially Schön's pragmatist design research. I briefly return to Schön's work in this introduction, before I elaborate on the notion in Chapter 5. Several theoretical concepts in pragmatist design research build on Schön's work, who studied how *professionals* work in design processes (Schön 1987). Participants engaging in accelerated design processes in game jams and hackathons are often *non-professionals*, such as students, amateurs, hobbyists, and people curious on these particular design processes. There are, however, hackathons and game jams conducted in professional settings within companies. As mentioned in Chapter 1, Microsoft, Google, and DoubleFine are professional settings in which hackathons and game jams are conducted internally for the employees. Depending on how teams choose to work, or are required to work, in these hackathons or game jams, participants may stick to their professional competencies, or they may move out of their comfort zone, and use the hackathon or game jam as a learning opportunity, and thereby become a non-professional for a short while. In a professional context, this can be beneficial as people learn new competencies, and become acquainted with colleagues' competencies. This aspect of hackathons and game jams as enabling the development of technology for both professionals, and non-professionals resonates well with *maker culture*.

According to Halverson and Sheridan, maker culture, or the maker movement, can be defined as referring: "[. . . ] broadly to the growing number of people who are engaged in the creative production of artifacts in their daily lives and who find physical and digital forums to share their processes and products with others." (Halverson and Sheridan 2014). According to Dougherty, one reason for the maker movement's emergence is because of people's need to: "[. . . ] engage passionately with objects in ways that make them more than just consumers." (Dougherty 2012b). Dougherty has been credited with facilitating the popularity of the maker movement via his company, Maker Media, which publishes the Make magazine, and conducts annual Maker Faires around the world. He compares today's makers with the enthusiasts from the early days of the Silicon Valley computer industry, who were essentially playing with technology: "They didn't know what they wanted computers to do and they didn't have particular goals in mind. They learned by making things and taking them apart and putting them back together again, and by trying many different things." (Dougherty 2012b). Today, makers benefit from several factors, which have emerged as a happy confluence, and which enables making for a broader audience:

▶ A greater interconnectedness driven by the Internet (Dougherty 2012b)
▶ Access to cheap and powerful hardware [1]
▶ Access to digital fabrication [2]
▶ Shared software and designs (Halverson and Sheridan 2014) [3]

Together with a renewed interest in local goals and resources (Halverson and Sheridan 2014), the emergence, and accessibility of these factors contribute to the democratising aspect of the maker culture, since making of technology are no longer only possible for experts with engineering backgrounds. In Hatch's words: "The real power of this revolution is its democratizing effects. Now, almost anyone can innovate. Now almost anyone can make. Now, with the tools available at a makerspace, anyone can change the world." (Hatch 2013).

## Hackathons and Game Jams

In the rise of the maker movement, or maker culture, hackathons have been described as a common *maker* activity (Jenkins and Bogost 2015). Game jams are typically linked to communities engaging with game development, and are not often referred to explicitly as a maker activity, like hackathons. This may be because game jams are implicitly assumed as a subcategory of hackathons, and are sometimes described as a hackathon for game development (see for instance Briscoe and Mulligan 2014), or because game development in general is not perceived as a maker activity. Even though game jams are not often described explicitly as a maker activity, like hackathons often are, I describe maker culture as a relevant context related to both game jams and hackathons. A reason for connecting not only hackathons but also game jams to maker culture is because hackathons and game jams generally include the above mentioned factors, and furthermore share the ideal of disseminating the practices of making, and game making to a broader audience. In paper 1 and 2 we contextualise how some of these factors dynamically come into play as creativity constraints during respectively a game jam and a hackathon. As we discuss in paper 1 and 2, the kind of resources which are available and can be accessed within a game jam or hackathon are

1: Such as Arduino, Raspberry Pi, Makey Makey etc.

2: Such as 3D printers, laser cutters etc in maker spaces or public spaces such as libraries.

3: Thingiverse, for example, serves as a database where people can upload and download files for 3D printers and laser cutters. Software such as the game engines Unity 3D, GameMaker, and Unreal Engine has also played an important role for making game development accessible (O'Donnell 2014).

important factors which can shape both the accelerated design process and the outcome. In paper 4 we explore which tools and materials game jam organisers provide to participants, with the specific purpose of supporting their participants' creativity. As a minimum, just by having internet-access via a computer participants can collaborate with others online and access a plethora of open platforms, toolkits, open datasets, guides, and fora for how to develop software. The vignette at the end of this section includes a number of recent examples on open online game jams and hackathons, which are enabled by the minimum of an internet connection and a computer.

## Criticising Maker Culture

Despite the democratising promise of the maker culture, critique has been pointed at the maker movement. In paper 5 page 7 I summarise this critique. Pointing back at for example Hatch's description of the promising democratising effects of the maker movement, and how it can change the world, several have raised concerns of an underlying techno-solutionism of the maker movement.[4] Jenkins and Bogost even describe making as "[. . .] a self-serving meta-hobby that refers to a realworld activity but does not interact with it." (Jenkins and Bogost 2015). Despite the visionary promise that the maker movement enables almost anyone to change the world, the movement is not seen as inviting by everyone. For instance, the movement has been criticised as maintaining a "white male nerd dominance" (Halverson and Sheridan 2014). This critique of a lacking diversity is supported by a survey conducted by a collaboration between Dougherty's Make company and Intel in an attempt to provide in-depth knowledge about the maker community. The study showed that 81% are male with a median age of 44 based on a sample of Maker Faire Exhibitors, MAKE magazine subscribers and MAKE newsletter subscribers (Dougherty 2012a). Tara Whelan further investigates this critique by interviewing women and exploring barriers for the women's adoption of a maker identity, even though they engage in maker activities.[5] It is beyond this PhD project's scope to exhaustively study the issues of among other things techno-solutionism and participant diversity of game jams and hackathons, however the above outline of these critique points serves to challenge the maker movement as a purely democratising and empowering movement. As we conclude in paper 5's outline of criticism against the maker movement, there is definitely potential in the maker movement in general, and in game jams and hackathons as well, to leverage a democratisation and broader dissemination of technology development, though it is not an inherent quality that can be assumed (Kaiying and Lindtner 2016). I argue that knowing how game jam and hackathon formats shape and accelerate design processes can be a step on the way to accommodate for some of these issues, since such a knowledge can enable better organising of game jam and hackathon formats, including providing appropriate tools, materials and other constraints. As we also argue in paper 2, much research focus on *what* hackathons can do, with little research on *how* they unfold (Trainer et al. 2016). In this context, my PhD project aim at contributing with the first steps of formulating how accelerated design processes unfold, and how the formats may influence them. In Section 7.4, I return to a critical discussion of hackathons and game jams.

4: Silvia Lindtner, Shaowen Bardzell, and Jeffrey Bardzell summarise how HCI researchers are concerned with techno-solutionism in making, which they define as: "[...] the view that technology can unilaterally solve difficult social problems" (Lindtner, S. Bardzell, and J. Bardzell 2016).

5: One example from Whelan's study is a woman who did not identify herself as a maker, however her description of her own practice involve several of the factors mentioned before: "Well I make computer games currently. So I do a lot of coding, stuff like that. I like doing electronics. I did computer and electronic engineering and then for my postgrad I was doing electronic engineering mostly in telecoms. So I liked making little boards for transceivers and stuff like that. I have like, a Raspberry Pi, I have an Arduino, I like playing around with them. I like doing little bits of robotics. I like, just silly things with flashing LEDs that look stupid. I liked making flashing LED things for sticking on my Christmas tree for a long time. I knit. I like making costumes and stuff like that as well. I LARP, so there's costumes and equipment and stuff like that for that. I made my own LARP arrows because they need to be safe, so they need to have sponge and shit at the top, so you don't... impale someone. So I guess that's the shit I make."" (Whelan 2018).

**Vignette: COVID-19 and online game jams and hackathons**

In the time of writing this dissertation, the novel corona virus pandemic has caused countries all over the world to take drastic measures in order to prevent the virus from spreading. Many people are forced to distance themselves socially and even isolate themselves in their homes. In the context of this, several examples on both hackathons and game jams have emerged in order to: entertain and connect people in isolation; develop educational games about the pandemic; create data visualisations; generate ideas and design prototypes which can aid people in different ways. The following particular game jams and hackathons illustrate how not much is needed to facilitate accelerated design processes besides having access to a computer and internet connection:

- ▶ The Quarantine Game Jam: "This event is about making a game, in the timespan of three days, to keep ourselves safe, happy and productive during this weekend!" (Jam 2020). The game jam received 275 entries on itch.io.
- ▶ The game jam Combating COVID-19: Videogames that Empower Kids: "These videogames will do more than help stop the spread of COVID-19, they also will empower school-aged children. The coverage of COVID-19 can be quite frightening and potentially triggering, especially to children who are trauma-sensitive. Empowerment is the antidote to feelings of helplessness." (Empower Kids 2020).
- ▶ COVID19 Global Online Hackathon is a collaboration with the World Health Organization (WHO), and is supported by Facebook, Giphy, Microsoft, Pinterest, Slack, TikTok, Twitter and WeChat: "The COVID-19 Global Hackathon is an opportunity for developers to build software solutions that drive social impact, with the aim of tackling some of the challenges related to the current coronavirus (COVID-19) pandemic." (Lessin and Heerwagen 2020). Seven themes are suggested to spark the participants' ideas: 1) **Health:** Address and scale a range of health initiatives, including preventative/hygiene behaviors (especially for at-risk countries and populations), supporting frontline health workers, scaling telemedicine, contact tracing/containment strategies, treatment and diagnosis development. 2) **Vulnerable Populations:** The set of problems facing the elderly and the immuno-compromised, such as access to meals and groceries, and supporting those who are losing jobs and income. 3) **Businesses:** The set of problems that businesses are facing to stay afloat, collaborate effectively, and move parts of their business online. 4) **Community:** Promoting connection to friends, family, and neighbors to combat social isolation and the digitizing of public services for local governments. 5) **Education:** Alternative learning environments and tools for students, teachers, and entire school systems. 6) **Entertainment:** Alternatives to traditional forms of entertainment that can keep the talent and audiences safe and healthy. 7) **Other:** The above themes are just suggestions. Feel empowered to get creative!" (Lessin and Heerwagen 2020).
- ▶ The CODEVID-19 hackathon: "The purpose of the hackathon ultimately isn't to win prizes, it's to build useful apps that people can use to manage and survive during the COVID-19 pandemic (C.-1. hackathon 2020).
- ▶ A hackathon by Machine Hack: "The objective of the hackathon is to gauge COVID-19 on three metrics- confirmed cases, recovered cases and death events for the next day using historical data as on a given date. As sad as it is to analyse the data around COVID-19 events, it is critical to keep a tab on the disease metrics to track the outbreak. The hackathon will be based on the data published by various agencies and the Johns Hopkins University Center for Systems Science and Engineering (JHU CSSE)" (M. H. hackathon 2020).
- ▶ The Hack for Wuhan hackathon (Wuhan hackathon 2020): "In a tradiotional [sic] hackathon, "basically you have a room, you have a location for 48 hours, and then you have developers hacking," said Xiaoyin Qu, Run the World's co-founder. But for Hack for Wuhan, all that was needed was the third ingredient: developers, plus online platforms. [...] The event drew 345 contestants, some of them working while quarantined, along with the participation of more than 6,000 developers at GitHub, an open-source tech platform. Instead of in-person judging, teams hosted online office hours where they demonstrated products for judges. One of the advantages of an online hackathon, compared a physical event held in a certain geographic location, Qu noted, is that it can more easily draw specialized expertise - experts in AI, for example - from around the world." (Palmer 2020).

## 4.2 Game Studies

There is a growing body of research literature on game jams in the research field of game studies, and I draw on several of these contributions in my PhD project. In this section I outline game studies as a research field and related works pertaining to this research field.

Espen Aarseth declared the year of 2001 as Year One of computer game studies, where he describes a shift from "the old field" of game studies to the "new field" of computer game studies, hence emphasising the medium of computers (Aarseth 2001). Aarseth further describes computer game studies as a research field where all the researchers: "[. . . ] enter this field from somewhere else, from anthropology, sociology, narratology, semiotics, film studies, etc, and the political and ideological baggage we bring from our old field inevitably determines and motivates our approaches." (Aarseth 2001). In Mäyrä's words, game studies can be described as a "polyphonic discipline" (Mäyrä 2014).

Traditionally, the research field of game studies has focused on notions of *game, player* and *playing*, whereas notions of *design, designer, process* and *practice* have been peripheral (Kultima 2015b). This is despite the fact that *game design* is the most used keyword in research on games, according to Melcer et al. who conducted a quantitative analysis of 8207 research papers from 48 core publication venues of games research from 2000 to 2014 (Melcer et al. 2015). It may seem contradictory that the game studies research field traditionally does not focus on notions of design, designer, process and practice while the keyword game design is the most frequent keyword of papers. This contradiction can be explained by the ambiguity of the word *design* itself, as it is both a noun (for example a final product) and a verb (the process of designing). Furthermore, the multi- and interdisciplinarity of the field might explain the wide use of the keyword game design despite the lack of focus on design, designer, process and practice, because multi- and interdisciplinarity entails many different kinds of approaches to and understandings of what constitutes game design (Kultima 2015b).

HCI is also a research field which has embraced studies on play and games as part of a shift of focus from: "[. . . ] traditional usability studies towards user experience studies, came a shift of focus to emotionally engaging experiences at CHI." (Nacke et al. 2016). Carter et al. reviewed 178 papers on topics of games, gaming and play published at the CHI conference between 2003 and 2013, and they identify four paradigms of research:[6]

6: Carter et al. follows Preece et al.'s (Preece, Sharp, and Rogers 2011) definition of a paradigm as a: "[...] general approach that has been adopted by a community of researchers and designers for carrying out their work, in terms of shared assumptions, concepts, values, and practices." (Carter et al. 2014)

▶ "**Operative games research** that leverages knowledge gained from the study of games or play to exert control upon the world, such as encouraging exercise or learning.
▶ **Epistemological games research** that uses games as a vehicle for understanding the use of all technologies, rather than only in the context of the unique modes of interactions or affordances of games and play, such as virtual embodiment or interfaces.
▶ **Ontological games research** that is concerned with the design and understanding of the ontology of games; rules, aesthetic, interfaces, fiction and game design patterns.

▶ **Practice games research** that is concerned with the emergent practices and experiences that occur as a result of interaction with games or toys, or when interacting with technology with a lusory attitude [...]" (Carter et al. 2014; emphasis added).

Though HCI embraced research on games and play, these paradigms illustrates that notions of design, designer, process and practice are not well represented within HCI research on games and play, which Carter et al. also notices: "[. . . ] but there are notably some topics mentioned that we did not find in our review; studies of developers, casual game design studies, and tools for game creation." (Carter et al. 2014). My PhD project does not fall into the four paradigms but does contribute to the gap which Carter et al. identified. Specifically, I address this gap by researching how game jam formats shape how people develop games. In recent years a more design-oriented research field of game studies have been argued strongly for, and in paper 1 I respond to and continue a call for bridging the fields of design research as conducted within HCI and game studies to: "[. . . ] support game studies with a well-established field of research in order to advance our understanding of game designers and their creative processes in designing games." (Olesen and Halskov 2018).

Research into game design processes has often been encumbered by several limitations, such as non-disclosure agreements which makes sharing of practices largely impossible (O'Donnell 2014). Game design in the game development industry has been discussed as largely: "[. . . ] invisible, hidden behind the famous names of publishers, executives, or console manufacturers." (O'Donnell 2014). This "secrecy culture" which pervades the industry (O'Donnell 2014) further entails mystical perspectives on creativity, idea generation, and management of creative processes (Kultima and Alha 2010). The opaqueness of game design processes is not only a challenge for research, but also for people looking to enter the game development industry: "Students are expected to make games without any knowledge of how they are produced in the actual game industry." (O'Donnell 2014). Where researchers have struggled to study game design practices, game design practitioners have likewise struggled with deploying research into their practices, perhaps because research into game design is still in its infancy (O'Donnell 2014). In other words, there is a clear gap between academia and the industry (Kultima 2015b). Though people develop games in a very particular way in game jams, Turner, Owen and Thomas argues that game jams have obtained an important role as a rite of passage for those wanting to enter the game development industry (Turner, Thomas, and Owen 2013). As game jams have increased in popularity, many use these game creation events to gain a first-hand experience of a full game creation cycle (Pirker, Kultima, and Gütl 2016). Game jams therefore attract a broad audience of: "[. . . ] game industry professionals and amateurs, students and non-industry professionals." (Turner, Thomas, and Owen 2013). At the same time, game jams have been embraced by the game development industry where companies are organising internal game jams in order to explore new game ideas and technology (Turner, Thomas, and Owen 2013). In this perspective, game jams can, to some extent, function as a bridge between academia and practice.

As a research topic and object of study, researchers saw value in game jams as a research platform and established the Global Game Jam

Research Committee in order to: "[. . . ] promote, facilitate, organize, and conduct scientific and technical research activities related to innovation, experimentation and collaboration." (Fowler, Khosmood, and Arya 2013). Game jams have thus given rise to a growing body of research on game design as deployed in game jams, since the design process of making games becomes accessible for researchers. Since 2015 there has even been an annual conference devoted to research revolving around game jams (Fowler and Khosmood 2015).

In this section, I have outlined how my research connects to the aspect of game design research of my PhD project which is implied by my focus on game jams, a particular form of game development. The research in my PhD project does not fall within traditional aspects of game studies, however it does address a gap in the research on game design. Hence, I contribute to an underdeveloped and nascent orientation of game design research, that is the studies of the notions of design, designer, process and practices, rather than notions of game, player and playing.

## 4.3  State on Research Context

So far in this chapter I have described maker culture and game studies as important and overarching contexts, or prior work, for the PhD project, and I have argued that there is little focus on how design processes in hackathons and game jams unfold. This lack of research has motivated my PhD project, and has at the same time influenced the methodology of the PhD project. Edmondson and MacManus argue that the state of prior work can be an aid in: "[...] identifying unanswered questions, unexplored areas, relevant construct, and areas of low agreement." (Edmondson and McManus 2007), and thereby be a key determinant of appropriate research methodology. They suggest that prior work can be evaluated on a continuum between mature to nascent theory (Edmondson and McManus 2007). The following three items are archetypes of theory, and illustrates what usually characterises theory that are more or less mature, and for each I highlight the type of research question and what generally characterises data collections:[7]

7: For a more extensive description of the three archetypes, see (Edmondson and McManus 2007).

▶ *Nascent* theory are generally emerging from exploratory and often qualitative work, while the relationships between phenomena are unexplored (John Zimmerman, Erik Stolterman, and Forlizzi 2010). Nascent theory seeks to answer questions of how and why, and research questions are open-ended inquiries about a phenomenon. The type of data collected is usually qualitative, and open-ended.
▶ *Intermediate theory* is when researchers confirm, refute, refine, and extend the works of others. Research questions are provisional explanations of phenomena and proposed relationships between new and established constructs. Intermediate theory generally moves from qualitative to quantitative approaches, and the type of data collected can therefore be both qualitative and qualitative.
▶ *Mature theory* is well-developed and has been studied over time with repeated and increasingly precise observations by a variety of scholars. Research questions are focused questions and/or hypotheses which relates existing constructs. The type of data is usually quantitative.

Edmondson and MacManus acknowledge that it is not an easy task to determine the state of prior work informing a research question (Edmondson and McManus 2007). The state of prior work can however provide a basis for evaluating the methodology used in order to approach a research question. In other words, the state of prior work can be used to evaluate the *methodological fit*, which is defined as the: "[...] internal consistency among elements of a research project" (Edmondson and McManus 2007), and depends on whether methods are *appropriate* in relation to where on the continuum prior work is.

In order to make the picture of the state on prior work more complete, I elaborate on the context of maker culture and game studies outlined in this chapter by briefly reviewing how researchers in recently published papers have studied hackathons and game jams, focusing on their research questions, which data was collected, and how it was collected. While these paper do not necessarily share the research interest of *design processes* in hackathons and game jams, they study the same phenomena of game jams and hackathons and hence constitute a part of the research context of the PhD project.

## Game Jams

A search in the ACM digital library for recently published research papers mentioning 'game jam' in their abstracts, reveals 12 papers published in 2019:

- ▶ Kankainen, Kultima and Meriläinen study the motivations for organising game jams which they point out is only little to no research about. They conducted an interview study where N=13 (Kankainen, Kultima, and Meriläinen 2019).
- ▶ Ferraz and Gama study female participation in game jams which they argue is a topic that is not researched much (Ferraz and Gama 2019). They use a mixed methods approach including a survey, interviews, and observations during a game jam. They focus on the survey and interview results, and the findings are generally about the women's experience of participating in a game jam.
- ▶ Arya et. al. present a report paper discussing the GGJ-Next game jam for youth established in 2018 (Arya, Gold, et al. 2019). The GGJ-Next seek to be the first globally organised event for experiential educational learning by playing and making games.
- ▶ A paper by Lai et. al. is a "conversation starter" about an investigative proposal for establishing a game jam license which better covers the specific needs of game jam outcomes than other alternative licenses such as the Creative Commons (CC) license (Lai et al. 2019).
- ▶ Pirker, Punz and Kopf present a short paper discussing the first steps for developing a group forming tool based on a quantitative analysis of data from the Global Game Jam 2014-2018 (Pirker, Punz, and Kopf 2019). They argue that research into team forming processes in game jams are limited.
- ▶ Kultima presents a small survey study with six participants who she defines as "superjammers", people who have participated in a significant number of game jams (Kultima 2019). She mentions that there are no studies exposing behaviour patterns of game jammers in general, which motivates the study.

- ▶ Faas et. al. motivate a study on how online game jams support self-development with noting that little is known about *online* game jams compared to physically co-located game jams (Faas et al. 2019). They conducted interviews and online observations.
- ▶ In our paper from 2019 Rasmussen, Halskov and I present an exploratory study of a real-time annotations prototype to document design processes in a game jam (Rasmussen, Olesen, and Halskov 2019). We sent a survey to the participants and conducted a semi-structured group interview with participant representatives.
- ▶ French et al. present a report paper on three "ZooJams" including zoo keepers, representatives from the Royal Society for the Prevention of Cruelty to Animals (RSPCA), the non-profit corporation The Shape of Enrichment, engineers, computer scientists, game developers, dog trainers, animal welfare experts, UX practitioners and networking specialists (French et al. 2019). The authors describe the rationale for their approach of game jam organisation and reflect on this.
- ▶ White et al. present a report paper on the first international game jam held in a university in Japan. They studied language barriers in a game jam context and conducted a survey with N=19 (White et al. 2019).
- ▶ Based on their potential for collaborative learning, creativity and play, Fowler emphasises and discusses game jams as a way to promote 9-11 year old girls' perceptions about computer science and Science, Technology, Engineering, and Mathematics (STEM). He argues that: "While the majority of game jams are targeted towards adults, there is a growing interest in creating game jams for children." (Fowler 2019)
- ▶ Ho and Tomitsch review brainstorming toolkits and explore game jams as a research environment for this (Ho and Tomitsch 2019). They seek to answer the research questions: What affordances do brainstorming toolkits provide, and what is the relationship between brainstorming and idea association. They furthermore conducted a survey with 812 jammers responding to the question of whether they used tools to brainstorm for ideas and if so, how.

From this cursory review of recent publications on game jams it appears that the authors generally use a mix of quantitative and qualitative approaches with several papers contributing to nascent research areas. Some papers propose a topic as a "conversation starter", or discuss topics with little prior research, such as online game jams or game jams specifically organised for children. Some papers explore design processes in game jams, however only a smaller part of the accelerated design process, such as team forming or brainstorming. While the report papers aim at providing a more complete picture of the full game jam process, they do this often from a top-down perspective of the organiser and focus on salient aspects such as outcome or participation experience.

**Pre- and Post Situ Studies of Game Jams**

An aspect particularly regarding research on game jams, is that several studies uses pre- and post situ methods such as surveys and interviews, that rely on the respondents' expectations, memory and ability to recall details about events after the fact. In her review of how 20 research papers define game jam, Kultima notes that over half of the papers used data collected from the Global Game Jam or discussed a topic in reference to the Global Game Jam (Kultima 2015a). This may be due to the fact that the Global Game Jam Research Committee has been established exactly to: "[...] promote, facilitate, organise and conduct scientific and technical research activities related to innovation, experimentation and collaboration [...]" (Fowler, Khosmood, and Arya 2013). The committee provide: "[...] global surveys that include questions by approved research projects, inviting all GGJ participants to respond, collecting and passing the data to researchers, and finally organizing means of disseminating the research findings." (Fowler, Khosmood, and Arya 2013). Contributions which utilise for example surveys can contribute with valuable insights on aspects of a design process and reveal overarching trends. However, in discussing the limitations of their own Global Game Jam survey study, Zook and Riedl reflect that: "Survey responses are limited to the most salient aspects of an experience, preventing detailed processual information gathering." (Zook and Riedl 2013). In addition to preventing detailed processual information gathering, there is a risk of respondents answering with after-the-fact rationalisations of design decisions, thereby overlooking subtle nuances of influential aspects in the situation, or that respondents are simply mistaken.

In the next chapter, I describe how being aware of these limiting aspects of pre- and post situ studies of game jams, and being informed by pragmatist design research have inspired parts of the methodology: I supplement contributions gained from survey studies, such as the ones in paper 3 and paper 4, with in situ autobiographical accounts of a game jam and a hackathon in respectively paper 1 and paper 2.

## Hackathons

A search in the ACM digital library for research papers mentioning 'hackathon' in their abstracts, reveals 13 papers published in 2019:

- ▶ Nolte presents the findings of a quantitative exploratory study focusing on the temporal connection between hackathon participation and start-up founders (Nolte 2019). The study used data sets which were compared. He note that despite an increasing interest in hackathons, there is little research on: "[...] whether and how hackathons contribute to fostering entrepreneurship and how they can be purposefully integrated into existing entrepreneurial practice." (Nolte 2019).
- ▶ Chua, Chua and Soo present a report paper on a hackathon structured as a five month long pilot program to bring data science into a high school's curriculum (H. X. Chua, E.-L. Chua, and Soo 2019). They report participants' survey responses and lessons learned.
- ▶ In this short paper, Kos conducted an exploratory first pass at documenting and cataloguing three collaboration styles in a female-focused hackathon which goes beyond competing: team-based, cooperative group, and individual participation (Kos 2019). Four types of event goals were identified: competition, exploration, dabbling, and observing. She collected narratives and anecdotes from participants of a hackathon via informal discussions.[8]

8: In paper 2 we identify the format as one factor which shapes the accelerated design process, and although Kos' study is based on preliminary findings from anecdotes, it further explores the format factor by suggesting alternatives (exploration, dabbling, observing) to competition, which is otherwise often used. The question for future research then is, how do elements of competition in hackathon and game jam formats shape the participants' way of designing and prototyping, and which other elements can be used as alternatives in organising hackathons and game jams? What organisational elements of the format could support the different kinds of event goals? How is creativity influenced by respectively competition, and other alternative elements of the format factor?

▶ Ito and Zettsu present a report paper on a hackathon for traffic engineering professionals who worked with traffic risk data in order to create car navigation prototypes (Ito and Zettsu 2019).

▶ Gama presents an experience report on a hackathon as part of an undergraduate course (Gama 2019). Gama report on participants' survey responses and lessons learned for the particular hackathon.

▶ Damen et al. present a case study exemplifying how: "[...] multidisciplinary HCI approaches in a hackathon setting can contribute to real life urban health challenges." (Damen et al. 2019).

▶ Gama et al. present a report paper on how data created in a "mapathon" can be used for applications in a subsequent hackathon, and reflect on lessons learned from this circular approach (Gama et al. 2019).[9]

9: Mapathons are hackathons specifically for generating map data through remote mapping.

▶ Hope et al. report on an iteration of an exploratory study which contributes to "a growing literature" on designing hackathon formats as more inclusive and accessible (Hope et al. 2019).

▶ Webb et al. present a case study on a hackathon format which they argue is a novel initiative: "[...] that adapts the conventional hackathon and draws on insights from the Open Hardware movement and Responsible Research and Innovation (RRI)." (Webb et al. 2019).

▶ Konopacki, Albu and Steibel use methods such as a literature review, observations, self-experiences, and interviews to showcase a "legal hackathon" which they developed in order to: "[...] develop draft bills collectively addressing a single issue and within a timeframe." (Konopacki, Albu, and Steibel 2019).

▶ Salinas, Emer and Neto present two case studies of a hackathon format for developing exploratory data analysis and visualisation skills in participants (Salinas, Emer, and Neto 2019). They reflect on lessons learned and report on participants' survey responses.

▶ Cutts et al. report on a "thinkathon" where students work through coding comprehension exercises (Cutts et al. 2019). They report on quantitative and qualitative data from a survey sent to students, and lessons learned.

▶ The last result of the search is a poster by Linnell et al. where they among other things mention a hackathon as part of an effort to improve diversity in participation of K-12 girls and underrepresented minorities in CS (Linnell et al. 2019).

These recent publications on hackathons generally use a mix of qualitative and quantitative approaches. Several of the papers roughly follow the same methodology of reporting on participants' survey responses and lessons learned from a particular hackathon format organised for specific, and often novel, purposes. Several of the papers are furthermore reports from the perspective of an author with an organiser role. Using pre- or post situ survey studies, the research methodology generally used for studying hackathons largely resembles the methodology generally used for studying game jams. Particularly for hackathons we have done an extensive review of how researchers conduct research *with* and *on* hackathons, and the above more recent, albeit more cursory, review reflect the findings in paper 5. In paper 5 we among other things demonstrate that the majority of publications conduct research *with* hackathons. In other words those publications use hackathon formats as means for exploring or answering a research question, but does not research how the formats influence and accelerate design processes.

Several of the recent publications on both game jams and hackathon *explore* different aspects of hackathons and game jams, while a few papers confirm, refute, refine or extend the work of others. While acknowledging that it is not easy to completely determine the state of prior work as emphasised by Edmondson and MacManus, I have in this chapter aimed at outlining the context of maker culture, game studies and how other researchers generally have studied hackathons and game jams. Based on this outline of related contexts and works, it should be equitable to claim that the state of the context of research on hackathon and game jams generally are in the intersection between *nascent* and *intermediate* theory. This is reflected in the ways hackathons and game jams are studied, which *generally* lean towards exploratory studies which employ both qualitative and quantitative studies. Furthermore, while several of the papers experiment with different ways of organising hackathons and game jams, and study participants', often self-reported, *experiences* of participation, few engage with the question of how the formats shape participants' design processes.

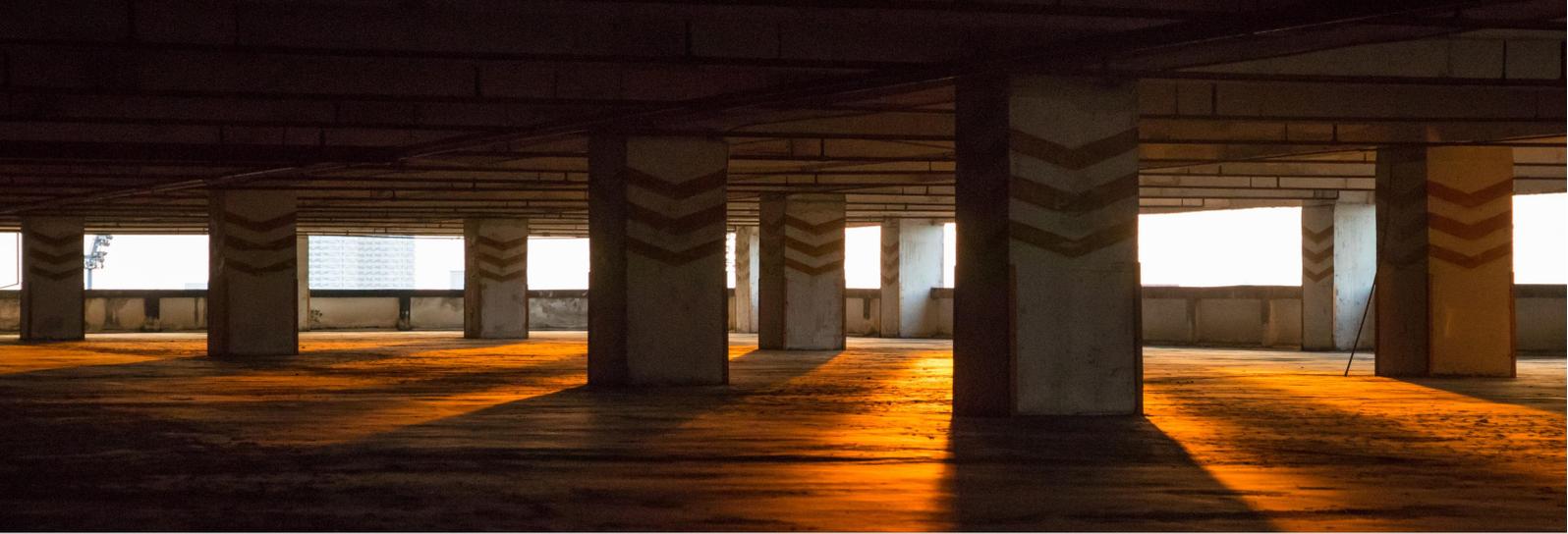

# 5 THEORETICAL FOUNDATION

The research environment and field, which I outlined in Chapter 3, emphasise among other things design processes as a central research topic. As a researcher in this research environment and field, the theoretical foundation of my PhD project has been informed by theory on predominantly *design processes*, and also creativity. In this chapter, I outline the theoretical foundation for my PhD project, which was a significant factor for shaping the methodology. I start by outlining design theory, focusing on pragmatist design theory. Subsequently, I briefly outline creativity research, focusing on the notion of creativity constraints. I conclude the chapter by summarising how the five papers connect to the theoretical foundation.

## 5.1 Design Theory

An essential pillar of the theoretical foundation for my PhD project is the philosophical tradition of pragmatism, in particular pragmatism as defined by John Dewey (Dewey 2005). The benefits of pragmatism in relation to design has been argued for meticulously (Dalsgaard 2014), and has inspired several prominent researchers such as Donald Schön (Schön 1987). Schön's pragmatist-based theories have been very influential for design theory in general, and hence also for Interaction Design (Löwgren and Erik Stolterman 2004). Therefore, I begin the chapter with a general introduction to pragmatist-based interaction design research including selected key concepts. As it is well established that creativity is acknowledged as an integral part of design (Biskjaer, Dalsgaard, and Halskov 2017), I furthermore select and describe key concepts from creativity research which in particular has informed the theoretical foundation.

### Three Accounts on Design

In this section, I first describe how design in general can be conceptualised from three different accounts: a conservative, romantic and a pragmatic account (Fallman 2003), since the PhD project's theoretical foundation is positioned within one of the three accounts, the pragmatic account. These three accounts on design was first conceptualised by John Christopher Jones (Jones 1992), an important contributor to design methodology (Löwgren and Erik Stolterman 2004). Table 5.1 show a brief overview of the three accounts.

**Table 5.1:** An overview of the three accounts on design, for a more elaborate overview see (Fallman 2003).

|                    | **Conservative Account** | **Pragmatic Account**    | **Romantic Account** |
| ------------------ | ------------------------ | ------------------------ | -------------------- |
| **Designer**       | Glass box                | Self-organising system   | Black box            |
| **Design Process** | Rationalism              | Reflective conversation  | Artful inspiration   |

### The Conservative Account

The conservative account on design is philosophically based in rationalism, and borrows methods and concepts from natural sciences, mathematics, and systems theory. A design process in the conservative account assumes a problem which can be solved, and progresses gradually from the abstract to the concrete via rational and structured methods. To capture the conservative account, the metaphor of a *glass box* is used to describe how the logic of the designer is seen as transparent and explainable (Fallman 2003).

### The Romantic Account

The metaphor of a *black box* is used to capture the romantic account. In the romantic account on design, the designer is given the role of a creative genius whose logic of a design process is not easily accessible. Instead of approaches inspired by science as in the conservative account on design, the romantic account on design is inspired by art, where the creative designer possess unusual talents The design process is shrouded in mystique and is opaque for scrutiny and explanations (Fallman 2003), almost as if design is the result of divine inspiration.

### The Pragmatic Account

Lastly, the less figurative metaphor of a *self-organizing system* is used to capture the pragmatic account on design. This account emphasises the particular *situatedness* of design processes. The designer iteratively interprets and creates meaning of the effects of their designs during design processes which are acknowledged as deeply situated in a world: "[...] which is already crammed with people, artifacts, and practices, each with their own histories, identities, goals, and plans." (Fallman 2003). Schön's notion of a reflective conversation with the materials of the design situation is a useful metaphor for the design process in the pragmatic account on design.

In the next section I unfold the pragmatic account on design by elaborating on several concepts from pragmatist-based design theory.

## Pragmatist Design Theory

Pragmatism is a philosophical movement which was largely developed by the American philosophers Charles Sanders Peirce, William James, John Dewey, Jane Addams and George Herbert Mead (Dalsgaard 2014; Legg and Hookway 2020). Particularly, pragmatism as conceptualised by Dewey has had a significant impact on design research, with Donald Schön being the most prominent design researcher building on Dewey's work (Dalsgaard 2014; Dixon 2020). Dewey's pragmatism and design theory converges in several themes and concepts (Dalsgaard 2014), and in the following text I outline key theoretical concepts in pragmatist design theory, focusing on the work of Schön. These theoretical concepts form the background for how I understand and approach design processes.

In general, pragmatist philosophy acknowledges both the existence and importance of the physical world while at the same time acknowledging and valuing: "[...] the inner world of human experience in action." (Johnson and Onwuegbuzie 2004). Thereby, pragmatism rejects traditional dualisms such as: "[...] rationalism vs. empiricism, realism vs. antirealism, free will vs. determinism, Platonic appearance vs. reality, facts vs. values, subjectivism vs. objectivism." (Johnson and Onwuegbuzie 2004). Theory and practice is instead considered closely intertwined and mutually susceptible: On one hand, theory emanate from and must be evaluated on the basis of practice, while on the other hand, practice is formed both explicitly and implicitly from theory, conceptual frameworks, and experience, which the designer brings with them (Dalsgaard 2014).[1] Hence, the world is seen as an emerging phenomenon, that is never stable. Therefore, a designer must have an experimental approach to the world, since they cannot trust conceptualisations defined beforehand, as they will likely change meaning over time (Schön 1992). A pragmatist perspective acknowledges the unspoken and immediate knowledge in everyday practice as the primary elements in our knowledge creation, and as essential for the understanding of design (Fallman 2003). This unspoken and immediate knowledge is the spontaneous and intuitive actions in everyday practice, also called *knowing-in-action* (Schön 1992).

The designer's knowledge and experience that they bring with them to form practice is the designer's *repertoire*, and this may be tacit for the designer. The repertoire then constitute the point of departure for the designer to obtain a preliminary understanding of the design situation, and to form the first steps of their *inquiry* (Löwgren and Erik Stolterman 2004). Temporary stability can be established in a given design situation by the designer's inquiry, an experimental what-if approach, where the designer intervenes and transforms the design situation via different kinds of experimental moves or inquiry strategies. The transformation of the design situation happens when the designer *frames* the design situation by selecting what should be in focus, and what should be in the periphery (Schön 1992). This articulation of the design situation then shapes future inquiry (Hansen and Halskov 2014). Inquiry consist of iterative phases, where the designer's actions provide input for their reflections, while the reflections in turn shape the subsequent actions that resolve design problems or open up new opportunities (Dalsgaard 2014). Metaphorically, Schön describes this as a *conversation with the materials of the situation* (Schön 1992). Schön further designates the iterative phases

1: While it is now well established that pragmatism constitute an advantageous framework and perspective for especially design, pragmatist philosophy can also entail some shortcomings for research. One example is that basic research may be valued less than applied research, since the latter can appear to produce more immediate and practical results. See (Johnson and Onwuegbuzie 2004) for a more elaborate discussion on this. For the purpose of my PhD project, pragmatist philosophy, however, constitute a well established and elaborate theoretical foundation as well as framework for understanding and approaching design processes, particularly from the works of Schön.

of inquiry: *reflection-in-action* and *reflection-on-action*. Reflection-in-action is the construction of new understanding, which informs the designer's decision of actions in situ, while reflection-on-action is reflection on actions after they are conducted, and after the design situation has ended.

Drawing on the designer's repertoire depends on the designer's *judgement* in order to conduct inquiry as a reflective conversation with the materials of the situation via reflection-in-action and framing processes. In Nelson and Stolterman's words, design judgement is the ability to: "[...] gain or project insight, through experience and reflection, into situations which are complex, indeterminate, indefinable and paradoxical." (Nelson and Erik Stolterman 2003). This ability to make quality design judgement is informed by the "[...] accumulated experience of choices made in previous situations [...]" (Olesen, Hansen, and Halskov 2018) or in other words: it is informed by the designer's repertoire. Schön juxtapose design judgements as acts of *seeing*, and refers to this judgement process as a kind of *seeing-moving-seeing* (Schön 1992). The designer continuously has to judge what appropriate actions are in response to the things they see. In the first instance of seeing in seeing-moving-seeing, the designer sees what is already there. The second instance of seeing, conveys a judgement about what the designer first saw, and how an act of inquiry or experiment, *moving*, has changed what was first seen. Such a design judgement is subjective and depends on the designer's ability to make judgements of quality. Paper 2 focus specifically on this aspect, design judgement, of pragmatist based design research, and how being in a hackathon situation informs and drives judgements about design decisions.

Technology plays a significant role in the designer's inquiry and is broadly defined as encompassing theory, methods, tools, and materials (Hansen and Dalsgaard 2012), since technology is seen as the use of tools or as a means to achieve a certain result (Dalsgaard 2014). The pragmatic perspective on design acknowledges that technology is not passive, but plays an active role in how designers frame their understanding of a design situation, and that technology can be means to transform the design situation (Dalsgaard 2014). Specifically, technology can support the inquiry-process by letting the designer: experience essential parts of the situation, develop hypotheses, conduct specific actions in order to transform the situation by for example conducting experiments or create externalisations of design concepts (Hansen and Dalsgaard 2012). This aspect of pragmatist design theory motivates and supports the analysis of *technology* in accelerated design processes in order to contribute to our understanding of them. In paper 1 and paper 2 we analyse how factors such as participants' repertoires and the available tools and materials shape the accelerated design processes.

In Chapter 1 I introduced a definition of *design processes* based on Löwgren and Stolterman's definition (Löwgren and Erik Stolterman 2004). On the basis of the outlined pragmatic design theory above, the working definition of design processes can be extended by introducing an idealised design process as initially conceptualised by Schön (Löwgren and Erik Stolterman 2004). The idealised description of a design process below clarifies how the theoretical concepts based on pragmatic design theory come into play:

▶ The designer approaches a design situation via *inquiry*, where the designer conducts experimental actions, where the designer *frames* what needs to be in focus of the situation. These actions are conducted on the basis of the designer's *repertoire*, and the designer may not be able to articulate why or how actions are conducted, as they may be the result of the designer's *knowing-in-action*.

▶ The result of an action may be surprising for the designer, who reflects on what happened and why (*reflection-in-action*).

▶ Via reflection, which may not be articulated in words, the designer dispute their own repertoire and considers the next action. The designer has to *reframe* the design situation in terms of new opportunities and limitations, by which the situation then transforms.

▶ The designer obtain a new understanding of the design situation. They initiate new experimental actions which again may be surprising.

**Design Space**

The notion of *opportunities* and *limitations* that are encountered and framed in a design process can be expanded with the concept of a *design space* (Biskjaer, Dalsgaard, and Halskov 2014). During my PhD project I have based my framing of design spaces on Biskjaer, Dalsgaard and Halskov's definition:

"[...] we define a design space as a conceptual space, which encompasses the creativity constraints that govern what the outcome of the design process might (and might not) be." (Biskjaer, Dalsgaard, and Halskov 2014).

Biskjaer, Dalsgaard and Halskov's definition is informed by creativity research into *constraints*. Traditionally, creativity research has studied the limiting role of constraints on creativity, however in recent years creativity researchers have studied how constraints at the same time can both limit and enable creativity (Biskjaer, Dalsgaard, and Halskov 2014). As we explain in paper 1, this definition is based on Schön's pragmatic perspective: "[...] in which a design space denotes a conceptual space of opportunities constructed and developed by the designers via inquiry." (Olesen and Halskov 2018). Opportunities can be limited in some areas of a design space, while they may be abundant in other areas. In accordance with the pragmatic account on design, a design space transforms dynamically during a design process, as it is constructed via the designer's inquiry. In the next section, Section 5.2, I elaborate on *creativity constraints* as a theoretical concept.

## 5.2 Creativity Research

Creativity research is a large field of research, and since I mainly draw on *creativity constraints* in my PhD project, I focus on this area of research. I begin the section with a brief outline of the development of creativity research. I then unfold the definition of creativity which is used in the PhD project, where particularly paper 4 draws on this. Lastly, I define creativity constraints and outline their relation to hackathon and game jam formats.

**Four Waves of Creativity Research**

Sawyer described a development of creativity research through three waves which were concerned with different foci (R. K. Sawyer 2011). *The first wave* of creativity research was concerned with personality and characteristics of creative individuals as well as with measuring creativity through for example divergent thinking tests. *The second wave* was characterised by an increased interest in the underlying processes when people across different domains in general engaged in creative tasks. *The third wave* was concerned with sociocultural themes and interdisciplinary approaches, and turned towards studies of creativity in groups of people. Recently, there has been what has been suggested to be a *fourth wave* of creativity research, which is characterised by an increased interest in how creativity unfold within *digital creativity* (Frich, Biskjaer, and Dalsgaard 2018). This fourth wave entails a focus on how a growing digitisation of creative practices, tools, and materials facilitate creativity. As we specifically state in paper 4, our work is informed by the third and fourth wave of creativity research, since game jams and hackathons generally are interdisciplinary group efforts, and the design work is typically profoundly aided by digital tools and materials.

## Definition

Exactly how creativity should be conceptualised is still a topic of debate (Falk et al. 2021). Plucker, Beghetto and Dow offer the following broad definition of creativity:

"Creativity is the interaction among *aptitude, process and environment* by which an individual or group produces a *perceptible product* that is both *novel and useful* as defined within a social context." (Plucker, Beghetto, and Dow 2004).

The definition underlies how creativity is generally conceptualised in the PhD project. In paper 4 we specifically orient the research question toward general creativity research, and therefore we explicitly state this definition in paper 4. I elaborate on the definition in the following:

### Aptitude

Aptitude refers to the dynamic characteristic or skill-set which can be influenced by experience, learning, and training. This opens up for creativity as something that can be facilitated and trained, and refutes the myth that creativity is an inherent quality only possessed by a few talented people.

### Process

Process refer to the ways in which people approach situations and come up with solutions. An obvious parallel, which again testifies to the close connection between creativity and design, is design processes and how people approach design problems.

**Environment**

Environment is construed broadly (Plucker, Beghetto, and Dow 2004), while environmental *affordances* are the possibilities for *action* which the environment offers to individuals (Barab and Plucker 2002). In regards to hackathon and game jam formats, I understand environment as entailing among other things the physical surroundings, and the tools and materials which are available for participants. As I will elaborate in the next section, these environmental factors are analysed especially in paper 1, paper 2 and paper 4 as part of the creativity constraints which governs the process.

**Perceptible Product**

Plucker, Beghetto and Dow acknowledges that creativity involves: "[...] latent, unobservable abilities and processes" (Plucker, Beghetto, and Dow 2004), however they argue that a perceptible outcome has to be created as it serves as: "[...] necessary evidence from which the presence of creativity can be inferred, determined and evaluated." (Plucker, Beghetto, and Dow 2004).

**Novel and Useful**

Novel and useful is the most prevalent aspect of how creativity is defined, and is the standard on which the perceptible product is measured. Novelty can be understood as "[...] a measure of how unusual or unexpected an idea is as compared to other ideas." (Kolko 2007).

**Social Context**

Whether something is evaluated as creative depends on the social context. In this way, creativity studies of 4th-grade science projects can be evaluated as valid as creativity studies of Nobel Prize winners. Since the evaluation of creativity is done within its own social context the definition excludes: "empty relativistic claims that a 4th-grade science project necessarily is as creative or significant as a Nobel Prize-winning discovery." (Plucker, Beghetto, and Dow 2004).

This definition of creativity is useful in regards to conceptualising creativity within game jams and hackathons. The emphasis on creativity as something that can be facilitated and trained, gives some validity to the popular claim that hackathons and game jams can potentially facilitate creative processes, and lead to novel and useful outcomes. The question is then *how* the game jams and hackathons facilitates creativity, and how the environment can be organised to support creativity? Furthermore, paper 4 supports that at least *novelty* is a prevalent understanding of creativity among game jam organisers. Novelty is likewise a concern in hackathons, but perhaps more characteristic for hackathons compared to game jams is the evaluation of *usefulness* in the developed outcome. Paper 5 shows that a prevalent concern regarding research with and on hackathons, is the sustainability of the developed outcomes of hackathons. Sustainability in this context refers to both longevity of the developed outcomes, and also to whether the developed outcomes satisfies the needs of the people they have been developed for, or in other words how *useful* the developed outcomes are within a certain social context.

## Creativity Constraints

As described in the beginning of this first part of the chapter, creativity is regarded as integral to design. *Constraints* are furthermore part of how interaction design (Löwgren and Erik Stolterman 2004) and design spaces are defined (Biskjaer, Dalsgaard, and Halskov 2014). In accordance with the pragmatic account on design, Löwgren and Stolterman argue that design is always carried out in unique contexts which entail particular limitations and restrictions. Therefore a designer must aim for developing a design in the most: "[...] suitable and creative way given the existing conditions." (Löwgren and Erik Stolterman 2004). Löwgren and Stolterman further argues that having limited resources and time can even stimulate creative and innovative thinking. This acknowledgement of constraints as an essential part of designing, and as both limiting and stimulating at the same time, resonates well with research on *creativity constraints* as conducted within the field of creativity research.

In this PhD project I conceptualise creativity constraints as:

"[...] all explicit or tacit factors governing what the agent/s must, should, can, and cannot do; and what the output must, should, can, and cannot be" (Onarheim and Biskjær 2013).

Specifically, creativity constraints can therefore include among other things: environmental factors, tools, materials, requirements, conditions, conventions, rules, and demands (Biskjaer, Dalsgaard, and Halskov 2014). Elster has proposed the following tripartite typology of creativity constraints (Elster 2000):

- ▶ **Intrinsic** constraints are dictated by qualities of the used tools and materials, such as hardware limitations. In paper 1, this constraint was exemplified by a software limitation in the game engine which was used during the game jam.
- ▶ **Imposed** constraints are external and imposed from the "outside", such as game jam organisers' choice of location, a time frame or a theme as discussed in paper 1.
- ▶ **Self-imposed** constraints are imposed by the designers themselves. This type of constraint was exemplified in paper 1, where the graphic designer chose 2D graphics instead of 3D graphics, as he reasoned that 2D graphics would be quicker to design, and therefore more feasible to do within the time frame of the game jam.

Creativity constraints then have generative qualities in a design process, and designers can use them decisively as resources (Biskjaer and Halskov 2014). Creativity constraints is a useful concept since it distils and clarifies the factors which shapes accelerated design process and thereby the outcome.

Creativity is a multifaceted phenomenon and therefore it should be studied with different approaches (Frich, Mose Biskjaer, and Dalsgaard 2018). The concept of creativity constraints is one theoretical approach through which I have viewed hackathons and game jams. In paper 4 we broaden the lense slightly and include additional key creativity aspects to creativity constraints: Novelty, risk-taking, and combinational creativity.

In paper 4 we base our understanding of *novelty* on Kolko (Kolko 2007), as described in the definition of creativity by Plucker, Beghetto and Dow (Plucker, Beghetto, and Dow 2004). We base our understanding of intellectual *risk-taking* in the context of creativity on Bheghetto's notion of "beautiful risks" which are about doing: "[...] something that has a chance to actually have a benefit to others." (Henriksen, Mishra, and Group 2018). According to Bheghetto, the only way in which something can be judged as creative is if one deals with uncertainty (Henriksen, Mishra, and Group 2018), thereby creativity inherently deals with taking risks. *Combinational creativity* is stressed as important by Boden and denotes: "[. . . ] unfamiliar combinations of familiar ideas" (Boden 2004).

As we acknowledge in paper 4, these four theoretical creativity aspects may not capture all parts of how creativity takes place in game jams, however we identify and discuss them as aspects that capture some of the most *prominent* and *central* themes of how organisers understand and promote creativity. Thereby, paper 4 is a step towards identifying how creativity in game jams can be conceptualised, and furthermore how creativity may be studied in future research.

## 5.3 Theory and Papers

Until now, I have summarised pragmatist design theory, focusing on key concepts as formulated and inspired by Schön, as well as creativity research, focusing on creativity constraints. These theories constitute the theoretical foundation for my PhD project, and I have mainly referenced to these theories as background for understanding and framing the work of my PhD project. In the following, I clarify some details on the relation between the theoretical foundation and the five papers of the dissertation. Table 5.2 presents an overview of the relations.

**Table 5.2:** An overview of the relation between the five papers of the dissertation and the theoretical foundation.

|  | Paper 1 | Paper 2 | Paper 3 | Paper 4 | Paper 5 |
|---|---|---|---|---|---|
| **Pragmatism** | Constitutes theoretical background | Constitutes part of theoretical background | Motivates emphasis on *in situ* design knowledge | Motivates orientation towards design process | Motivates future research on relation between hackathon format and process |
| **Design Space** | Constitutes theoretical and analytical scaffolding | Clarifies design processes as situated design judgements |  | Constitutes part of theoretical background |  |
| **Creativity Constraints** | Scaffolds analysis of design space transformations | Constitutes part of theoretical background |  | Constitutes part of theoretical background |  |

## Pragmatism

Pragmatism positions design processes as deeply situated in an ever-unfolding world. This connection to and emphasis on the design situation inspired me to conduct autobiographical design case studies in paper 1 and 2, in order to gain insights on the inner workings of accelerated design processes. In paper 1 we write specifically about pragmatist design theory as the theoretical background for the research question. In paper 2 we build on Schön's metaphor of a conversation in order to scaffold our study of how *design judgement* situated in an accelerated design process unfold. As I described in the section above, this particular metaphor of Schön is an essential aspect of pragmatist design theory.

Pragmatist design theory underlies paper 3, 4, or 5 as well. In paper 3 we focus on design knowledge as generated *in situ*, and we explore how this can be documented for subsequent analysis. This emphasis on the immediate situation, and our exploration of how to capture design knowledge generated in the situation demonstrate our pragmatic account on design, where we acknowledge and emphasise the situatedness of design.

In paper 4 and 5 we neither refer to pragmatism theory directly, or build on pragmatist design theory, such as in paper 2 and 3. However, paper 4 retain an orientation toward the *design process* as we study how game jam organisers understand creativity, and how they organise their events to potentially support participants' creativity *during* a game jam.

In paper 5 the theoretical foundation of pragmatist design theory is not immediately obvious. However, the pragmatic account on design is underlying our motivations for arguing how a greater awareness of how hackathon formats can shape processes, and outcomes can ultimately lead to more systematic research with and on hackathons.

## Design Space

In paper 1 we directly apply the notion of design spaces as a theoretical and analytical scaffolding for the autobiographical design case study of a game jam. Specifically, we study how the inner workings of a game jam can be captured and analysed as a dynamic transformation of a conceptual constraint-based design space. In order to answer this research question, we explore the use of *design space schemas*, a notation technique that seeks to document options and limitations of a design space at a given point in time in a design process (Biskjaer, Dalsgaard, and Halskov 2014). I elaborate on this technique in Chapter 6.

In paper 2 we apply the notion of design spaces as we conceptualise design processes as situated judgements that guide choices between options in a shared design space. Building our conceptualisation of design processes on the notion of design spaces, enable us to study how design judgement is influenced and driven by the hackathon format. Specifically, we study how design judgement were made on how options and limitations were framed during the hackathon design process.

In paper 4 we furthermore apply the notion of design spaces, however in a more peripheral way. In this paper we focus on the integral role of creativity in design processes, and highlight theoretical concepts from creativity research. We continue the notion of design spaces as an essential part of designers' design processes, where the goal is to: "[...] explore the limits of, and ideally transcend, the immediate design space of a given creative situation or task." (Falk et al. 2021). In the paper, we focus on, among other theoretical creativity concepts, *creativity constraints* which essentially constitute a design space.

## Creativity Constraints

During my PhD project I have mainly drawn on research regarding creativity constraints because they are part of the applied definition on design spaces. Since paper 1, 2 and 4 directly or indirectly applies the notion of design spaces, they therefore also draw on the concept of creativity constraints as part of their theoretical backgrounds.

In paper 1 we explicitly identify, analyse and discuss the three types of creativity constraints as they emerged during the four events in the game jam.

In paper 2 we conceptualise a hackathon as a design space which is constituted by certain constraints. We focus on designers' judgement of how to navigate a design space and how judgement is driven by the particular constraints of a hackathon format.

Furthermore, in paper 4 we specifically base the theoretical background on creativity research focusing on selected key concepts. Compared to paper 1 and 2, paper 4 expands the theoretical background on creativity research, by discussing selected key theoretical aspects from creativity research in addition to creativity constraints.

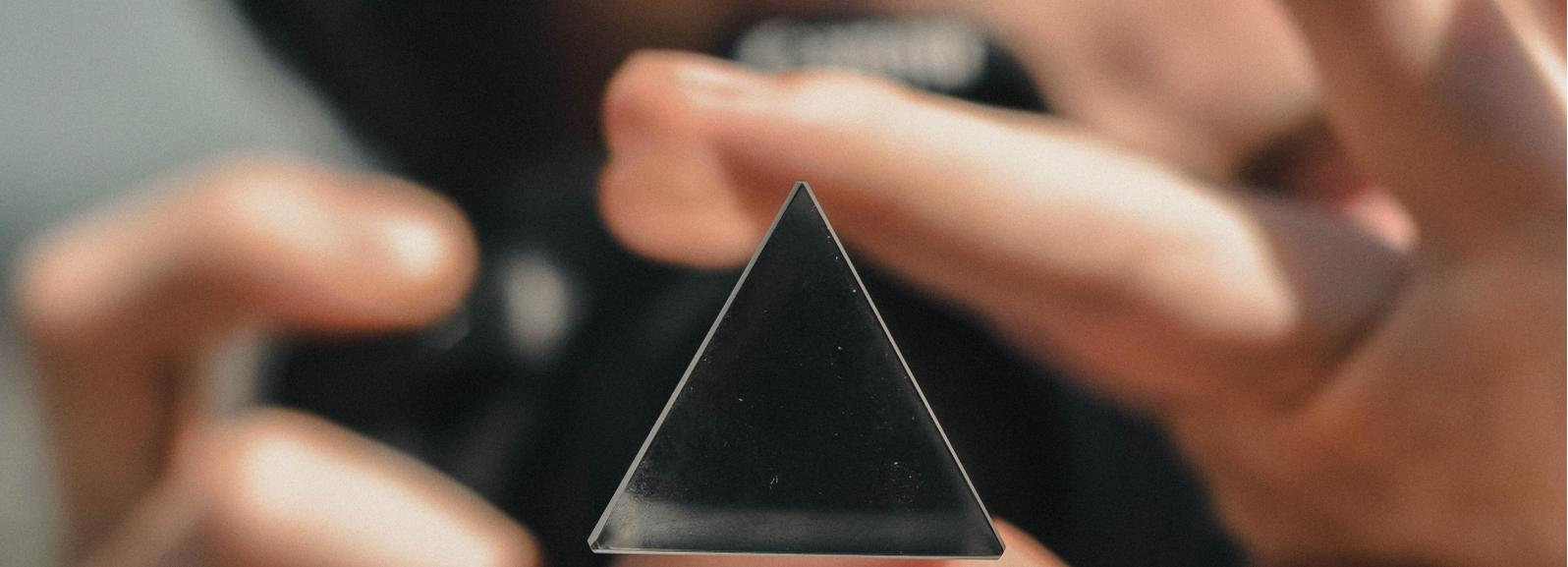

# 6 RESEARCH METHODOLOGY

In this chapter I discuss the reasoning behind the choice of the different methods used during the PhD project.

The research methodology of the PhD project is informed by the theoretical foundation as well as motivated and inspired by the state of prior work on hackathons and game jams. When conducting research, the researcher brings a certain *worldview* with them which influence the practice of research (Cresswell 2014). One consequence of a worldview based on pragmatist design theory, is that the world is viewed as a dynamic and emerging phenomenon, which is never stable. With this worldview, a pragmatic researcher does not focus on methods but on the *research question* and use all the available approaches to understand and approach the research question (Cresswell 2014). Instead of subscribing to only one way of researching (for example quantitative or qualitative), pragmatic researchers look to several different approaches to find the ones that best suit the inquiry of the research problem, which in practice entail a mixed methods approach (Cresswell 2014). Informed by the pragmatic perspective on design, which entails that human cognition and activity are deeply situated, I have used a mix of qualitative and exploratory methods which engages with accelerated design processes in situ. Paper 1, 2, and 3 exemplifies this. As part of improving the understanding of accelerated design processes, I broaden the perspective in paper 4 and 5 and move from the situatedness of accelerated design processes to a top-down perspective, focusing on reported conceptualisations and usage of game jams and hackathons.

Another reason why the methodology of the PhD project is primarily exploratory and qualitative, is the *state* of the research context of game jam and hackathon research in general. As I argued in Chapter 4, there is little research on the inner workings and processual aspects of hackathons and game jams in general. According to Edmondson and MacManus, exploratory and qualitative approaches are generally appropriate when the state of prior work is nascent and intermediate: "[...] when little is known about a research topic or question, initial steps must be taken to explore and uncover new possibilities before useful quantitative measures can be informative." (Edmondson and McManus 2007).

**Table 6.1:** Chronological overview of selected appended papers.

|  | Type | Method Description |
|---|---|---|
| **Paper 1: The Dynamic Design Space During a Game Jam** | First person research | Autobiographical design case study documented by design space schema |
| **Paper 2: Four Factors Informing Design Judgement at a Hackathon** | First person research | Autobiographical design case study |
| **Paper 3: Co-notate: Exploring Real-time Annotations to Capture Situational Design Knowledge** | Intervention | Case study of real-time annotations, using surveys and interviews |
| **Paper 4: How Organisers Understand and Promote Participants' Creativity in Game Jams** | Top-down Organiser perspective | Online survey study |
| **Paper 5: 10 Years of Research With and On Hackathons** | Top-down Researcher perspective | Literature review |

The purpose of the PhD project has been to contribute to a better understanding of accelerated design processes. In order to obtain a better understanding, I have approached the research questions from three different research approaches:

▶ First-person research (paper 1 and paper 2)
▶ Intervention (paper 3)
▶ Top-down organiser and researcher perspective (paper 4 and paper 5)

In the following two subsections of this introduction to the methodology, I describe two approaches which have been used across my PhD project: thematic analysis, and case studies. Secondly, I discuss the autobiographical method used in paper 1 and paper 2. These papers take on a bottom-up perspective on participation in a game jam and a hackathon. Thirdly, I discuss the use of an exploratory prototype in paper 3. Next, I discuss the methods used to leverage insights into top-down perspectives on game jams and hackathons in the form of game jam organisers' understanding of creativity (paper 4) and researchers' usage of hackathons (paper 5). Lastly, I discuss advantages and limitations of the methodology of my PhD project. Table 6.1 shows a chronological overview of the methods employed in each paper. Figure 6.1 shows a chronological overview of the relative starting point of the studies of each paper (grey dots) and the relative publication date for the four published papers (black dots).

## Thematic Analysis

During the PhD project and across the papers, I have used thematic analysis as a method for analysing the data which was gathered via the above-mentioned methods: autobiographical design case studies, an intervention, an online survey and a literature review. Thematic analysis has become a recognised method to qualitatively approaching data analysis, and is a method: "[...] for systematically identifying, organizing, and offering insight into patterns of meaning (themes) across a data set." (Braun and V. Clarke 2012). The method offers a distinct approach to analysis via six phases:

1. Familiarising oneself with the data.
2. Generating initial codes.
3. Searching for themes.
4. Reviewing potential themes.
5. Defining and naming themes.
6. Producing the report.

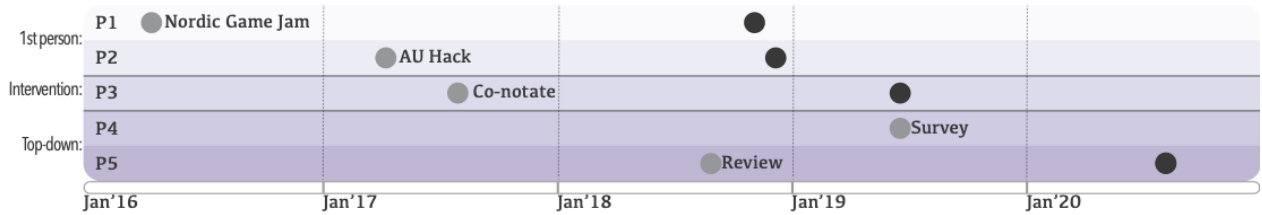

**Figure 6.1:** A timeline of studies (grey dots) and publication time (black dots) for paper 1, paper 2, paper 3, paper 4, and paper 5 (illustrated by purple bands. The grey dots illustrate the relative starting point of the study.

This iterative approach to developing codes and themes has been useful for structuring the identification of patterns in the data sets in a systematic way. For supporting and further structuring the thematic analyses I have used programs such as NVivo and Excel spreadsheets, which are useful for digitally assisting the recording, storing, indexing, sorting, comparing, and coding of qualitative data (Leech and Onwuegbuzie 2011). The purpose of thematic analysis is to identify patterns across a data set relevant to answering a particular research question, since numerous patterns could be identified across any data set (Braun and V. Clarke 2012). Therefore, I have aimed for transparency in the research papers by clarifying the particular research interests, perspective, and questions pertaining to each paper. In practice, thematic analysis served as a flexible analysis method, as it structures and ensures an iterative process. The iterative analysis process enabled a collaborative process with my co-authors, which ensured that the data sets were analysed and discussed thoroughly from multiple perspectives.

Thematic analysis can furthermore be approached inductively or deductively (Braun and V. Clarke 2012). An inductive approach is a bottom-up approach, and is driven by the data. The coding of the data therefore closely matches the content of the data. Conversely, deductive thematic analysis is a top-down approach, where the researcher draws on knowledge about concepts, ideas, or topics, which are used to code the data. Compared to the inductive approach, the developed codes in a deductive approach are more connected to the researcher's knowledge. As Braun and Clarke explain: "[...] here, what is mapped by the researcher during analysis does not necessarily closely link to the semantic data content." (Braun and V. Clarke 2012). However, in practice one can never be purely inductive or deductive, as it is often a combination of the two approaches: "[...] we always bring something to the data when we analyse it, and we rarely *completely* ignore the semantic content of the data when we code for a particular theoretical construct – at the very least, we have to know whether it is worth coding the data for that construct." (Braun and V. Clarke 2012). Paper 4 is an example where we had a deductive approach to the thematic analysis. The benefit of this was that we could compare the organisers' rather brief answers with theoretical aspects, and thereby support our analysis with more substance. In paper 5, we had a more inductive approach when conducting the thematic analysis, as we wanted an inclusive review of how researchers use hackathons. In both papers, we discuss which approach we predominantly use while emphasising that one can never be purely inductive or deductive. Therefore, stating for example one's research interest is important in this context.

## Case Studies

Another common denominator methodologically is how I approached the game jams and hackathon in paper 1, 2, and 3 as *case studies*. A case study is a detailed examination of a single example (Flyvbjerg 2006) in its natural context (Ramian 2007). Case studies are particularly suitable for inquiring complex phenomena (Ramian 2007), such as how game jams and hackathons shape and accelerate design processes. The kind of knowledge which case studies generate are context-dependent (Flyvbjerg 2006). The case studies conducted in paper 1, 2 and 3 can therefore be seen as generating *context-dependent knowledge*. The kind of knowledge they generate are therefore not generalisable, as the case studies are exemplars of the phenomena of game jams and hackathons. The case studies in paper 1 and 2 were furthermore approached as *autobiographical design*, a type of first person research, which I describe in the next section.

## 6.1 First Person Research

Paper 1 and paper 2 address the first research question of the PhD project: *How can we understand the accelerated design processes during hackathons and game jams?* In order to answer this, I wanted to obtain and capture detailed insights into accelerated design processes in a game jam and a hackathon format. In line with pragmatism, Holtzblatt and Beyer argues for observing *in the context* of a phenomena if this kind of detailed knowledge is wanted, as opposed to for example survey studies where people's memory can be unreliable (Holtzblatt and Beyer 2014). They elaborate:

"The everyday things people do can become habitual and unconscious, so they are generally unable to articulate their practices. People can say what they do in general terms and can identify critical problems; they can say what makes them angry with the tools they use. But they usually cannot provide day-to-day details about what they do. They cannot describe inner motivations such as the need to express a particular identity or to feel connected with people they care about. They are likely to forget about the workarounds they had to invent to overcome problems in their current products." (Holtzblatt and Beyer 2014).

Motivated by this perspective, I focused on capturing knowledge in situ of a game jam and a hackathon, and accordingly I applied a method for this in the particular context of an accelerated design process. To a great extent, paper 1 and 2 follow the same kind of method in order to leverage more detailed insights and contribute with rich exemplars on the inner workings of game jams and hackathons. The studies reported in both paper 1 and paper 2 can be described as first-person research, where the researcher's own perspective and experience is acknowledged and granted a central place (Desjardins and Ball 2018). In HCI this methodological standpoint has become more prominent concurrently with a shift of research foci from public/work spheres to private spheres of life (Chamberlain, M. Bødker, and Papangelis 2017). The particular methodological approach I have used for the first-person research in paper 1 and paper 2 is *autobiographical design*, which I elaborate in the section below.

## Autobiographical Design

Neustaedter and Sengers define autobiographical design research as: "[...] design research drawing on extensive genuine usage by those creating or building the system." (Neustaedter and Sengers 2012). Autobiographical design is similar to *autoethnography* which also revolves around the study of the *lived experience* with a system in a sociological context or situation. Subjectivity and the co-shaping of the research and researcher is emphasised and supported in autoethnography (Desjardins and Ball 2018). The main difference between autobiographical design and autoethnography is the additional focus on the designer/researcher's *designing* and *building* of the system in autobiographical design (Desjardins and Ball 2018). The design process is hence central to autobiographical design research.

Prior to the start of the PhD project I had considerable experience with participating in game jams and 24 hour design competitions.[1] As Desjardins and Ball notes, personal interests, curiosity and imagination as well as research purposes are important factors in initiating autobiographical design projects (Desjardins and Ball 2018). The initiation of my PhD project was similarly driven by my own personal interest in game jams, and I could furthermore draw on my knowledge as a participant in accelerated design processes. Because of my experience as a participant in accelerated design processes, I can be considered an *insider* of the phenomena I research. Researching a phenomena as an insider can be beneficial, as the researcher shares a language and experiences with other participants of the phenomena, who in addition potentially trust the researcher more (Hodkinson 2005). These aspects of an insider perspective can contribute to a greater understanding of the phenomena for the researcher. My experience was furthermore essential for carrying out the autobiographical design case studies in practice, as the double role of being both an active participant and observant researcher can be difficult. Knowing what to expect in an accelerated design process can accommodate for managing this difficult double role (Blomberg et al. 1993), as experience enables insight on for example: knowing when to document and interview without disturbing the ongoing activities and thereby compromising the authenticity of the design process, and when to, in a timely manner, design, develop and deliver, since the progression of the design process is dependent on this.

When employing autobiographical design research in the context of accelerated design processes there is an even greater focus on the re-

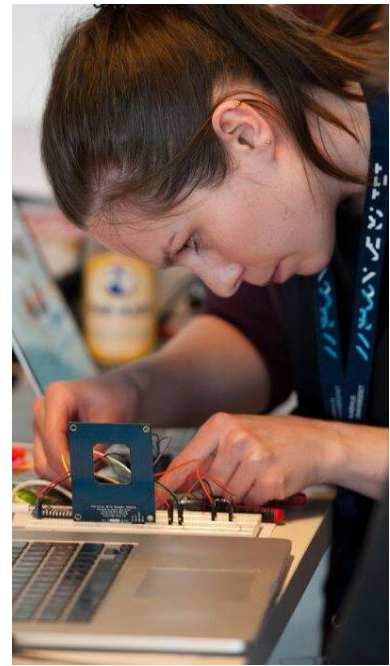

**Figure 6.2:** In the studies in paper 1 and paper 2 I participated in the game jam and hackathon as an active participant contributing to the accelerated design processes. Photo by courtesy of (*Major League Hacking* 2020).

1: On my personal website I have a portfolio of some of the game jams, hackathons and design competitions which I have participated in: `http://www.jeanettefalk.com/`.

searcher's design process since actually living with the designed system is often absent, because the longevity of the developed prototypes or systems is often short (Falk Olesen and Halskov 2020). However, an imagined experience of living with the designed system may be present as a consideration or guiding principle for the designer/researcher in the immediate accelerated design process. One example of an imagined experience of living with the designed system, is in paper 2 where we designed a system for a team member's partner, who was lactose intolerant. I use *autobiographical design case study* as a methodological framing for the case studies conducted in paper 1 and paper 2, in order to articulate and leverage my own experience with and insider perspective on accelerated design processes. I apply the term *case study* to the methodological framing to further emphasise the study of the game jam and the hackathon as single examples in their natural contexts. For both papers, we furthermore seek to encompass not only the perspective of the participating designers/researchers (In paper 1 the participating designer/researcher includes myself, while in paper 2 that includes Nicolai Brodersen Hansen and myself), but also the perspective of the group members. Both accelerated design processes in respectively paper 1 and paper 2 was namely *collaborative* efforts to design and build functioning prototypes within the time frames.

Documentation plays an important role in autobiographical design (Neustaedter, Judge, and Sengers 2014), and in the sections below I describe how documentation was done to support the reporting of the autobiographical design case studies in paper 1 and paper 2. Documentation supports recalling design details and rationales of design decisions, which may otherwise have been forgotten (Desjardins and Ball 2018). Documentation supports that in situ findings are captured for later analysis and reported authentically in a clear and honest way (Desjardins and Ball 2018). These in situ findings can for example be insights into the project's underlying rationale such as choices made, breakthroughs, challenges or even paths not taken (J. Bardzell et al. 2016). In both paper 1 and paper 2 we aimed at reporting details from the documentation in order to ensure authenticity and present the reader for the premises of the studies. Documentation of the accelerated design processes in paper 1 and paper 2 was conducted slightly different from each other. Reflecting on the method from paper 1, we decided to further develop the method used for documenting the design process in paper 2. In the two sections below I describe the documentation methods.

**Design Space Schemas**

In paper 1 we used a notation technique called *design space schemas* (Biskjaer, Dalsgaard, and Halskov 2014). These schemas are designed to document how a design space transforms during a design process. A design space schema is a simple table where a column represents an *aspect* of a design space, and the cells underneath a column represents *options* or *alternatives* for that particular aspect, see Figure 6.3. In practice, I started out with empty design space schemas, and filled in *aspects* and *options* during the game jam, when I assessed that the design space had developed and transformed. The documentation of the accelerated design process was further supported by field notes, recordings of group discussions, screenshots from the group members' computers, and photographs. A design space schema captures the opportunities of the *design object* which is being designed for and the documentation serve to answer *how* the design space dynamically transformed. As we also make clear in paper 1, the design space schemas were filled out whenever I as the participating researcher/level designer judged that the design space had transformed. These transformations were most clear after group discussions. In this sense, the collected data captured in the design space schemas is *reflected*, as the data collection is based on my in situ judgement.

**Status Updates**

In paper 2, we were interested in analysing how *design judgement* is influenced by certain factors during a hackathon. In order to analyse the in situ design judgements of why certain approaches were chosen over others during the hackathon we introduced frequent and semi-structured *status updates*. These status updates were prompted in the group regularly during the hackathons, in contrast to the group discussions in the game jam in paper 1 which emerged dynamically in the social context of the group. As our research question in paper 2 was focused on design judgement, we structured the status updates around: what each team member was working on and why, what challenges they were currently facing, and next steps. We asked these questions to be able to inquire *why* a certain approach was chosen over others, and thereby be able to analyse what factors influenced the design judgement of which design moves should be done during the hackathon. Three such status updates were recorded, and in addition to the status updates we recorded the initial idea generation and development until one idea was elaborated and planned. The status updates turned out to gain a double role: Initially, the status updates were intended as a resource for us as researchers, however they were also used in situ as a way for our group to know what each other were working on.

## 6.2  Intervention

In this section I discuss the method in paper 3. The method in paper 3 was an intervention where we facilitated an academic game jam, where 25 participants were required to analyse and reflect on design decisions during the game jam. The paper addresses RQ3: *How may we explore alternative ways of organising hackathons and game jams for academia and for supporting creativity?*, particularly with the purpose of capturing in situ knowledge for later analysis. We explored this in paper 3 by imposing a specific process structure on the participants and further encourage the documentation of their accelerated design processes using the exploratory prototype Co-notate. I elaborate how the findings from paper 3 contributes to answering the research question in Chapter 7. In the section below, I outline the method of conducting an intervention, or an *in-the-wild study* (Rogers 2012), with an exploratory research prototype.

| Character Design | Gameplay | Props | 9:50 |
|---|---|---|---|
| Flat on top:Carry something | Balancing box on top | Container with fluid | |
| Head: Carry underneath | Holding box underneath (hooks) | Container with glowing stuff | |
| | Holding box underneath (magnetism) | | |
| | Cooperative | | |
| | Cooperative with possibility of screwing it up for each other | | |
| | Singleplayer | | |

**Figure 6.3:** A design space schema with some aspects (top row), and options (column under each aspect) filled in. A version of the figure is published in paper 1.

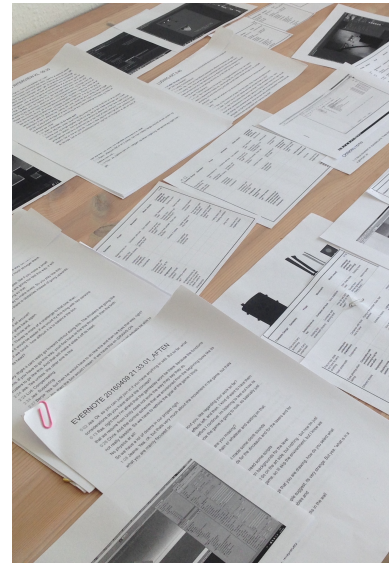

**Figure 6.4:** Collected documentation from the autobiographical design case study during the Nordic Game Jam: Transcribed interviews, screenshots, and design space schemas. A version of the figure is published in (Olesen 2017)

## Co-notate

Since documentation plays an important role for capturing in situ insights during accelerated design processes, the research prototype Co-notate, described in Chapter 2, proved to be an opportune and useful way for further exploring how to document accelerated design processes. Co-notate enables for multiple people collaborating on real-time *tagging* of interesting moments during an accelerated design process. The functionality of Co-notate is particularly interesting as it can capture in situ knowledge potentially without interrupting the accelerated design process considerably. Especially in a game jam setting where participants are highly pressured on time, documentation of the design process should be easy to do in order to not take time from other high priority tasks directly related to the game jam. If documentation takes too much time or becomes a hassle to do, the process of documenting and potentially capture in situ knowledge risk being given a lower priority. Furthermore, the prototype enables immediate searching and filtering of the annotated data, which is not only useful for post hoc analysis, but also for in situ endeavours, such as settling discussions on previous design decisions.

The setup of the study can be described as an in-the-wild study (Rogers 2012), which differs from ethnographic approaches, which the methods in paper 1 and paper 2 can be described as albeit they follow a particular ethnographic approach: autobiographical design. *Wild approaches* do not focus on observing existing practices, like ethnographic approaches do, but instead they focus on augmenting people, places and settings by installing interventions and encouraging different ways of behaving (Rogers 2012). As we emphasise in paper 3, the study is exploratory in nature. This follows the in-the-wild approach which does not seek to develop solutions that fit into existing practices, but where: "[...] the trend has been to experiment with new technological possibilities that can change and even disrupt behavior." (Rogers 2012). Therefore we were interested in learning about the participants' reactions to and experiences with using the Co-notate prototype in an accelerated design process which can be difficult to document.

The role as a researcher in an in-the-wild study, such as the one we conducted in paper 3, entails a shift of the locus of control from the researcher to the participant (Rogers 2012). As researchers, we facilitated the process however we did not have control over how the participants actually used Co-notate during the game jam. We were for example not present during large parts of the game jam and could not explain the purpose of the tool, remind participants to use it, or fix things if they did not go according to the plan. As a result, which we also discuss in paper 3, there were some examples on surprising situations with using Co-notate during the game jam. For example, even though the prototype is intended to enable all participants to tag in real time, many groups appointed one of their members to be the main tagger. Another interesting example was a group who used a tag to annotate moments with a 'Eureka!' tag, in order to capture immediate impressions of an interesting situation without having to decide how to describe it further. These surprising results were valuable for the insights on using a real-time annotations prototype to capture in situ knowledge.

## 6.3 Top-Down Perspective

Paper 4 and paper 5 addresses RQ2 *How have people organised hackathons and game jams in academia and for supporting creativity?*, and based on this, the two papers further addresses RQ3: *How may we explore alternative ways of organising hackathons and game jams for academia and for supporting creativity?* In Chapter 7, I address how paper 4 and 5 contribute to answering the research questions. In this section, I explain the methods used in paper 4 and paper 5. The methods in paper 1, 2 and 3 aim at *depth* in their contributions, by leveraging approaches from first person research and an exploratory in-the-wild study. In order to balance the findings of the PhD project, I wanted to contribute with *breadth* in the findings as well. Hence, the motivation for paper 4 and 5 is triangulating the research contribution of the PhD project with approaches, where the broader perspective of people's general experiences and conceptualisations of accelerated design processes are sought for. Paper 1, 2 and 3 revolve around generating deep insights from single game jams and a hackathon, whereas paper 4 and 5 accumulates insights from multiple game jams and hackathons: Paper 4 researches the perspective of how game jam organisers' understand creativity in regards to game jams, and what kinds of initiatives they take to potentially support participants' creativity. Paper 5 inquires how researchers have utilised hackathons as part of their processes, such as means for teaching, learning or research, or even as the subject for their research.

### Survey Study

Similar to hackathons, game jams have been commended for their potential to facilitate creativity. However, as we point out in paper 4 only few researches the relationship between game jams and creativity. Therefore, we wanted to research this particular relationship. Furthermore, we wanted to inquire the game jams and creativity from an *organiser* perspective, because an organiser decides on some of the constraints of a game jam format, which can potentially influence the participants' creativity. These constraints can for example be: the game jam venue, who is invited, what kind of material and support is provided, what events or workshops are planned during the game jam, what kind of theme, or other design constraints are provided.

In order to get *preliminary and structured* insight into how a broad range of game jam organisers conceptualise creativity in relation to game jams, we decided on conducting an online survey. Though surveys may not be able to provide insights on detailed processual aspects, they can still provide preliminary insights on overarching trends and salient aspects of a phenomena (Zook and Riedl 2013). One of the reasons for choosing a survey approach is that surveys can be useful for providing insight into for example respondents' attitudes, experiences, intents, and demographics (Müller, Sedley, and Ferrall-Nunge 2014), which we were particularly interested in in regards to game jam organisers. Furthermore, we based our decision on an online survey study on the assumption that an online survey can: "[...] effectively and efficiently reach widely distributed respondents [...]" (Law et al. 2009), meaning we could obtain insights into attitudes, experiences and intents from a *broad range* of game jam organisers. The survey was responded by 27 game jam organisers from Austria, Denmark, Finland, Germany, Ireland, Japan, Netherlands, Norway, Poland, Sweden, UK and US, and the survey consisted of a mix of open-ended and closed-ended questions. The open-ended questions were designed to gather insights on qualitative aspects of the organisers' perspective on creativity, while the closed-ended questions were multiple choice and rating questions. The latter questions were useful for addressing and comparing the organisers' attitudes. For example, the rating questions, based on a 5-point Likert-scale, were useful for comparing the organisers' attitude towards creativity against learning.

In general, the survey was useful for analysing the game jam organisers' attitude towards creativity in relation to game jams. The survey was designed primarily by Annakaisa Kultima and myself because of our own experience with participating in and organising game jams. To accommodate for potential *blindspots* in the survey design, we provided open-ended questions and further encouraged the responding organisers to provide detailed answers in order to enable potentially unexpected answers. Being aware that the survey provide preliminary and salient insights into how organisers understand and organise for creativity in relation to game jams, we highlight that the paper is a *first step* into studying a complex phenomena such as creativity in relation to game jam organising, and we furthermore suggest directions for future research on this particular relationship.

## Literature Review

Motivated by an interest in balancing the findings of the PhD project with *breadth*, and an interest in learning more about how researchers have used and shaped hackathon formats in academia, we conducted a review of research publications in the Association for Computing Machinery (ACM) Digital Library.

As we describe in paper 5, we aim for an inductive review of the 381 papers. Though we aim for *breadth* in paper 5 by conducting an extensive and inductive review and hence identify patterns emerging from the review sample, it is however impossible to be purely inductive (Braun and V. Clarke 2012). In the words of Hart: "There is no such thing as the perfect review. All reviews, irrespective of the research question and audience, are written from a particular perspective or standpoint that belongs to the reviewer. This perspective often originates from the school of thought, vocation or ideological standpoint the reviewer has chosen." (Hart 2018). This bias is not necessarily bad for the review findings, it merely implies a particular kind of reader for whom the findings will have relevance and interest (Hart 2018). We therefore highlight our research perspective in the methods section in paper 5, by stating our particular research interest in how formats shape accelerated design processes and furthermore by providing a detailed description of the review method. As a result of our research perspective and interest, the discussion section in paper 5 revolves around different *kinds* of hackathon formats which the reviewed publications described as part of research processes. Also motivated by our research perspective and interest, we argue that reflecting on the use of hackathons as part of research can advance research processes and lead to more systematic research with hackathons.

## 6.4  Methodology Reflections

As a consequence of the pragmatic worldview which have influenced my research methodology, the focus has been on the different research questions of the five papers, after which I have tailored suitable methods to appropriately approaching the different research questions. As I have outlined in this chapter, this has resulted in the employment of a range of different methods which have enabled me to address the three research questions of the PhD project. I have used these mixed methods to approach the different research questions based on the assumption that collecting diverse types of data provide a more complete understanding and answer of a research question (Cresswell 2014).

In the sections above I have discussed the different methods, and how we sought to accommodate for potential weaknesses of the methods. Therefore, in the following two sections I sum up on some of the salient advantages and limitations of the overall methodology.

### Advantages

**Autobiographical Design Case Studies**

The first person research method, autobiographical design, enabled me to gain rich insights into the inner workings of game jams and hackathons. Furthermore, by using this method I could utilise my own experience with game jams, since this enabled me to optimise how I conducted the study in practice. This approach ensured that the contributions to RQ1 were based on in situ knowledge.

**Top-Down Methods**

The top-down approaches enabled me to inquire accelerated design processes from a multi-perspective as well as the above mentioned first-person perspective. Here, we sought to leverage knowledge from a broad range of researchers and experienced game jam organisers. This approach enabled me to supplement the detailed, contextual knowledge from paper 1, 2, and 3 with insights on salient aspects of game jams and hackathons.

**Appropriate Methods for a Nascent and Intermediate Research Topic**

As I have argued in the beginning of the chapter, the state of prior work may be described as generally being nascent and intermediate. Therefore a mix of exploratory and qualitative methods, as have been conducted during my PhD project, can be seen as appropriate for the study of game jams and hackathons

**Continuous Peer Review**

A part of my PhD project's methodology, which I have not mentioned above, is the submission of the papers for publication, and the peer reviews which follows. This in itself has contributed to advancing my reflections on the methodology, and enabled me to clarify and strengthen the arguments in the papers iteratively and continuously during the PhD project.

## Limitations

### Blind Spots as a First Person Researcher

Though first person research can enable insights into the inner workings of accelerated design processes, there can also be bias connected with this kind of method, since the account of how the accelerated design processes unfold rely on the researcher. I have sought to accommodate for this, by providing transparent descriptions of how the method was conducted, and in that way clarify the circumstances for the reader. In paper 1, for example, I highlight that the data is *reflected*, meaning the data was captured when I assessed it was appropriate. Furthermore, I aimed at conducting meticulous documentation of the accelerated design processes in order to support the subsequent analysis.

### Order of Studies

The order of the studies arguably have had impact on the collective insights from conducting the studies. Being an insider of the phenomena I have studied, a natural step seemed to continue my participation in game jams and hackathons, however now with a researcher role and not just as a participant. Building on pragmatist design theory, I have emphasised the value of engaging deeply with practice. This engagement with practice first, contributed with detailed and contextual insights into some of the aspects which seem to characterise accelerated design processes. Building on this knowledge, I was experienced with the challenges of documenting in accelerated design processes, which further inspired the intervention with Co-notate in paper 3. The order of the conducted studies could have been reversed, so that the top-down approaches of the literature review in paper 5 and the survey in paper 4 could have been conducted before the first person research methods in paper 1 and 2, which then led to paper 3. In this way, I could have started my PhD project with a broader perspective on accelerated design processes. Starting with a situated knowledge of accelerated design processes have enabled me to identify a prominent gap in the research literature during my PhD project: Only few engages with the question of *how* the formats shape participants' design processes. Perhaps beginning with a literature review, such as the one in paper 5, would have motivated studies during my PhD project which focused more on approaching RQ3[2] , which I ultimately do not provide exhaustive and ultimate answers to. Instead, with the current order of studies, my PhD project contributes to a foundation for how we can understand accelerated design processes. This foundation, together with the collected insights on benefits and challenges of accelerated design processes, may instead then lay the ground for future research to explore alternative ways of organising hackathons and game jams.

2: How may we explore alternative ways of organising hackathons and game jams for academia and for supporting creativity?

**Particular Game Jam and Hackathon Formats**

The game jams and hackathon which were studied in paper 1, 2, and 3 were formats organised in particular settings. Therefore, the findings are not necessarily generalisable to game jam and hackathon formats organised in other settings, such as in purely online formats or in corporate settings. However, the Nordic Game Jam, which was studied in paper 1, have been used as a template for other game jam formats, such as the Global Game Jam (Fowler, Khosmood, and Arya 2013). The game jam in paper 3 was also organised after this template. Therefore, I argue that the formats are *representative* for many game jam and hackathon formats, the latter also following a typical pattern (Lodato and DiSalvo 2016). There are, however, still a need to further explore particular *adaptions* of game jam and hackathon formats.

**Asymmetrical Focus on Game Jams and Hackathons**

Though paper 1 and 2 are *both* autobiographical design case studies of respectively a game jam and a hackathon, paper 3, 4 and 5 are not in the same way symmetrical studies of hackathons and game jams. This means: I have not conducted an exploratory intervention with a real time annotations technology in both a game jam and a hackathon; I have not conducted a survey with organisers of both game jams and hackathon; I have not conducted a systematic and extensive literature of how research have engaged with both hackathons and game jams. However, conducting these studies symmetrically for both game jams and hackathons would likely not have been possible to do within the time frame of the PhD project. By conducting studies which alternately focused on game jams and hackathons I aimed at generating varied and multi-perspective insights on their common denominator of accelerated design processes. Via my PhD project I thereby lay the foundation for more extensive and symmetrical comparisons of game jams and hackathons for future research.

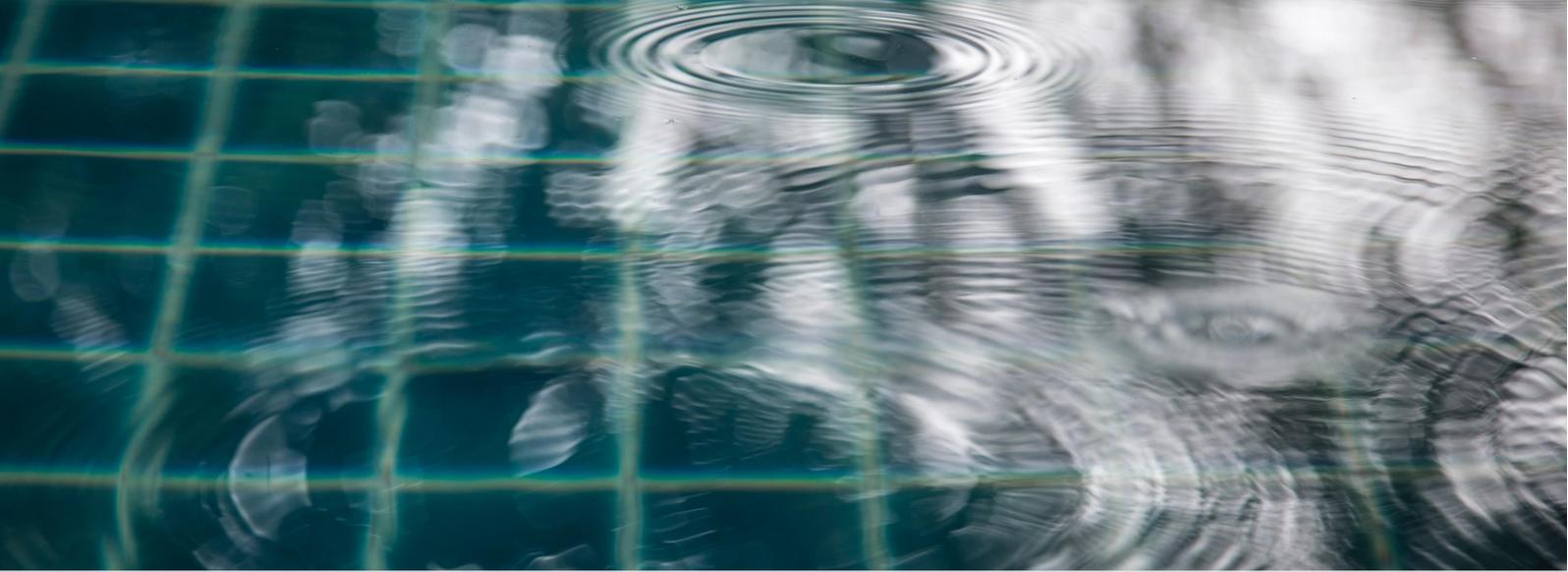

# 7 CONTRIBUTIONS

This chapter summarises the contributions of my research from the five papers and is structured around the three research questions:

**RQ1: How can we understand the accelerated design processes during hackathons and game jams?**

**RQ2: How have people organised hackathons and game jams in academia and for supporting creativity?**

**RQ3: How may we explore alternative ways of organising hackathons and game jams for academia and for supporting creativity?**

The PhD project contributes with *research on (or about) design* as well as *research for design* (Forlizzi, Zimmerman, and Stolterman 2009). Based on Frayling's original definition on *research into art and design* (Frayling 1993), Forlizzi, Stolterman and Zimmerman describe *research on design* as focused on: "[...] the human activity of design, producing theory that describes the process of design." (Forlizzi, Zimmerman, and Stolterman 2009). The goal is: "[...] to develop a detailed and unified understanding of the human activity of design or of design related activities such as creativity." (Forlizzi, Zimmerman, and Stolterman 2009). *Research for design* is intended to be applied in the practice of design, and can be described as: "[...] a *theoretical outcome* of many different activities that provides designers with theories in the area of *research for design*." (Forlizzi, Zimmerman, and Stolterman 2009). The contributions addressing RQ1 and RQ2 can be described as *research on design*, while the contributions approaching RQ3 can be described as *research for design*. Though I do not contribute with for example design guidelines and implications which are intended to support design directly, the contributions to RQ3 support the organisation of setups for accelerated design processes. In this way, the contributions to RQ3 are indirect research for design, where I argue that the setup for accelerated design processes can have implications for design.

The following three sections sum up the contributions. In the beginning of each section, I outline the contribution of a paper and how it addresses

Photo by Milada Vigerova on Unsplash

the RQ. In addition to the contribution of my work, I also identify some questions and areas of research on accelerated design processes for future research. I conclude each RQ section with a short sum up of the main contributions. I conclude the chapter with a discussion of some implications of my PhD project's contributions.

## 7.1  RQ1

**How can we understand the accelerated design processes during hackathons and game jams?**

In this section I refer to the findings in paper 1 and paper 2. Motivating their research of game jams, Turner, Thomas and Owen described that: "[...] 'something' very unique happens within a game jam and [...] this 'something special' is an important aspect of the potential future life of a vibrant games industry." (Turner, Thomas, and Owen 2013). This "special something" which happens during game jams is highly valued by the participants despite the fact that the developed games are evaluated as only average by the participants themselves (Preston et al. 2012). Similarly for hackathons, the developed prototypes are often not really usable (Zukin and Papadantonakis 2017) and rarely innovative: "What actually gets made in the way of apps or interfaces or visualizations is rarely innovative, however. The rough-hewn nature of the products keeps them from deployment or even rudimentary evaluation" (DiSalvo, Gregg, and Lodato 2014). Despite this, one of the main reasons for people to participate in hackathons is because of *learning* (Briscoe and Mulligan 2014), which indicate that participating in a hackathon is still perceived as beneficial.

As I demonstrated in Chapter 4, prior and related research have in general not studied the inner workings of accelerated design processes. Through my PhD project I contribute to this gap in research by contextualising and concretise this "special something". Informed by pragmatist design theory, and by working in a research environment which emphasised design processes, I wanted to use methods which enabled me to gain rich insights into the accelerated design processes in game jams and hackathons. I hypothesised that these kinds of insights could contribute to a deeper understanding of accelerated design processes. My aim is that a long-term effect of a greater understanding of accelerated design processes will better equip participants, organisers, researchers, and others who engage with these processes to *participate in*, *organise*, and *research* accelerated design processes for specific contexts and purposes.

In the following two subsections I sum up the main contributions and insights from paper 1 and paper 2 which contribute to a deeper understanding of accelerated design processes. As we also emphasise in paper 1, we do not claim that the findings of the study are generalisable and therefore valid for all instances of accelerated design processes in game jams. On the contrary, because the study was a single autobiographical design case study, the detailed accounts in the paper serve as *detailed examples* on design practice in the particular form of an accelerated design process. In highlighting that Schön's: "[...] account of reflective practice in design rests on a single account of a conversation between an architecture teacher and his student" (Wakkary 2004), Wakkary argues that: "[...] more accounts, case studies and examples of reflective practice in design are required." (Wakkary 2004). The first main contributions of my PhD project, as addressed in paper 1 and paper 2, should therefore be framed as rich exemplars which contextualise the phenomena of game jams and hackathons, and *how* game jams and hackathons may accelerate and shape design processes. These are important contributions of the PhD project, since not much research have been done on this, despite a widespread use of the formats. The exemplars hence contribute to a foundation for understanding accelerated design processes.

## Paper 1

For the study in paper 1 we identified and analysed four events, which significantly transformed the design space during the accelerated design process. We were interested in the *temporal* transformation of the design space, the interplay of design decisions, and how creativity constraints were created, encountered and managed. The four events were:

1. Initial establishment of the design space
2. Elaborating the design space
3. Inquiry into gameplay options
4. Breakdown of movement and gameplay in the digital prototype

The main contribution of paper 1 is the detailed written autobiographic design accounts, which seek to capture insights on *how* the design space during the accelerated design process transformed.

The written accounts in paper 1 demonstrate how oftentimes the game jam format had a significant influence on the accelerated design process. For the *detailed* written accounts on how the four events unfolded and transformed the design space I refer to paper 1. Here, I will highlight some examples from the four events which are characteristic of how the design process was accelerated.

Design decisions during the four events of the game jam were generally concerned with what was realisable within the limited time frame. **Event 1** was concentrated on aligning *expectations* and *preferences* of the group members in relation to developing a game prototype. One example of a preference was formulated by the graphic designer, who preferred 2D graphics instead of 3D graphics, since the former was faster for him to create and therefore more *appropriate* for the game jam. **In event 2** the design space was elaborated after one game idea had been decided. Even though an overall game idea was agreed upon, event 2 show that several *options* were in play and needed to be decided. Though this is not necessarily characteristic for accelerated design processes in particular, event 2 demonstrates how a design process in a game jam is a complex interplay of considering *aspects* and *options*. **In event 3** the complex interplay of considering aspects and options for the design is continued, especially with regards to the aspect of *gameplay*.

**Event 4** is particularly interesting in regards to understanding how a design process is accelerated by the format of a game jam. Here, we experienced a breakdown in the prototype which had significant repercussions for the design process. The breakdown meant that other parts of the design process was unable to progress. In the context of having to solve the breakdown within the short time frame of the game jam, a strategy of *reframing* the game design concept turned out to be a beneficial path in order to finish the game design in a timely manner. After having spent a considerable amount of time on developing and playtesting alternatives, we turned back to the defective part of the design and reframed how it could be understood. Reframing the breakdown was *necessary* as none of the alternatives worked, and time was running out.

**Table 7.1:** Transcripts from conversations during the second day of the game jam. (Not previously published.)

| Event 3, Discussion of gameplay, morning 2nd day | |
| --- | --- |
| 2:23 Level designer | Can we test it somehow? |
| 2:26 Programmer 1 | I can try and make a quick... |
| 2:31 Graphic designer | Yes, because what I think I'm afraid of, is that that will get really complicated really quickly, which might like towards the end, fuck us over, because we suddenly realise that something is like, broken, that we can't fix, because we are doing that, but its probably a more exciting gameplay. |
| 2:57 Level designer | Maybe. |
| 2:59 Graphic designer | Because you don't have to be as synchronised, if we loose the fluid in that thing, because on a string it can't really tip that much |
| 3:15 Programmer 1 | Yea, but you can still shake it, so if it falls over, if you have an open box at least |
| 3:21 Graphic designer | But, I mean we could bring it down to no fluid and just have the reactor core, and if you bang it or drop it, everything explodes into radiation and then. So i think yeah, on sort of narratively, fluid works better if there is, sort of holding a tray of things, and then the sort of don't bang the unstable reactor that kinda works better with strings. |
| 3:58 Level designer | Yea, I think you are right about that. |
| 4:02 Graphic designer | Im not at all against... |
| 4:04 Level designer | Me neither, but it could be nice if we could test it. |
| 4:05 Programmer 1 | Yes. Ill make a quick test. |
| Event 4, Design concerns, evening 2nd day | |
| 0:02 Level designer | Ok, so you can just join in if you have anything to add. But so far, what considerations do you have about the concept? |
| 0:13 Programmer 1 | right now I'm afraid that the gameplay won't be there, because the functions that we require from Unity does not work the way they say they do. |
| 0:28 Programmer 2 | And the control scheme that we envisioned from the beginning looks like its not really feasible. So we have to rethink the goal of the game I think. No physics at all! So we have a lot of downs in our project right. |

Table 7.1 shows excerpts from transcripts of two short discussions in my group during the second day of the game jam. The first transcript shows a discussion from event 3 on how design decisions revolved around what was *feasible* to make within the time frame vs. more exciting, but risky, gameplay. The second transcript shows a short conversation from event 4 on how the envisioned gameplay turned out to be unlikely to be able to make within the time frame because of intrinsic constraints of the game engine Unity.

The detailed accounts in paper 1 contributes to understanding *how* an accelerated design process unfolds in the context of a game jam. Paper 1 illustrates how an accelerated design process can be conceived as a complex interplay of co-dependent and mutually susceptible design decisions which dynamically transform the design space, which is much alike design spaces in non-accelerated design processes. As we argue in the paper, several of the same strategies occur in the context of an accelerated design process in a game jam as in the non-accelerated five-week design process studied by Biskjaer, Dalsgaard and Halskov (Biskjaer, Dalsgaard, and Halskov 2014). Paper 1 also illustrates how *certain* design decisions make sense and are *prioritised* in the specific context of an accelerated design process, such as continuing the design process with a "defective" prototype after having reframed the "defective" part by turning it into a feature. Had the design process been non-accelerated more effort may have been put into developing a better-functioning alternative.

**Future Research**

Based on the contributions from paper 1, we suggest additional studies of how the format of a game jam may shape how a design space is transformed. This could be done by exploring and comparing how the design space and strategies change during respectively accelerated design processes and non-accelerated design processes. We furthermore suggest future studies of the role of creativity constraints in the context of game jams, and how these creativity constraints may shape strategies of navigating design spaces. Studies like these can contribute to further developing what it is that characterises accelerated design processes. In addition, these kinds of studies can support the organisation of accelerated design processes by providing greater insight into how different factors may impact the design process.

## Paper 2

For paper 2, we were interested in how design judgements were *influenced* and *driven* by a hackathon situation. As we hypothesise in the paper, design judgements made in a hackathon are supposedly different from design judgements made in other contexts since the former must be made rapidly due to a short time frame and limited resources. We identified four factors which in particular impacted design judgements during the accelerated design process:

1. "**The hackathon format:** Frames and imposes certain elements to the design process. Prompts prioritizing of resources.[1]
2. **Tools and materials:** Play a central role in idea generation, since the design process and development of a prototype depends on what tools and materials are available. [2]
3. **Domain knowledge:** A main driver for idea generation and user insights. Sought internally and highly dependent on the composition of team members. [3]
4. **Technical knowledge:** The team depends on their internal technical knowledge and what is actually realisable within the team." (Olesen, Hansen, and Halskov 2018).[4]

Even though the factors may be found in non-accelerated design processes as well, we argue that *the way* in which the factors impacted design judgements was *characteristic* of the accelerated design process.

1: **The hackathon format** includes: "The short time frame, the case selections, the availability of the Tech Wizard booth, the judging criteria, the scheduled presentation session at the end of the hackathon, where judges also tried the prototypes." (Olesen, Hansen, and Halskov 2018).

2: **Tools and materials** includes: Version control software such as GitHub, hardware development kits like the Particle Photon, a selection of hardware from the Tech Wizards like DC motor, RFID tags and reader.

3: We define **domain knowledge** as: "[...] participants' knowledge and former experience with respect to the domain being designed for." (Olesen, Hansen, and Halskov 2018).

4: We define **technical knowledge** as prior knowledge about working with software or hardware.

## Hackathon Format

Comparable to the accelerated design process in paper 1, it became clear how design judgement was oriented towards the immediate situation of the hackathon in paper 2 because of the *format of the hackathon*. When our group encountered a problem late on the second day of the hackathon, the concern was how we could develop a *presentable* and *functioning* prototype rather than for example a secure one, as expressed by one of the other group members: "...we (P1 & P2) found a problem between the backend and the frontend where they can't communicate, because of some internet security basically, which means we can't run the javascript from the frontend to the backend. . . because GitHub updated their pages in July 2016, so that it's safe to use them. . . And it's not good for us that it's safe. It's good for the rest of the world." (P2, second day, status update, T:23:25:34).

## Tools and Materials

The *availability and selection of tools and materials* were significant drivers for our design judgement. For example, even though certain concepts were envisioned in our group, we did not have access to the proper tools and materials to prototype those concepts, which were then discarded. As we were committed to developing a functioning prototype within the time frame, we were therefore dependent on what tools and materials were available in for example the tech booth at the hackathon site. One type of envisioned, however unavailable, output was therefore replaced with a different but available output. These types of *exchanges* happened frequently, where vision were *constrained by* and *aligned with* the immediately available building blocks at the hackathon site.

## Domain Knowledge

The factor of domain knowledge is interesting in the context of accelerated design processes, as this was how we identified potential *target users* and *user needs* for our prototype. However, though we as *designers* and *developers* ourselves of course possess varied experiences which constitute our individual domain knowledge, we are not necessarily *target users*. Though we aimed at being user centred in our accelerated design process, our knowledge about the domain being designed for came from a *second hand perspective*. Our domain knowledge in this case can then be described as being based, at best, in empathetic anecdotes about living with a lactose-intolerant person, or, at worst, in stereotyped assumptions about the user.

## Technical Knowledge

As feasibility within the time frame was a main concern for our accelerated design process, we had to align ambitions regarding the prospective prototype to the *technical skills* each of us possessed, as expressed by a participant: "[...] But. . . it is much easier to build with the technical competences that we have if it's physical. Because I have some experience with augmented reality. But if it is only me who can build that, it might not be the most optimal thing to do for our team...we should adjust to our team." (P2, Friday, idea generation, T:23:09:46). The technical knowledge which our group possessed about certain tools and materials would also be prioritised over the availability of tools and materials. This meant that even though there may have been tools or materials immediately available at the hackathon site which were related to our concept vision, if we did not have the necessary technical knowledge, we would not choose that tool or material. This could have been a possibility if one wanted to learn new technical skills in relation to an available tool or material, however learning a new tool or working with a new material could be risky.

### Future Research

For future research, we suggest studies into the potential effect on the idea generation phase when a functional prototype has to be developed almost immediately after coming up with ideas. This is based on the observation that the idea generation phase was highly focused on practicalities of giving form to the ideas. Additionally, these studies could explore the effect on the idea generation phase which different kinds of prototyping may have. Different kinds of prototyping could be for example low-functionality prototyping using video or paper vs. more high-functionality prototyping using Arduinos or Raspberry Pi's. We could for example ask: Does high-functionality prototyping entail a greater inwards oriented focus (thereby potentially disregarding concerns for real-world impact) than low-functionality prototyping? Perhaps developing low-functionality prototypes enable participants to focus less on the practicalities of a functional prototype, and more on exploring how their idea potentially may have an impact on people. Furthermore, future studies could explore whether an inward oriented focus occur less in hackathons which are organised with the intention and incentives for continuing the development of prototypes after the hackathon deadline: That is, do these incentives have an impact on participants' idea generation and prototyping? And if not, how can accelerated design processes be organised to support longer-term effects, if that is the purpose of the hackathon?

### RQ1 Contributions

Paper 1 and paper 2 contribute with *descriptive and detailed process-level knowledge* on how two accelerated design processes was shaped by the formats of respectively a game jam and a hackathon. The two papers serve as *exemplars* on accelerated design processes, which contextualise and solidify how a design process may be accelerated in a game jam and hackathon format. This is an important contribution of the PhD project, as it contributes to a foundation for understanding accelerated design processes. Specifically, we observed how group composition entail particular *preferences, expectations, domain knowledge*, and *technical knowledge* which affect what is feasible for the group to achieve within the short time frame of a hackathon or a game jam. Furthermore, the two papers demonstrate how the constrained format can entail a *heightened focus on the situation of the game jam or hackathon*, or in other words an *inward oriented focus*, and not necessarily a focus on impacts beyond the immediate design situation, because the limitation of resources such as time, tools and materials are *particularly* *pervasive.* This has impact for what can be expected to be designed in similar game jam and hackathon formats as the ones we studied.

Based on the contribution to RQ1, future research can explore the following research questions:

- ▶ How do the formats of respectively accelerated and non-accelerated design processes shape how a design space is transformed?
- ▶ Which creativity constraints occur in accelerated design processes and how do they shape strategies of navigating design spaces?
- ▶ How is the idea generation phase impacted by an almost immediately subsequent prototyping phase?
- ▶ How do different kinds of prototyping impact the preceding idea generation phase?
- ▶ Does incentives for continuing the development of prototypes after hackathons have an impact on participants' idea generation and prototyping?
- ▶ If not, how can accelerated design processes be organised to support and ensure longer-term effects?

## 7.2 RQ2

**How have people organised hackathons and game jams in academia and for supporting creativity?**

As the contributions to RQ1 show, the way in which design processes in game jams and hackathons are accelerated are influenced by the *format* of the game jam or hackathon, or in other words, by how the game jam or hackathon is *organised*. In this section I highlight the findings of paper 4 and paper 5, which contribute with insights on respectively how game jam organisers understand and promote creativity in game jams, and how researchers use hackathons as part of or as the focus of their research. The contributions with respect to RQ2 are hence retrospective and contemporary overviews of hackathons and game jams. Therefore, I do not suggest directions for future research in this section, however, in answering RQ3, which is more prospectively oriented and speculative in nature, I return to discussing directions for future research based on the findings from paper 4 and 5, and in addition to the findings of paper 3.

### Paper 4

Paper 1 and paper 2 contribute with insights into how accelerated design processes in a game jam and hackathon format are also creative processes. For example, we analysed how personal preference serve as *self-imposed* and *generative constraints* for how the accelerated design process unfolded. The four factors in paper 2 also had significant influence on how the ideation for the concept unfolded. These findings point towards that the way a hackathon or game jam format is *organised* can potentially influence creativity, which is our motivation for and focus in paper 4.

The contribution of paper 4 is a first step towards understanding and exploring creativity in the setting of a game jam, and is therefore not an exhaustive answer. One of the contributions is establishing that creativity is indeed a concern for game jam organisers, and more so than for example supporting participants' learning, which is otherwise a main motivation for participants to attend game jams (Arya, Chastine, et al. 2013; Fowler, Khosmood, and Arya 2013).[5] We furthermore found that organisers find it important to support participants' creativity, see Figure 7.1. This further motivated the study on how the organisers understand and promote creativity as well as how their best practices can be furthered, since organisers are in a position where they can greatly affect the format of the game jam.

As is fit for nascent research, the paper proposes tentative answers to how game jam organisers understand and promote creativity, as well as proposing how organisers may advance their practices of promoting creativity (Edmondson and McManus 2007). In the words of Edmondson and Manus, the contribution of such nascent research is: "A suggestive theory, often an invitation for further work on the issue or set of issues opened up by the study" (Edmondson and McManus 2007). The main audience of the paper is therefore game jam organisers who specifically wish to support and promote participants' creativity, but also researchers who wish to study creativity in the context of game jams. For the latter, researchers may further explore creativity in game jams by considering the four key aspects. In addition to the four aspects, we also identify several directions for future research on the relationship between creativity and game jams.

Based on our knowledge of and experience with creativity research, we propose four key theoretical aspects, which we argue play particularly significant roles in the context of game jams:

▶ **Novelty** is a definitional criterion of creativity. Referring back to my PhD project's working definition of creativity, novelty is "[...] a measure of how unusual or unexpected an idea is as compared to other ideas." (Kolko 2007)
▶ **(Intellectual) Risk-taking**, in the sense of flexibility, open-mindedness, and tolerance of ambiguity, is an important part of promoting a creative mindset (Harris 2004; Prabhu, Sutton, and Sauser 2008).
▶ **Combinational creativity** is seen as vital in creativity and refers to when one: "[...] discover new and potentially useful emergent properties; that is, properties not commonly seen in the component concepts, but that emerge only in combination." (Boden 2004).
▶ **Creativity constraints** can be defined as "[...] all explicit or tacit factors governing what the agent/s must, should, can, and cannot do; and what the output must, should, can, and cannot be" (Onarheim and Biskjær 2013). Constraints then have a double role as both enabling and limiting.

We unfold a more extensive theoretical outline of the four aspects in paper 4. These four aspects were used as guiding *propositions* for the thematic analysis of the organisers' responses. We therefore expected to see these aspects reflected in the organisers' responses.

Through the analysis, we saw that the organisers reflect the four aspects in *nuanced ways* in their responses to how they understand creativity in

5: Though there therefore may be a mismatch between organisers and participants in terms of what is important in a game jam, supporting creativity can still prove beneficial if the main goal is learning, because learning is tightly connected to creativity: for example, Wallas described the creative process as starting with a *preparation phase* where one learns about a problem and gather relevant information (Wallas 1926). More recently, Plucker, Beghetto, and Dow also framed creativity as an integral part of learning (Plucker, Beghetto, and Dow 2004).

game jams. For example, several organisers reflected the theoretical aspect of *creativity constraints* by responding how game jams can be liberating or restraining on participants' creativity. However, not many responses reflected the *dual role* of creativity constraints, where they can at the same time be both restraining and liberating on creative processes (Boden 2004). Consequently, we argue that creativity research can support and further organisers' practices for promoting participants' creativity, as creativity research can offer both new perspectives (e.g. the dual role of creativity constraints) and suggestions for initiatives which organisers can employ during game jams.

The following sections sum up how the organisers understood creativity in relation to game jams in terms of the four key aspects, and furthermore how they have sought to support their participants' creativity. For the detailed findings I refer to paper 4.

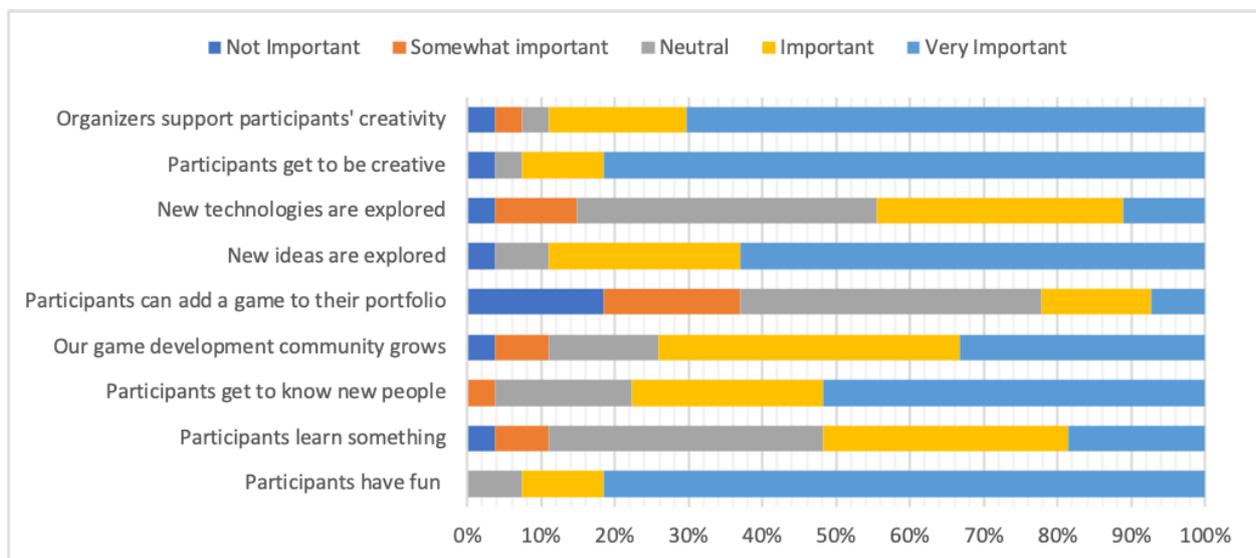

**Figure 7.1:** A five-point Likert scale showing the distribution of how important the organisers found each statement. Especially the statements: "Organisers support participants' creativity" and "Participants get to be creative" are interesting in this case, where we can determine that organisers indeed find creativity to be an important part of game jams (Falk et al. 2021).

**How the Game Jam Organisers understood Creativity**

**Novelty:** The aspect of novelty was a central concept for how the organisers understood creativity, and how they understood novelty as pertaining to both the developed game prototypes and to the participants' accelerated design process. Novelty was expressed as the making of games which had for example not been seen before and were in some way surprising. In creativity research, novelty is usually an aspect which concern the outcomes of a creative process. Some organisers' understanding of creativity in terms of novelty also concerned the participants' *process* in addition to creative outcomes. Regarding novelty in the accelerated design processes of participants, it was for example expressed as "[...] trying new things, stepping outside of your comfort zone as a game developer, finding solutions to problems in ways you hadn't thought about before [...]" (Organiser 26).

**Risk-taking:** This was often mentioned in relation to creativity in game jams, and was also mentioned as a way for participants to achieve novelty, by stepping outside of their comfort zone and explore new ideas that the participants: "[...] normally wouldn't when trying to make a commercial game." (Organiser 8). Many organisers understood a game jam as a time and place where the participants can experiment, and where it is accepted to take risks and fail.

**Combinational creativity:** This aspect was reflected by only a few organisers as the combination of game concepts, game mechanics, or just ideas in general in maybe "unorthodox ways" (Organiser 19) and in that way creating something new. As with risk-taking, this way of combinational creativity was understood as a way of obtaining novelty for participants.

**Creativity constraints:** Several organisers addressed the aspect of creativity constraints, by explaining how a game jam can both liberate and limit participants' creative accelerated design processes. One organiser expressed how game jams can be liberating for participants' creativity: "The freedom to do whatever you want with the technology at hand [...]" (Organiser 4), while some reflected how game jam formats can be restraining for creativity, as for example expressed by organiser 12: "game jam venues and budgets put various restrictions on what can be achieved" (Organiser 12).

One reason for conducting the thematic analysis deductively, was that the organisers' answers were generally quite shallow, as I also explain in Chapter 6. The organisers' brief responses on how they understand creativity can also point towards the possibility that the organisers operate with a more colloquial understanding of what creativity is. This again suggests that the organisers may be able to advance their practices to support participants' creativity by leveraging knowledge from creativity research.

### How Game Jam Organisers Promoted Participants' Creativity

When asked how the organisers attempted to promote their participants' creativity, several of the organisers reported a range of different types of initiatives, which I briefly describe in the following:

**Establishing the physical surroundings:** These initiatives included: having art installations at the game jam venue, providing games where the participants could play, having rooms at the game jam venue dedicated to relaxation, and having mobile escape rooms available. We frame these initiatives as closely related to the aspect of *creativity constraints*.

**Supplying tools and materials:** The organisers reported a range of different tools and materials which were provided to participants, for example paper, markers, assets for the Unity Asset store, motion capture suits, and even a dome, where a person instructed participants on how to use it. For a specific game jam where the game developers collaborated with researchers, the researchers were asked to bring: "[...] any items or media they feel is helpful for this process." (Organiser 21). Like the above-mentioned initiatives for establishing the physical surroundings, supplying tools and materials can be argued to be related to creativity constraints (Dalsgaard 2017). An organiser (O12) reported how they provided tools to participants to "play around and mess with" in order to create a space safe for failing, and, thereby, for taking risks. Highlighting this example, we argue that supplying tools and materials can be a way of encouraging *risk-taking* as well as a way to introduce *creativity constraints*. We elaborate on this in the Discussion in paper 4, and I return to this discussion in Section 7.3.

**Selecting a theme:** Themes are a recurring aspect of game jams, and has been described as definitional for game jams (Kultima 2015a). A theme in a game jam is intended to inspire the participants, and is perhaps the most used way for organisers to support participants' creativity. 77.8% of the organisers reported using a theme when organising game jams. Organisers often reflected an understanding of creativity in game jams in terms of *creativity constraints*, and selecting a theme was a clear example on an initiative closely related to creativity constraints which the organisers took.

The following two initiatives, Providing talks and facilitating discussion and Organising different activities, were closely related to *inducing novelty*, as the organisers through the two initiatives sought to introduce elements of surprise, a key component of novelty (Gerten and Topolinski 2019).

**Providing talks and facilitating discussion:** Organising a talk as part of a game jam was a widely used practice intended to inspire the participants, and 55,6% of the organisers reported organising talks. These talks could take the form of expert talks, lightning talks, and, a more surprising form of talk; a round-table conversation with a monk for a game jam combining game design and religion.

**Organising different activities:** Besides talks, many different kinds of activities were reported as ways of supporting participants' creativity. This could be "show-and-tell" sessions, where participants during the game jam present their provisional concepts for each other before continuing the game jam. Some organisers would check in on how the participants were doing frequently during the game jam. One organiser organised physical warm-up games (Organiser 25), while another organised a: "small creativity exercise at the beginning of the event" (Organiser 19). Another example is Organiser 8 who organised a "speed dating" exercise where "[...] jammers got to brainstorm & know each other a bit more". A couple of surprising activities were organised by Organiser 20, who let the participants swim with toys at one game jam, and at another game jam let participants and refugee children play together.

Despite the possibility that the organisers may understand creativity in a colloquial way and may not necessarily be informed by creativity research as such, the organisers reported many interesting attempts of supporting participants' creativity. These initiatives may very well serve the purpose of supporting the participants' creativity, however we did not investigate whether those attempts actually have an effect on participants' creativity or not in paper 4. Rather, paper 4 contributes with an inquiry of how the organisers' themselves believe that they support the participants' creativity.

In informing my answer to RQ3: *How may hackathons and game jams be organised in academia and for supporting creativity?*, I build on our discussion from paper 4 on how game jam organisers may advance their practice of supporting participants' creativity by leveraging insights from recent creativity research. I therefore return to this contribution of paper 4 in Section 7.3.

## Paper 5

In paper 5 we review 381 papers and synthesise the insights from the review by documenting multiple exemplars of different research practices using hackathons. The contribution is an *overview* of thematically representative exemplars, and the identification of *three overarching motivations* for using hackathons as part of research practice: *Structuring learning, structuring processes*, and *enabling participation*. The overview of the exemplars is divided into these three motivations. Furthermore, for each motivation we categorise the paper into whether it is research *with* hackathons (constituted approx. 88% of the reviewed sample) or research *on* hackathons (constituted approx. 12% of the reviewed sample), meaning that the research used hackathons respectively as *means* for research purposes or as the *focus* of their research. As we intend to provide an overview of the many different ways hackathons have been employed as part of research, we aim for *breadth* over *depth* in how the examples are reported. Based on the review, paper 5 further contributes with a *condensed overview* of how the researchers have framed both the *benefits* and the *challenges* of using hackathons in academia, structured after the three *motivations*, see table 7.2.

In the following, I sum up the overview from paper 5, and I highlight some key examples. For the *detailed* findings of the exemplars I refer to paper 5.

### Structuring learning

This motivation constituted approx. 22,5% of the reviewed sample. Research *with* hackathons furthermore constituted 21% of this motivation, while research *on* hackathons constituted 1,5%. Hackathons were used as both extracurricular activities, and as integral parts of formal education. Hackathons used for the purpose of structuring learning are aimed at for example improving the participants' technical, communicative, collaborative or creative skills. Students are even recommended to participate in hackathons, while learning has been found to be one of the main motivations for participating in hackathons (Briscoe and Mulligan 2014). The papers representing research *on* hackathons with the motivation of structuring learning focus on how hackathons can act as a learning instrument and how they can support learning. However, one challenge which was mentioned was a lacking formal structure and pedagogy in hackathon formats, which may hinder learning (Warner and Guo 2017).

### Structuring processes

This motivation constituted approx. 43% of the reviewed sample. 35% of the papers used hackathons as a *means* for structuring processes, while 8% of the papers were research *on* how hackathons shape processes. The focus for research *with* hackathons with the motivation of structuring processes, was on the *outcome* of the hackathons or the *facilitation of* *specific outcomes*. Outcomes in this contexts were for example *prototypes*, or *empirical data* like observations. The exemplars for the papers which represented research *with* hackathons were further divided into the following four topics:

- ▶ **Producing and studying hackathon outcomes:** Several of the reviewed papers presented a hackathon outcome as part of their research, where the focus in the paper was on the outcome.
- ▶ **Hackathon-inspired workshop formats:** Many researchers have embraced hackathons as inspiration for structuring workshops at for example conferences, or as part of their research method.
- ▶ **Increasing collaboration between stakeholders:** A motivation often mentioned for structuring processes with hackathon formats was the close collaboration which could be promoted between stakeholders, such as students, researchers, developers and practitioners.
- ▶ **Evaluating prototypes:** Hackathons were also used to evaluate prototypes which had been developed beforehand.

The papers which represented research *on* hackathons were further divided into the following five topics:

- ▶ **Shaping processes and outcomes:** These papers focused on how hackathons in different ways affected processes and outcomes. Paper 2 is an example on this. Different ways of organising hackathons were discussed in the papers, such as a format for an "innovation-driven" hackathon (Frey and Luks 2016). An interesting discussion in this context is DiSalvo, Gregg and Lodato who discuss that the technical outcomes of hackathons are secondary to outcomes such as the ways in which participants imagine and perform citizenship (DiSalvo, Gregg, and Lodato 2014). However, they further discuss that hackathons may reinforce a *certain technological citizenship*.
- ▶ **Mismatch in expectations:** Since many attend hackathons for the sake of learning, this may clash with for example organisers and stakeholders' expectations to the outcomes of the hackathon.
- ▶ **Use of hackathons in scientific communities:** Some papers revolved specifically around the use of hackathons in scientific communities with the purpose of for example creating scientific software features.
- ▶ **Hackathons and other accelerated design processes:** Papers within this topic researched formats related to hackathons, such as game jams. Paper 1 is an example on this. Some papers within this topic compare hackathons and game jams, and we argue that these kind of comparisons can be worthwhile to for example clarify similarities and differences.
- ▶ **Sustainability of hackathon processes and outcomes:** A concern which was mentioned often in several papers was the sustainability or the longevity of the hackathon processes and outcomes. For example, Ferrario et al. discusses that despite the admirable ambition of some hackathons to do "social good", it can be problematic when oftentimes hackathon participants' understanding of complex social issues are obtained informally, for example via discussions with representatives at the hackathon venue (Ferrario et al. 2014).

**Enabling participation**

This motivation constituted approx. 34% of the reviewed sample. 32% of the papers used hackathons *as settings* for engaging citizens and communities in different ways. 2% of the papers were research *on* the particular kind of participation hackathons promote. The papers representing research *with* hackathons were divided into the following three topics:

- ▶ **Engaging and empowering citizens and communities:** Many papers used hackathons to engage citizens and communities by for example using open government data as a material for the hackathon. Within this context, another paper revolved around how hackathons should not be judged on whether the technical outcomes were picked up for later development, but rather on how hackathon prototypes can be understood as public speculation about possible futures (Boehner and DiSalvo 2016).
- ▶ **Increasing diversity:** Hackathons have widely been used to promote diversity in different contexts. There can however be some problems with hackathons and promoting diversity; Dombrowski, Harmon and Fox found that if participants were for example perceived as too "political" by their peers, the "political" participants risked being excluded from groups as conflicts were seen as disruptive for the accelerated design process (Dombrowski, Harmon, and Fox 2016).
- ▶ **Hackathons as maker activities:** Hackathons were often mentioned as an example on a typical *maker* activity. For this reason, we included some key points of discussions on maker culture in general, since these also apply to hackathons. For example, though the maker movement, including the activity of hackathons, can be seen as an effort to democratise technology development, one concern is an underlying *techno-solutionism*.[6] However, Jenkins, Dantec and DiSalvo argue that even though hackathons may form around an idealistic premise, they should not be dismissed on this account as there is value in how hackathons assume that technology can play a role in social issues and effect change in: "[...] contexts where human action alone has not succeeded in doing so." (Jenkins, Le Dantec, et al. 2016).

6: *Techno-solutionism* refers to the perhaps naive view that technology can unilaterally solve difficult social problems (Lindtner, S. Bardzell, and J. Bardzell 2016).

Papers representing research *on* hackathons were divided into the following two topics:

- ▶ **Diversity in participation:** Though hackathons have been mentioned as a way to increase diversity, several papers discuss how hackathons are however often attended by a homogeneous crowd despite a "come-one-come-all" ethos. In order to accommodate for this, Richard et al. argue for designing hackathon formats to appeal more broadly by following recommendations from the National Center for Women and Information Technology (Richard et al. 2015).[7]
- ▶ **Organising hackathons for broad participation:** Related to the former topic, the papers sorted into this topic were more focused on the question of *how* to involve different groups.

7: The recommendations are: 1) including promotional materials that feature women and a diverse range of students, 2) actively recruit women, 3) provide ongoing encouragement, 4) allow participants to create projects that appeal to them, 5) encourage mixed teams with experienced and inexperienced members, 6) host a tutorial or how-to event, 7) focus on learning and different ways to win, 8) include female mentors, educators, and judges. 9) make sure the space is accessible to all, and 10) educate others involved (Richard et al. 2015).

Some of the concerns which were voiced in the reviewed papers were about how participants at hackathons may obtain information via infor-mal conversations with representatives of target users, hence ignoring formal social science methodologies specifically for gaining a *deep understanding* of certain topics, which can last months (Ferrario et al. 2014). These concerns reflect the contributions to RQ1, where I argue that the acceleration of design processes may entail an *inward oriented focus*, where knowledge about potential target users is gained for example via the immediate domain knowledge of the participants themselves.

**Table 7.2:** A condensed summary of benefits and challenges for both research *with* and *on* hackathons, sorted after the three motivations: Structuring Learning, Structuring Processes, Enabling Participation.

| BENEFITS | CHALLENGES |
|---|---|
| **Structuring Learning** | |
| ▶ Engaging participants in user research, requirements establishment and evaluation stages<br>▶ Immerses participants into ideation<br>▶ Prompting rapid decision-making<br>▶ Collaborative and peer learning<br>▶ Hands-on creativity<br>▶ Real-world problems<br>▶ Collaboration and interdisciplinarity<br>▶ From the highly theoretical to the concrete<br>▶ Learning and skill-development | ▶ Lacking formal structure and pedagogy<br>▶ Lacking feedback from instructors |
| **Structuring Processes** | |
| ▶ Hands-on assistance from experts<br>▶ Hands-on approach and solutions<br>▶ New research skills and theoretical perspectives<br>▶ Lightweight, empirical research products<br>▶ Supporting refinement of ideas and research questions<br>▶ Facilitate new research projects and publications<br>▶ Negotiation of trade-offs between what is desirable and feasible in practice<br>▶ Setting common task in a motivating way to participants<br>▶ "Fun and easy" way to understand participatory design and code development<br>▶ Technical expertise<br>▶ Design process experience<br>▶ Imagining citizenship in new ways<br>▶ Commitment through group dependence and appreciation of work<br>▶ Contributes to array of HCI research methods | ▶ Privileges flashy demos over careful deliberation<br>▶ Uncertain outcomes<br>▶ Variety of communities of practice<br>▶ Technologies best understood in the end of hackathon, leading to late experimentation<br>▶ Pace can be problematic for case collection, user testing, and interaction design in general<br>▶ Not suitable development format for all solutions<br>▶ Deep, creative insights about theoretical topics cannot be expected while also dealing with unfamiliar tools and materials<br>▶ Utility and quality issues with prototypes<br>▶ Participants' understanding of issues are gained informally and from personal experience<br>▶ Ignores otherwise lengthy methodologies for gaining a deeper understanding<br>▶ Limited sustainability and implementation of hackathon outcomes<br>▶ Methodological hegemony<br>▶ Reinforcement of certain technological citizenship<br>▶ Mismatched expectations between collaborators |
| **Enabling Participation** | |
| ▶ Networking and meeting like-minded people<br>▶ Public speculation about possible futures<br>▶ Diversifying participation in computing<br>▶ Assuming technology having a role in social issues | ▶ Techno-solutionism<br>▶ Inclusivity and diversity challenges<br>▶ Participants perceived as too "political" can be excluded from groups<br>▶ Negative experiences in hackathons can turn off minority participants from future similar activities |

The contributions in paper 5 also informs RQ3, and provide a critical foundation for how we may understand both the benefits and challenges of hackathons, and for considering how and when to organise a hackathon. In other words, learning from the many different experiences of using hackathon formats in paper 5, we may begin moving beyond adopting a "typical" hackathon format, and instead adapt the format to better suit situated needs. After summing up the contributions to RQ2 in the following, I outline the contributions to RQ3.

## RQ2 Contributions

Paper 4 and paper 5 contribute with retrospective and contemporary *overviews*, respectively, of how game jams have been organised for supporting creativity, and of how hackathons have been organised within academia. Specifically, paper 4 contributes with a report on what game jam organisers find important when organising game jams, and a nuanced overview of how they understand creativity in terms of novelty, risk-taking, creativity constraints and combinational creativity. Additionally, we report different initiatives which the organisers believed support their participants' creativity. Paper 5 contributes with an overview of how hackathons have been used as means and as research focus in academia for: structuring learning, including teaching; structuring different processes, including for facilitating research in various ways; and for enabling participation, including engaging citizens and minorities in developing with technology. Another important contribution of paper 5 is the overview of researchers' experiences of benefits and challenges of hackathons. This contribution prepares the ground for how to critically reflect on when and how to use hackathons.

Providing and analysing empirical data on how people have organised and used game jams and hackathons before support the discussion on how game jam and hackathon formats may be organised and used reflectively and critically in the future. The outlined contributions of paper 1, 2, 4 and 5 so far points towards how some game jams and hackathons have been organised so far. However, the way in which the accelerated design processes have been organised may not always have been the most optimal, which especially paper 5, but also the insights from paper 1 and 2, indicate. In approaching RQ3, I outline how I have approached the question of how we may explore alternative ways of organising and using accelerated design processes.

## 7.3 RQ3

**How may we explore alternative ways of organising hackathons and game jams for academia and for supporting creativity?**

After conducting the studies for paper 1 and paper 2, I was inspired by the challenge of documenting accelerated design processes in order to leverage in situ knowledge. When a design process is accelerated it becomes important that documentation does not slow down the process, if the documentation of an *authentic* accelerated design process is desired. I therefore wanted to explore how to organise an academic game jam, where documentation would be implemented as part of the accelerated design process in an intentionally *non-intrusive* way. For this purpose, Co-notate, which paper 3 revolve around, seemed promising. My contribution to RQ3 is further informed by paper 5, which similar to paper 3 explores hackathons for academic purposes, and paper 4, which explores game jams and creativity. As RQ3 is more speculative and prospectively oriented, my PhD project does not provide exhaustive and ultimate answers to it. However, based on the contributions to RQ3 which are informed by the findings of paper 3, 4, and 5, I discuss how we can move towards a more critical perspective on when and how to organise accelerated design processes, such as game jams and hackathons, in Section 7.4.

In the following subsection I focus on the contributions of paper 3, since these contributions most directly approach RQ3. Subsequently, I discuss how paper 5 and paper 4 also inform RQ3.

### Paper 3

The contribution of paper 3 is an exploratory study of *how a real time annotation tool* provided to participants in an academic game jam can support the participants' documentation of their in situ reflections, and thereby enable the participants to pre-empt subsequent analysis. Based on the study, we discuss the potential that real time technology pose for *capturing, producing and communicating* knowledge in design research and practice, and compare real time technology with other kinds of similar tools for documenting design processes. Co-notate is distinguished from these other tools in the sense that it operates with user generated real time annotations from several people at the same time, and using a range of different tags.[8]

We distributed a survey to the participants, and interviewed some of the participants, in order to analyse their *experience* of using Co-notate in an academic game jam setting. Based on our findings, we argue that real time annotations may inform: "[...] mediated representations with the immediacy of situational knowledge for subsequent reflection and knowledge construction." (Rasmussen, Olesen, and Halskov 2019). Real time annotations may accommodate conventional documentation being *over- or misinterpreted*, or even *disregarded*, which we argue that non-annotated documentation may risk.

In the following I will highlight findings of the paper which concern the use of Co-notate in accelerated design processes.

8: For example, the tool Assistant (FirstAgenda 2020) uses automatically generated tags, and the tool ReflecTable (Hook et al. 2013) uses a binary approach to tagging, as it only registers whether the user has pushed a button or not.

**Annotating Accelerated Design Processes in Real Time**

From the thematic analysis of the data we found that certain *aspects* of Co-notate showed potential for documentation in accelerated design processes:

- ▶ **Supporting Capture of Immediate Impressions:** This aspect was particularly interesting regarding the documentation of accelerated design processes. Moments which in situ was assessed as having immediate importance, without the tagger necessarily being able to explain why, could be indicated with a more expressive, and less descriptive, tag. From our study, an example on this was a "Eureka!" tag used every time: "[...] something fell into place, every time something came together." (Interviewee 3, 50:33). Capturing immediate impressions like this may enable the annotation of impulses and moods, such as frustration, excitement, stagnation, or inspiration, without having to reflectively decide on a tag that described the moment as such. In this way, we observed a distinction between *expressive* and *descriptive* tags, where expressive tags may be particularly useful in an accelerated design process, where it can be even more difficult to take the time for in situ reflection.

- ▶ **Distributing Responsibility:** The act of tagging may be difficult in an accelerated design process, as participants can get absorbed in the process, and forget to tag interesting moments: "[...] because when the conversation was in an important moment, no one had time to tag." (survey). The participants reported many different strategies of tagging: Some groups distributed the responsibility of using specific tags between the group members; Some groups assigned one person to conduct all the tagging; One group chose a decentralised strategy to tagging, where all members of the group shared responsibility of tagging by using a single device dedicated to the tagging. This shared device may have entailed a sense of distributed awareness and shared responsibility of tagging among the group members, because the act of tagging become more visible. When everyone is more aware of sharing the responsibility of tagging, we hypothesise that the taggers can then become more immersed in the design process as well.

- ▶ **Instantly Searchable Documentation:** Though Co-notate is intended to support later analysis, the tool showed potential in settling discussions during the accelerated design processes as well. The annotated recordings are immediately searchable, which enables participants to quickly search for a specific, tagged moment. For example, revisiting earlier discussed ideas could serve as inspiration or clarify what was agreed upon.

- ▶ **Supporting Retrospective Analysis:** The above mentioned aspects of Co-notate indicate the potential for supporting the participants' reflections in situ. The participants also reported how Co-notate was helpful for their analysis of the accelerated design process afterwards. Some participants, however, reported difficulties regarding the post situ analysis due to usability issues, and differing tagging strategies resulting in inconsistently tagged data. Despite some participants experiencing issues with the tagged data for the post-situational analysis, we saw a potential for the real-time annotations as they invite participants to make cursory reflections in situ, and in this way *pre-empt* the subsequent analysis.

Besides some reported usability issues and difficulties in differing tagging strategies within groups, some participants found it challenging to *predefine appropriate tags* to use with Co-notate in their accelerated design processes. Some participants found that predefined tags focused their conversations during the accelerated design process, while others found that the predefined tags were restraining. Since the tags can be defined by the participants themselves it can be a balancing act to define appropriate tags before a design process without knowing how the design process will unfold. The challenge is then to define tags that are *specific* enough to describe in situ moments and thereby support later analysis, and at the same time define tags in a way so that they are *general* enough to be applicable in the moment. Instead of predefined tags, we suggest that real-time annotations tool such as Co-notate should allow for changing predefined tags and defining new tags during a design process.

We acknowledge that retrospective analysis of documented design processes may involve biases, however via real time annotations we argue that analysis can move closer to the messy situation of design processes and may thereby accommodate for biases. As expressed by a participant:

"We transcribed and used photo documentary of the Co-notate sessions. This worked very well for describing some of our key decisions and ideas flourishing right from the source individual spurring it form the top of his head. This way, we did not need to clumsily explain things backwards – everything was right there for the taking." (Survey.)

Based on our findings we therefore argue that real time annotations indicate potential in informing intermediate forms of knowledge by annotating mediated representations, such as audio and video recordings, with in situ and immediate impressions.[9]

**Future Research**

In addition to findings related to how future real-time annotation tools may be further explored in terms of the key issues we identified, we also discuss some directions for future research:

▶ Exploring real-time annotation tools in empirically driven fields of study and practice.
▶ Developing mobile applications of real-time annotation tools.
▶ Developing real-time annotation tools which combine manual tagging with automated capturing of situated knowledge.

These directions are mostly related to the further development of real-time annotation tools, and are not as such directly related to future research on accelerated design processes. However, as we have argued, real-time annotation tools have shown potential to be particularly suited for exploring accelerated design processes. Therefore, from that perspective, it is relevant in relation to documenting accelerated design processes to further explore real-time annotation tools.

## Paper 5

Both paper 3 and paper 5 contribute to the understanding of how hackathons can be organised in academic settings. Based on insights

9: In paper 3 we use the terms *intermediate forms of knowledge* and *mediated representations* as coined by Höök and Löwgren (Höök and Löwgren 2012). Intermediate forms of knowledge are: "[...] more abstracted than particular instances, without aspiring to be at the scope of generalised theories." (Höök and Löwgren 2012), and can for example be in the form of patterns, guidelines, annotated portfolios, methods and tools.

from reviewing 381 papers conducting research with and on hackathons, and from synthesising benefits and challenges of hackathons, we discuss some directions for future research with and on hackathons. We argue that since hackathon formats are not necessarily a suitable development format for all solutions (Frey and Luks 2016), it becomes key to *adapt* the hackathon format in order to challenge hackathon formats as a methodological hegemony.[10] How the hackathon format is then adapted and organised depends on for example: What kind of outcome is wanted, how important it is that the prototypes are turned into products with a greater degree of completeness, and in what context the hackathon is to be organised within. A hackathon format may entail a pace which is too fast to obtain deep understandings of certain topics, however the hands-on approach and rapid prototyping may be more suitable for exploring the first steps of going from abstract concepts and theories to concrete prototypes. The key examples selected for the broad overview of research with and on hackathons may then serve as a *reference work* where inspiration for different ways of adapting hackathon formats can be explored. In the above mentioned section on Page 73 I summed up some of these key examples.

Furthermore, the synthesis of the benefits and challenges of hackathons in Table 7.2 can make researchers and organisers aware of what benefits to emphasise and organise for, while looking out for how to accommodate for the challenges which hackathons can carry. Ultimately in the paper we call for more reflections on how hackathon formats are used and adapted to suit different purposes, as well as more emphasis on describing the particular hackathon circumstances research was conducted in. In this framework article, I take some of the first steps to unfold this discussion in Section 7.4.

**Future Research**

We believe that the overview of benefits and challenges of hackathons can be beneficial for future research with and on hackathons. In paper 5, we also highlight several examples which may serve as inspiration for how to adapt hackathon formats for different purposes. We call for future research with and on hackathons, to critically reflect on which kinds of outcomes can be expected to be achieved from hackathons. Specifically, we argue that studying the relation between how hackathons are organised and the outcomes is a valuable direction for future research.

## Paper 4

Paper 4 informs the contribution to RQ3, as we discuss how game jam organisers may advance their organisation of game jams in regards to supporting participants' creativity. In the paper we outline how organisers may explore alternative creativity research-based initiatives for supporting participants' creativity. The paper may furthermore serve as a basis for future research into the complex role of creativity in game jams. Because of the preliminary nature of the study in paper 4, the suggested initiatives for furthering the organisers' practices still need to be explored in the context of game jams, and we therefore suggest several directions for future research.

10: Avle, Lindtner and Williams discuss how certain methods, often originating from Silicon Valley and including hackathons, constitute a methodological hegemony: "[...] a dominating approach towards design and technology production." (Avle, Lindtner, and Williams 2017).

**Table 7.3:** The left column of the table shows the four creativity aspects, the middle column shows which creativity aspects the organisers' initiatives reflected, and the right column show how organisers may further their practices regarding novelty, risk-taking, creativity constraints and combinational creativity. We did not identify organiser initiatives specifically aimed at supporting combinational creativity, therefore the corresponding field is empty.

| Theoretical aspects | Organiser Initiatives | Research-based recommendations |
|---|---|---|
| Novelty | Providing Talks and Facilitating Discussion | Provide talks on: how everyone is capable of being creative, the underlying concepts of creativity and how these may be implied in practice (Onarheim and Friis-Olivarius 2013). |
| | Organising Activities | Encourage divergent thinking (Acar and Runco 2019; Finke, Ward, and Smith 1992; Guilford 1967) |
| | | Spend time on problem searching rather than problem solving (K. Sawyer 2013; Scott, Leritz, and Mumford 2004) |
| | | Introduce elements of surprise (Gerten and Topolinski 2019) |
| | | Unusual uses for common objects (K. Sawyer 2013) |
| Risk-Taking | Supplying Tools and Materials | Provide new and unfamiliar tools and materials to challenge prototyping processes |
| | | Explicitly communicating the value of taking risks (Dewett 2006) |
| | | Stand up check ins to verbally encourage taking risks during game jam (Sternberg 2010) |
| | | Provide prizes for most risky idea (Sternberg 2010) |
| | | Organise intermediate level competition (Baer et al. 2010) |
| | | In non-competitive game jams, consider change in group memberships (Baer et al. 2010) |
| Creativity Constraints | Establishing the Physical Surroundings | Consider how tools and materials may act as creativity constraints |
| | Selecting a Theme | Encourage participants to articulate their own constraints (Baker et al. 2020) |
| | Supplying Tools and Materials | Consider highly visually interesting environments (McCoy and Evans 2002; Richardson and Mishra 2018) |
| Combinational Creativity | | Ideation tool kits (Ho 2017) |
| | | Provide creativity exercises for combination (combining ideas which does not |
| | | fit together naturally, making analogies, meeting people unlike oneself) (K. Sawyer 2013) |
| | | Provide theme encouraging combination |

Table 7.3 shows an overview of the four creativity aspects, the organiser initiatives which reflected the four creativity aspects, and additional and alternative suggestions for initiatives based on insights from creativity research. For detailed descriptions and discussions of the initiative, I refer to paper 4.

## Future Research

Future research can further explore the relationship between the different attempts by the game jam organisers and how these support participants' creativity. We furthermore suggest several directions for future research on creativity in game jams in this regard. The directions are the relationship between creativity and the role of: "the time frame, the location (whether it is online, physical or a hybrid between online and physical), the available tools and materials, group forming, collaborative ideation, divergent and convergent processes, decision making, evolution and application of themes and topics as ideation prompts, the influence of stakeholders, how to support different experience levels of the participants, competitive and non-competitive game jam formats, group hierarchies and turn-taking." (Falk et al. 2021).

## RQ3 Contributions

Paper 3, 4 and 5 contribute to the discussion of how hackathons and game jams may be organised, particularly within academia and, furthermore, with the purpose of supporting creativity. Paper 3 contributes with the identification and analysis of benefits and challenges of using a real time annotation tool to capture in situ knowledge in accelerated design processes. The game jam we organised in paper 3 can be framed as an *adaption* of a game jam format with the purpose of exploring an academic game jam, in the sense of supporting the participants' reflections and knowledge generation. In paper 5 we broaden the perspective, and explore how other researchers have adapted hackathon formats for specific purposes as well. Paper 5 can then be used as a form of reference work for researchers wishing to conduct research with or on hackathons, as the paper contributes with a broad overview of how others have organised hackathons in academia as well as an overview of benefits and challenges of hackathons. Paper 4 contributes to the discussion of how game jams may be organised for supporting creativity, by discussing how the game jam organisers' already existing practices may be advanced with insights from recent creativity research in terms of the four creativity aspects.

Based on the contributions to RQ3, future research can explore the following research endeavours:

- ▶ Exploring real-time annotation tools in empirically driven fields of study and practice.
- ▶ Developing mobile applications of real-time annotation tools.
- ▶ Developing real-time annotation tools which combine manual tagging with automated capturing of situated knowledge.
- ▶ Exploring how to critically adapt hackathon formats for different purposes, including exploring how to support the benefits of hackathons formats while accommodating for the challenges of the formats.
- ▶ Studying the relation between how hackathons are organised and the outcomes.
- ▶ Exploring the relationship between organiser initiatives for supporting creativity in game jams, and participants' creativity.
- ▶ Exploring the role of several different factors, for example the time frame, and their impact on participants' creativity.

## 7.4 Discussion

By addressing the three research questions, my PhD project contributes with findings which cover and explore how: accelerated design processes can be understood; how game jam and hackathon formats for accelerated design processes have been organised and used before; and how we may explore alternatives for how to organise the formats for particular purposes. In this section I discuss some implications of the contributions from addressing RQ1, RQ2, and RQ3.

### Understanding

An implication of my PhD project is contributing to how accelerated design processes can be understood as a particular kind of design process. All design processes, accelerated or not, are always constrained somehow, and, in the words of Löwgren and Stolterman: "[...] there is only a certain amount of resources and time at the designer's disposal." (Löwgren and Erik Stolterman 2004). However, I argue that the degree of how much the limitation of resources (like the intended short time-frame, but also for instance the access to end-users) impacts accelerated design process is relatively *high* and *pervasive* compared to how this limitation impacts non-accelerated design process. This particularly *high* and *pervasive* impact of the imposed and intentionally limited resources of game jam and hackathon formats may influence participants' design judgement to prioritise certain more feasible design decisions over others, which would otherwise have been desirable. In other words, these particular design processes are then accelerated in the sense that tradeoffs and shortcuts are frequently taken in order to ensure prototype functionality in a timely manner. This impact of the formats which accelerates design processes can potentially lead to a *highly inward oriented focus on the immediate design situation of the format*, where the relatively immediate feasibility of prototypes is prioritised over thoroughly identifying and designing for authentic user needs.

While this observation is based on two case studies in paper 1 and paper 2, and can therefore not be generalised, the extensive review in paper 5 identify several researchers who voiced concerns about particularly hackathons, and how the kind of fast innovation for example risks devaluing: "[...] the slow work of creating infrastructures." (Irani and Silberman 2014). Continuing on this concern, Gama notes how developers in hackathons use their personal experience for creating technology which they think will be appealing to users (Gama 2017). Ferrario et al. raises another concern in regards to how knowledge is gained in hackathon formats (Ferrario et al. 2014). Hackathon participants' knowledge are often gained informally via for example representatives of target users at the hackathon venue. This can be problematic as it ignores formal social science methods for gaining deeper understandings of social issues (Ferrario et al. 2014). Irani also note how the pace in hackathons rarely left time for "real footwork" to build trust with citizens (Irani 2015). Especially paper 2 exemplifies how and why these concerning practices may unfold in a hackathon setting. Together, paper 1 and paper 2 *contextualise* how an accelerated design process unfold, and help solidify why these types of design processes may lead to short-lived outcomes.

Summarising from the insights from particularly paper 1, paper 2 and paper 5, accelerated design processes are accelerated in the sense that participants may *prioritise* for example: immediate and maybe even defect technical functionality as long as it serves the purpose of visualising an idea, and cursory and informal methods for obtaining immediate understanding. At the same time participants may for example disregard: formal and lengthy methods for obtaining deeper understanding, and deliberate technical development including ensuring technical security measures.

**Other Accelerated Design Processes: Design Sprints**

While I have only explored hackathon and game jam formats in my PhD project, the discussion of how we may understand accelerated design processes can be broadened to for example design sprints as well, which I would argue are formats for accelerating design processes. Contrary to game jams and hackathons, which are usually quite unstructured, design sprints are highly structured five-day design processes, where each day have a different goal and plan for design activities. Design sprints are meant for: "[...] answering critical business questions through design, prototyping, and testing ideas with customers." (Knapp, Zeratsky, and Kowitz 2016). As the quote suggests, the design sprint is meant for startups, companies and organisations. The first day of a design sprint is dedicated to build a map, or a focus, for the sprint days. This is about gathering as much information as possible "[...] while preventing the usual meandering conversations." (Knapp 2020a). This information is gathered within the team doing the design sprint, and from experts and guests from the outside, where each expert or guest is interviewed for fifteen to thirty minutes, according to the design sprint plan (Knapp 2020b). The second day is dedicated to solving the problem identified on the first day. It is also on this day potential customers will be recruited for the prototype test conducted on the fifth and last day of the sprint. The third day is dedicated to choosing one of the solutions generated during the second day. The fourth day is about building a prototype which can illustrate the solution and be used for the test during the fifth day. Finally, the solution, illustrated by a "fake-it-till-you-make-it" prototype (Knapp 2020a), is presented to and tested by the customers who were recruited on the second day of the design sprint.

From the description of the design sprint process, we could raise concerns similar to the ones pointed at hackathons: will the design sprint teams rely on their personal experiences and what they think will be attractive for target users? While the design sprint includes inviting *experts* to the design sprint team, they may still not be the actual target end users (or in their words, *customers*) but may be representatives, who possess a second-hand insight on end users. Furthermore, are three days enough to reach out to relevant customers and have them do a test of the prototype as well as an interview?

Accelerated design processes such as hackathons, game jams and design sprints have drawn a lot of attention, and are today used in many different contexts.[11] While there may be potential benefits, as identified and summarised in paper 5 and in Table 7.2, there are important concerns raised against accelerated design processes. Therefore, an important question which we should ask is then: When is it *appropriate* to accelerate design processes, and when should we *not* accelerate design processes? I discuss this in the section Organising on Page 86.

11: Design sprints have been used at companies such as: Slack, Uber, Airbnb, Medium, Dropbox, Facebook, McKinsey, IDEO, LEGO, the United Nations, the New York Times, etc. (Knapp 2020a).

**Non-Accelerated Design Processes: Soma Design**

In order to approach this question, we can first consider a design process which present itself as opposed to accelerated design processes. One example on such a design process is *soma design*, which I will briefly describe in the following. Soma design processes build on somaesthetic theory (Shusterman 2008; Höök, Eriksson, et al. 2019). For a soma designer, this entails a heightened focus on the first-person experience of their own body, bodily experiences and senses - their *somas* (Höök, Eriksson, et al. 2019). A part of soma design is acknowledging that these bodily experiences are all very different. In order for a soma designer to truly develop empathy and understanding for users, the designer must further acknowledge that they cannot experience being another body (the users), and must develop empathetic engagement with the users through *slow, long-lasting* partnerships. Furthermore, a soma designer "[...] must constantly question their own political position, identifying what preconceptions and ideals they are imposing on the users." (Höök, Eriksson, et al. 2019). These soma design aspects stand in contrast to accelerated design processes, where there is little, if at all any, time to for example develop and explore long-lasting partnerships with users. Instead, as exemplified in paper 2, accelerated design processes risk prompting designers to impose their own preconceptions on users, or what I have described as an inward oriented focus. Another key aspect of soma design which stand in contrast to accelerated design processes, is the encouragement to keep returning to exploring interactions with a design through a first-person engagement. This is also referred to as *staying in the undecided* to properly engage with the somatic experience of a design material and its aesthetic potential before deciding on a design concept (Tsaknaki et al. 2019). In contrast, in accelerated design processes there is a certain pressure and need to quickly decide on a design idea and then move into developing with design materials, which potentially leaves little if any time to exploring aesthetic possibilities. As such, accelerated design processes and soma design processes are two very different ways to design.

In order to begin comparing design processes which are respectively accelerated and slow, we can draw on a concept from cognitive psychology: dual-system theory (Kahneman 2011). Dual-system theory is a well-established model of human cognition, and describes two systems: *system 1* is fast, intuitive, and effortless thinking, and *system 2* is slow and analytical thinking, which requires greater cognitive effort (Kannengiesser and Gero 2019). System 1 is believed to govern much of our everyday behaviour and thereby constitute a large part of human reasoning (Kannengiesser and Gero 2019). System 1 thinking can, however, entail cognitive biases, as the system is "overconfident" (Kahneman 2011). Kannengiesser and Gero connects dual-system theory to design thinking and argues that system 1 thinking plays a role in parts of design processes (Kannengiesser and Gero 2019). They further argue that even though system 1 may entail cognitive biases, this way of thinking can be a tool for responding to uncertain and ambiguous situations, where analytical reasoning would be impossible or impractical (Kannengiesser and Gero 2019).

Based on this, I suggest a possible *hypothesis*, which would need further study: accelerated design processes potentially prompt more system 1 thinking, while slow design processes, such as soma design, strive for mostly system 2 thinking. While it is beyond the scope of my PhD project to answer this hypothesis and compare accelerated and non-accelerated design processes exhaustively, the dual-system theory may provide a framework for beginning to understand the differences of how for example design judgement is influenced in the two different kinds of design processes. I will now discuss some implications which my contributions may have on how we may more critically organise for accelerated design processes, and when the organisation of these design processes may be appropriate.

## Organising

An implication of my PhD project is how organisers of game jams and hackathons may consider when and how to organise these formats considering some of the challenges which accelerated design processes may entail and, for instance, how they may prompt an inward oriented focus for participants. Whether an inward oriented focus on the immediate design situation constitutes a concern which should be paid attention to or not depends on the *purpose* of the game jam or hackathon format. As I have described in this overview article, game jams and hackathons are employed for a wide range of purposes, where a highly inward oriented focus on the immediate design situation of the format may or may not prove problematic. In the following subsections, I reflect on this matter.

### When Accelerated Design Processes are Part of Longer Design Processes

Accelerated design processes may be incorporated as part of a longer design process, such as within an organisation or in a research process. If we assume that accelerated design processes may prompt mostly system 1 thinking (Kannengiesser and Gero 2019), accelerated design processes can then potentially be tools to quickly respond to uncertain and ambiguous situations, where analytical reasoning would be impractical. Accelerated design processes may then be appropriate when a hands-on and rapid prototyping approach is suitable for exploring the *first steps* of going from abstract concepts and theories to concrete prototypes. In those first steps of a longer design process, facilitating an accelerated design process could be used to quickly develop prototypes, which can then be explored and compared to each other.[12]

### When Outcomes are Valued in the Immediate Design Situation

The accelerated design processes which I studied in paper 1 and paper 2 largely resembled what we can frame as "typical" formats, in the sense that they follow a structure for game jams and hackathons which is frequently repeated. Some of the challenges regarding the limited sustainability and longevity of the developed outcomes, as described in Table 7.2, may not be a concern if the game jam or hackathon aim to encourage participants to create outcomes which have *value in the immediate design situation*. This could for example be a prototype with entertainment value created at a game jam, as was the case in paper 1, or in cases where the emphasis is on participants' process of speculation about possible futures manifested in experimental prototypes (Boehner and DiSalvo 2016). However, though a limited usefulness or short longevity of outcomes may not be an issue in some cases, there are still challenges in for instance *who* gets to be part of these accelerated design processes and who does not (DiSalvo, Gregg, and Lodato 2014). While there have been some attempts to rethink hackathon formats to be more inclusive, see for instance (Richard et al. 2015; Taylor, L. Clarke, et al. 2018), not much research have been conducted on how to adapt accelerated design processes to mitigate the challenges they face in terms of diversity.

### When Outcomes are Valued Outside the Immediate Design Situation

If the intention with organising hackathons or game jams is to create outcomes which have value *outside* of the immediate design situation of the format, then a potential inward orientation of the participants' accelerated design process is a matter of concern. This could for example apply to game jams where the desired outcomes are to be used in education, or outcomes with the purpose of solving real issues for communities. Many of the game jam and hackathon examples mentioned in the vignette in Chapter 4, which aim to help in different ways during the novel corona virus pandemic, are examples on where the outcomes are expected, or at least hoped, to have value outside the immediate design

12: An example on this, are the game jams by DoubleFine, mentioned in Chapter 1, where the game jam outcome which win the popular vote may be developed into a commercial game afterwards (Wallace 2017).

situation. When it is highly important that the developed outcomes of a design process are truly and deeply aligned to meet users and their needs, *typical* accelerated design processes can be problematic. In this context, long-term participatory design processes or soma design may be more preferable. Repeating a game jam or hackathon format without considering and accommodating for the challenges of accelerated design processes risk contributing to a methodological hegemony (Avle, Lindtner, and Williams 2017). Instead of repeating sub-optimal formats on the one hand or, on the other hand, dismissing these formats completely, we should instead critically reflect on how the formats can be adapted and tailored for situated needs, as there still are some benefits related to the formats. Some of the most important benefits, in my opinion, being the formats' potential to invite and engage a broad audience to explore technology and imagine possible futures, as argued by (Boehner and DiSalvo 2016).

### When Short-Term Design Process Engagement Can Be a Benefit

Accelerated design processes do not necessarily require long-term engagement from participants. This can certainly be a drawback in some situations, such as when it is important to truly and deeply engage with users' lived experiences (Höök, Eriksson, et al. 2019) and develop sustainable and long-lasting outcomes. However, this short-term commitment can, in some cases, be a benefit as well. In the cases where there is public access to participating in a game jam or hackathon, participants, who have never developed technology before and are curious on technology development, can explore technology and participate in imagining possible futures with the technology. Participants can also experience what it means to engage in a design process of generating ideas and attempt to manifest those ideas into prototypes.[13] In these cases, participants do not have to engage in long-term processes, such as pursuing an education or professional career, to experience technology development or a design process. An accelerated design process may also be a first step for some participants in pursuing such longer-term processes, as is the case for some game jam participants. This reflect the promise of the maker movement, that accelerated design processes can potentially contribute to democratising technology development. However, even in the maker movement this democratisation is not a given (Kaiying and Lindtner 2016), and, given the reported issues on diversity and inclusion in hackathons, accelerated design processes have to be carefully organised to be able to realise this potential. Additionally, while accelerated design processes can provide insights into some areas of designing and developing technology, they are still a particular kind of design process, and cannot directly be compared to how designing and developing technology in non-accelerated design processes happens. This is important to stress in contexts where accelerated design processes are used as ways to expose participants to what it means to design in other non-accelerated design processes. An example on this could be the use of game jams and hackathons in education to prepare students for the practical aspects of designing games and interaction design in industry. In these cases, it is important to clearly communicate the similarities, differences, benefits and challenges between these accelerated and non-accelerated design processes.

13: This is for example the case in professional game development, where much secrecy clouds and mystifies how games are developed (O'Donnell 2014), however game jam can be a way of demystifying at least some aspects of what it means to develop a game.

**When Participation is a Concern**

While hackathons and game jams can be opportunities for some participants to try out designing and developing technology and games for the first time, there are still ethical challenges in terms of who gets to participate, who does not get to participate (DiSalvo, Gregg, and Lodato 2014), as well as how people get to participate. There have been some examples on hackathons which in different ways seek to deliberately invite or recruit certain participants, who can be framed as end-users in relation to the theme of the hackathon. One example on a hackathon where certain participants were deliberately invited is the hackathon by Kopéc et. al. who invited older adults to participate in teams of young programmers (Kopeć et al. 2018). However, because there were several issues regarding the collaboration between the older adults and the young programmers, Kopéc et. al. reflect that it is not enough to only invite a target group to participate in a hackathon. The reason for this being that it cannot be assumed that all participants, whether they are end-users or developers, can participate on equal terms (Kopeć et al. 2018). End-users needs to not only be invited as an input to the hackathon, but to be actively supported and perceived as participants with valid insights during the process of a hackathon, if full collaboration and participatory design is envisioned. Two examples on hackathons which also invited certain participants, but facilitated several initiatives in order to support their participation, are the hackathons by Birbeck et. al. (Birbeck et al. 2017) and Hope et. al (Hope et al. 2019). The hackathon themes in these two cases revolved around self-harm (Birbeck et al. 2017), and breastfeeding (Hope et al. 2019), and the participants had some kind of connection to these two themes. In both of these cases, the invited people were supported during the hackathon through different initiatives.[14] These three examples on hackathons each illustrate different ways of participation. An important insight here is that participation is not only a question of *who* participates, but also *how* they participate. These insights point towards similar discussions in the research field of participatory design, where questions of ethics and participation have also been a topic for discussion (Robertson and Wagner 2012).

Looking towards participatory design can be a worthwhile endeavour for those organising for and researching diverse participation in accelerated design processes, since many of the concerns regarding participation in accelerated design processes relate to concerns about how participatory design are carried out in general today. For example, Bødker and Kyng have discussed how participatory design projects today are problematic as they often have a focus on here-and-now collaborations for here-and-now purposes, without much consideration towards long-term, sustainable outcomes such as learning and empowerment (S. Bødker and Kyng 2018). Particularly regarding how people participate in participatory design projects, Bossen et.al. argue how user involvement revolves around tapping their ideas and knowledge in a one-way process: "Requirements are elicited from, usability tested upon, and systems are delivered to users" (Bossen, Dindler, and Iversen 2012). While it may be difficult to improve for example the longevity of the direct hackathon and game jam outcomes (developed prototypes),[15] matters of participation should be payed attention to, as accelerated design processes in some cases can still present a stepping stone for something else for those who do participate.[16]

14: In the case of the hackathon by (Birbeck et al. 2017), the organisers among other things consulted with local mental health organisations to ensure the creation of a sensitive and tactful space. One initiative was to provide a space for participants who would need to withdraw if they for example became upset because of the particular hackathon theme (self-harm).

15: For the matter of sustainability in terms of hackathon outcomes, there are some notable recent attempts on hackathon formats where the authors have tried to approach this matter in different ways: (Webb et al. 2019; Gama et al. 2019; Chan et al. 2020).

16: A couple of examples on this could be: the development of proof of concept-prototypes which would need further iteration after the accelerated design process; the invitation of people, who have not previously had the chance to imagine with and develop technology, to do exactly that; inspiration for people to pursue an education or a career dealing with the development of technology or games; the development of greater collaboration skills.

While the above-mentioned sections discuss some situations where accelerated design processes could be appropriate, most of these situations still call for some adaptions to how typical game jam and hackathon formats are organised, in order to mitigate the challenges which the formats entail.

## Researching

A third implication of my PhD project's contribution is how accelerated design processes may be researched specifically in order to leverage in situ knowledge on how the game jams and hackathons accelerates design processes. In that perspective, further developing and exploring the real-time annotation tools can be a valuable endeavour.

As game jams and hackathons are being used in many different contexts today, I argue that it is a worthwhile pursuit to further research for example how game jams and hackathons promote the prioritising of certain design decisions over others, and how they are distinguished from non-accelerated design processes. Through the summarised contributions of my PhD project, I have outline several research questions and directions for future research on accelerated design processes. While my PhD project contributes substantially to how we may understand accelerated design processes, only preliminary steps were taken to explore alternative ways of organising hackathons and game jams. Exploring alternative ways of organising hackathons and game jams is an important endeavour, as these formats do entail some promising potentials, however their potential democratising aspects cannot be assumed.

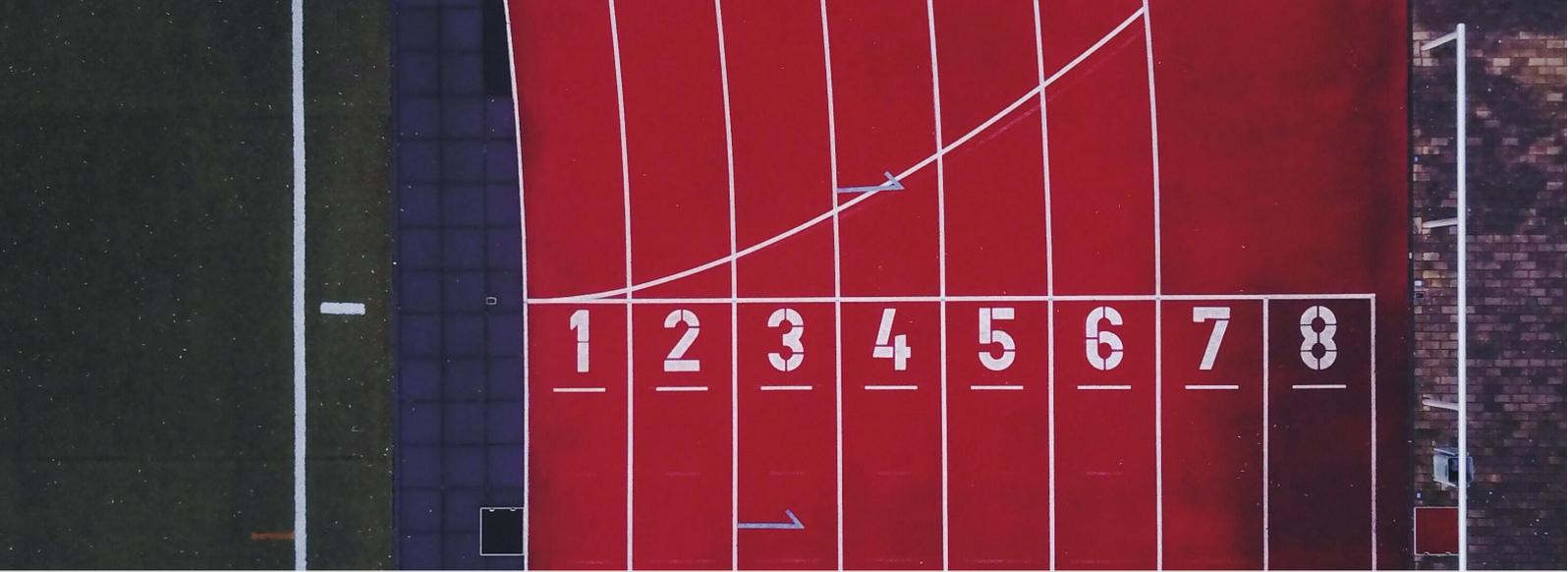

# 8   CONCLUSION

This chapter concludes the overview article, and summarises each chapter of the article. After this chapter and the bibliography of the overview article, the five papers are appended in chronological order.

## 8.1  Introduction

In Chapter 1 on page 7 I introduced the overarching research questions of the PhD project:

**RQ1: How can we understand the accelerated design processes during hackathons and game jams?**

**RQ2: How have people organised hackathons and game jams in academia and for supporting creativity?**

**RQ3: How may we explore alternative ways of organising hackathons and game jams for academia and for supporting creativity?**

I introduced the motivation for the research questions which is based on the observation that only few research contributions engage with a processual focus on the inner workings of accelerated design processes, despite an increase in research interest in general into game jams and hackathons. The proposition of my PhD project is based on the hypothesis that a profound and meticulously developed understanding of how game jam and hackathon formats accelerate design processes, provide a base for better adapting and utilising game jam and hackathon formats for specific purposes. I furthermore introduced how I understand hackathons and game jams, and how I define *accelerated design processes*.

## 8.2  Overview of Appended Papers

In Chapter 2 on page 16, I provided an overview of the five appended papers:

Photo by Marvin Ronsdorf on Unsplash

▶ **Paper 1: The Dynamic Design Space During a Game Jam**, Jeanette Falk Olesen and Kim Halskov
▶ **Paper 2: Four Factors Informing Design Judgement at a Hackathon**, Jeanette Falk Olesen, Nicolai Brodersen Hansen, and Kim Halskov
▶ **Paper 3: Co-notate: Exploring Real-time Annotations to Capture Situational Design Knowledge**, Søren Rasmussen, Jeanette Falk Olesen, and Kim Halskov
▶ **Paper 4: How Organisers Understand and Promote Participants' Creativity in Game Jams**, Jeanette Falk Olesen, Michael Mose Biskjær, Annakaisa Kultima, and Kim Halskov
▶ **Paper 5: 10 Years of Research With and On Hackathons**, Jeanette Falk Olesen, and Kim Halskov

I summarised the research questions for each paper, the main contributions of the papers, and how each research question was approached in the papers.

## 8.3  Research Environment & Field

In Chapter 3 on page 22, I outlined the research environment I have worked in, on a day-to-day basis, as well as the research field my work has been conducted within and contributing to. The research environment has a strong profile in regards to design processes, and I have benefited from this in terms of inspiration, feedback, discussions, and collaborations. I outlined the research field as consisting of mainly HCI and Interaction Design. My PhD project is within the overlapping area of HCI and Interaction Design, where the focus is on how people design and develop technologies, with a specific focus on game jams and hackathons.

## 8.4  Related Work

In Chapter 4 on page 25, I outlined *maker culture* and the research field *game studies* as relevant contexts for my PhD project. Maker culture has been commended as a democratising movement, where "[...] anyone can make." (Hatch 2013), but has at the same time been criticised for an underlying techno-solutionism, for example. This praise as well as criticism has been pointed towards hackathons as well, as a typical maker activity, and I argue that game jams reflect much of the same aspects.

Particularly relevant in regards to game jams, I outline game studies. Game studies has traditionally focused on the notions of game, player, and playing, while notions of design, designer, process, and practices have been overlooked. My PhD project contributes to this overlooked area of game studies.

I conclude the chapter with a cursory review of research on hackathons and game jams in order to provide a more complete picture on the state on prior work. Based on this, I argue that it should be equitable to claim that the state of the context of research on hackathons and game jams generally are between nascent and intermediate theory. This has implications for the methodology of the PhD project, which is generally exploratory and qualitative. I furthermore point out that few papers from the cursory review engage with the question of how game jam and hackathon formats shape participants' design processes.

## 8.5 Theoretical Foundation

In Chapter 5 on page 37 I describe the theoretical foundation of my PhD project, which is based on mainly Schön's pragmatist design theory. Via several theoretical concepts from pragmatist design theory, I unfold how I understand design processes, including the notion of design spaces. I furthermore unfold how I define creativity during my PhD project which is based on Plucker, Beghetto, and Dow's definition: "Creativity is the interaction among *aptitude, process and environment* by which an individual or group produces a *perceptible product* that is both *novel and useful* as defined within a social context." (Plucker, Beghetto, and Dow 2004). From creativity research I mainly draw on the notion of creativity constraints, defined as: "[...] all explicit or tacit factors governing what the agent/s must, should, can, and cannot do; and what the output must, should, can, and cannot be" (Onarheim and Biskjær 2013).

I concluded the chapter by summarising how the theoretical foundation, in terms of pragmatism, design space, and creativity constraints, relates to the five papers.

## 8.6 Research Methodology

In Chapter 6 on page 48 I discuss the methodology of my PhD project. The methodology was informed by several factors: The research environment, the related work, and the theoretical foundation. The research environment provided, as I described above, inspiration, feedback, discussions, and collaborations. The related work inspired and motivated an exploratory and qualitative approach. Building on pragmatist design theory as the theoretical foundation entailed a worldview where the world can be seen as a dynamic and emerging phenomenon, which is never stable. This worldview further entails that human cognition and activity are deeply situated. Therefore, I have used a mix of qualitative and exploratory methods which engaged with accelerated design processes *in situ*. This generally describes the methods used in paper 1, 2, and 3. In order to balance the findings and provide a better understanding of accelerated design processes, I broaden the perspective in paper 4 and 5, and move from the situatedness of accelerated design processes to a top-down perspective, focusing on reported conceptualisations and usage of game jams and hackathons.

First, I described thematic analysis which is how I generally have approached the analysis of data, as it is a flexible analysis method. Secondly, I described the different methods of the methodology: autobiographical design case studies, intervention, survey studies, and literature review. I conclude the chapter with summarising advantages and limitations of the methodology.

## 8.7 Contributions

In Chapter 7 I sum up the contributions of the five papers, and discuss how they relate to the three research questions. I conclude the chapter with a discussion of some implications of my PhD project's contributions for how to understand, organise and research accelerated design processes in hackathons and game jams.

**RQ1: How can we understand the accelerated design processes during hackathons and game jams?**

Paper 1 and paper 2 contribute with *descriptive and detailed process-level knowledge* on how two accelerated design processes was shaped by the formats of respectively a game jam and a hackathon. The two papers serve as *exemplars* on accelerated design processes, which contextualise and solidify how a design process may be accelerated in a game jam and hackathon format. This is an important contribution of the PhD project, as it contributes to a foundation for understanding accelerated design processes. Specifically, we observed how group composition entail particular *preferences, expectations, domain knowledge*, and *technical knowledge* which affect what is feasible for the group to achieve within the short time frame of a hackathon or a game jam. Furthermore, the two papers demonstrate how the constrained format can entail a *heightened focus on the situation of the game jam or hackathon*, or in other words an *inward oriented focus*, and not necessarily a focus on impacts beyond the immediate design situation, because the limitation of resources such as time, tools and materials are *particularly pervasive.* This has impact for what can be expected to be designed in similar game jam and hackathon formats as the ones we studied.

Based on the contribution to RQ1, future research can explore the following research questions:

- ▶ How do the formats of respectively accelerated and non-accelerated design processes shape how a design space is transformed?
- ▶ Which creativity constraints occur in accelerated design processes and how do they shape strategies of navigating design spaces?
- ▶ How is the idea generation phase impacted by an almost immediately subsequent prototyping phase?
- ▶ How do different kinds of prototyping impact the preceding idea generation phase?
- ▶ Does incentives for continuing the development of prototypes after hackathons have an impact on participants' idea generation and prototyping?
- ▶ If not, how can accelerated design processes be organised to support and ensure longer-term effects?

**RQ2: How have people organised hackathons and game jams in academia and for supporting creativity?**

Paper 4 and paper 5 contribute with retrospective and contemporary *overviews*, respectively, of how game jams have been organised for supporting creativity, and of how hackathons have been organised within academia. Specifically, paper 4 contributes with a report on what game jam organisers find important when organising game jams, and a nuanced overview of how they understand creativity in terms of novelty, risk-taking, creativity constraints and combinational creativity. Additionally, we report different initiatives which the organisers believed support their participants' creativity. Paper 5 contributes with an overview of how hackathons have been used as means and as research focus in academia for: structuring learning, including teaching; structuring different processes, including for facilitating research in various ways; and for enabling participation, including engaging citizens and minorities in developing with technology. Another important contribution of paper 5 is the overview of researchers' experiences of benefits and challenges of hackathons. This contribution prepares the ground for how to critically reflect on when and how to use hackathons.

**RQ3: How may we explore alternative ways of organising hackathons and game jams for academia and for supporting creativity?**

Paper 3, 4 and 5 contribute to the discussion of how hackathons and game jams may be organised, particularly within academia and, furthermore, with the purpose of supporting creativity. Paper 3 contributes with the identification and analysis of benefits and challenges of using a real time annotation tool to capture in situ knowledge in accelerated design processes. The game jam we organised in paper 3 can be framed as an *adaption* of a game jam format with the purpose of exploring an academic game jam, in the sense of supporting the participants' reflections and knowledge generation. In paper 5 we broaden the perspective, and explore how other researchers have adapted hackathon formats for specific purposes as well. Paper 5 can then be used as a form of reference work for researchers wishing to conduct research with or on hackathons, as the paper contributes with a broad overview of how others have organised hackathons in academia as well as an overview of benefits and challenges of hackathons. Paper 4 contributes to the discussion of how game jams may be organised for supporting creativity, by discussing how the game jam organisers' already existing practices may be advanced with insights from recent creativity research in terms of the four creativity aspects.

Based on the contributions to RQ3, future research can explore the following research endeavours:

- ▶ Exploring real-time annotation tools in empirically driven fields of study and practice.
- ▶ Developing mobile applications of real-time annotation tools.
- ▶ Developing real-time annotation tools which combine manual tagging with automated capturing of situated knowledge.
- ▶ Exploring how to critically adapt hackathon formats for different purposes, including exploring how to support the benefits of hackathons formats while accommodating for the challenges of the formats.

▶ Studying the relation between how hackathons are organised and the outcomes.

▶ Exploring the relationship between organiser initiatives for supporting creativity in game jams, and participants' creativity.

▶ Exploring the role of several different factors, for example the time frame, and their impact on participants' creativity.

I furthermore discussed some implications of the contributions of my PhD project in terms of three overarching themes: Understanding, Organising and Researching. In discussing how we may understand accelerated design processes, I also discuss design sprints as a form of accelerated design process, and soma design as a form of non-accelerated design process. In reflecting on when to organise accelerated design processes, I unfolded this discussion in terms of several different considerations:

▶ When Accelerated Design Processes are Part of Longer Design Processes

▶ When Outcomes are Valued in the Immediate Design Situation

▶ When Outcomes are Valued Outside the Immediate Design Situation

▶ When Short-Term Design Process Engagement Can Be a Benefit

▶ When Participation is a Concern